\documentclass[useAMS,usenatbib]{mn2e}

\usepackage{times}  % to get nice fonts for PDF
\usepackage{listings}
\usepackage{graphics}
\usepackage{subfigure}
\usepackage{graphicx}
\usepackage{amssymb}
\usepackage{hyperref}
\usepackage{aas_macros}
\usepackage{lscape}
\usepackage{rotating}

\def\simprop{ \lower .75ex \hbox{$\sim$} \llap{\raise .27ex \hbox{$\propto$}} }

\title[The CO emission of galaxies]{
Predictions for the CO emission of galaxies from a coupled simulation of galaxy formation and 
photon dominated regions}
\author[Claudia del P. Lagos]{
\parbox[t]{\textwidth}{
\vspace{-1.0cm}
Claudia del P. Lagos$^{1}$, 
Estelle Bayet$^{2}$,
Carlton M. Baugh$^{1}$,
Cedric G. Lacey$^{1}$,
Tom A. Bell$^{3}$,
Nikolaos Fanidakis$^{4}$, 
James E. Geach$^{5}$
}
\vspace*{6pt} \\
$^{1}$Institute for Computational Cosmology, Department of Physics,
University of Durham, South Road, Durham, DH1 3LE, UK.\\
$^{2}$Department of Physics, University of Oxford, Denys Wilkinson Building, Keble Road, Oxford OX1 3RH, UK.\\
$^{3}$Centro de Astrobiología (CSIC/INTA), Instituto Nacional de T\'ecnica Aeroespacial, 28850 Torrejón de Ardoz, Madrid, Spain\\
$^{4}$Max-Planck-Institut fuer Astronomie, Koenigstuhl 17, 69117 Heidelberg, Germany.\\
$^{5}$Department of Physics, McGill University, Ernest Rutherford Building, 3600 Rue University, Montreal, Quebec H3A 2T8, Canada.
\vspace*{-0.5cm}}

\begin{document}

%\date{Accepted ???. Received ???; in original form ???}

\pagerange{\pageref{firstpage}--\pageref{lastpage}} \pubyear{2011}

\maketitle

\label{firstpage}

\begin{abstract}
We combine the galaxy formation model 
{\texttt{GALFORM}} 
with the Photon Dominated Region code {\tt UCL$_{-}$PDR} to 
study the emission from the rotational transitions of $^{12}\rm CO$ (CO) in galaxies 
from $z=0$ to $z=6$ in the $\Lambda$CDM framework. {\texttt{GALFORM}} is used to predict 
the molecular (H$_2$) and atomic hydrogen (HI) gas contents  
of galaxies using the pressure-based empirical star formation relation of Blitz \& Rosolowsky.
From the predicted H$_2$ mass and the conditions in the interstellar medium, we 
estimate the $\rm CO$ emission {in the rotational transitions 1-0 to 10-9} 
by applying the {\tt UCL$_{-}$PDR} model to each galaxy. We find that 
deviations from the Milky-Way CO-H$_2$ conversion factor come mainly from 
variations in metallicity, and in the average gas and star formation rate surface densities. 
In the local universe, the model predicts a CO$(1-0)$ luminosity function (LF), 
 CO-to-total infrared (IR) luminosity ratios for multiple CO lines and 
a CO spectral line energy distribution (SLED)
which are in good agreement with observations of luminous and ultra-luminous IR galaxies.
At high redshifts, the predicted CO SLED of the brightest IR galaxies 
reproduces the shape 
and normalization of the observed CO SLED. 
The model predicts little evolution in the CO-to-IR luminosity ratio
for different CO transitions, in good agreement with observations up to $z\approx 5$. 
We use this new hybrid model to explore the potential of using colour selected 
samples of high-redshift star-forming galaxies to characterise the evolution of the 
cold gas mass in galaxies through observations with the Atacama Large Millimeter Array.
\end{abstract}


\section{Introduction}

The connection between molecular gas and star formation (SF)  
is a fundamental but poorly understood 
problem in galaxy formation. 
Observations have shown that the star formation rate (SFR) correlates with 
the abundance of cold, dense gas in galaxies, suggesting that 
molecular gas is needed to form stars. A variety of observational evidence supports this conclusion, 
such as the correlation between the  
surface densities of SFR and
$^{12}\rm CO$ (hereafter CO) emission and between the CO and infrared (IR) luminosities (e.g. \citealt{Solomon05}; 
\citealt{Bigiel08}). 
The CO luminosity traces dense gas
in the interstellar medium (ISM), which is dominated by
molecular hydrogen (H$_2$). The IR luminosity approximates the total luminosity emitted by
interstellar dust, which, in media that are optically thick to ultraviolet (UV) radiation, 
is expected to correlate closely with the SFR in star-forming galaxies. 

In the local Universe, high-quality, spatially resolved 
CO data show a tight and close to linear
correlation between the surface density of the SFR
and the surface density of CO emission, that
extends over several orders of magnitude and in very different environments: 
from low-metallicity, atomic-dominated gas to high-metallicity,
molecular-dominated gas (e.g. \citealt{Wong02}; \citealt{Leroy08}; Bigiel at al.,
2008, 2011; \citealt{Schruba11}; \citealt{Rahman11}). This suggests 
that SF proceeds in a similar way in these different environments. Support for this also comes 
from the IR-CO luminosity relation, with 
 high-redshift submillimeter galaxies (SMGs) and
quasi-stellar objects (QSOs) 
falling on a similar relation to luminous IR galaxies
(LIRGs) and ultra-luminous IR galaxies (ULIRGs) in the local
Universe (e.g. \citealt{Solomon97}; \citealt{Scoville03}; \citealt{Greve05};  \citealt{Tacconi06};
\citealt{Evans06}; \citealt{Bertram07}; \citealt{Bayet09b}; \citealt{Tacconi10};
\citealt{Genzel10}; \citealt{Riechers10}; \citealt{Daddi10}; \citealt{Geach11}; \citealt{Ivison11}; 
\citealt{Bothwell12}). 
\citet{Bayet09b} studied the correlation between different CO transitions and IR luminosity 
 and found that this correlation holds even up to $z\approx 6$.

The CO emission from galaxies is generally assumed to be a good indicator of molecular gas content. 
However, to infer the underlying H$_2$ mass from CO luminosity 
 it is necessary to
address how well CO traces H$_2$ mass. This relation is usually parametrised 
by the conversion factor, $X_{\rm CO}$,
which is the ratio between the H$_2$ column density and the integrated CO line intensity.
Large efforts have been made observationally to determine 
the value of $X_{\rm CO(1-0)}$ for the CO$(1-0)$ transition, and it 
has been inferred directly in a few
galaxies, mainly through virial estimates and measurements of dust column density.
Typical estimates for normal spiral galaxies lie in the range 
$X_{\rm CO(1-0)} \approx (2-3.5) \times 10^{20} \rm cm^{-2} / K\,km\,s^{-1}$ (e.g.
\citealt{Young91}; \citealt{Boselli02}; \citealt{Blitz07}; \citealt{Bolatto11}). However, 
systematic variations in the value of $X_{\rm CO(1-0)}$ have been inferred in galaxies whose ISM conditions 
differ considerably from those in normal spiral galaxies, 
favouring a larger $X_{\rm CO(1-0)}$ in low-metallicity galaxies and 
a smaller $X_{\rm CO(1-0)}$ in starburst galaxies (e.g. 
\citealt{Leroy07}, 2011; \citealt{Magdis11}; see \citealt{Solomon05} for a review).

Theoretically, most studies of $X_{\rm CO}$ are based on 
Photon Dominated Region (PDR; e.g. \citealt{Bell06})
or large velocity gradient (LVG; e.g. \citealt{Weiss05}) models.
Such models have been shown to be an excellent theoretical tool, 
reproducing the emission of various
chemical species coming from regions where the CO emission dominates
(i.e. in giant molecular clouds, GMCs, where most of the gas is in the atomic
or molecular phase, with kinetic temperatures typically below $100$~K,
and densities ranging from  $10^3\,\rm cm^{-3}$ to $10^5\,\rm cm^{-3}$). 
These models have shown that $X_{\rm CO}$ can vary considerably
with some of the physical conditions in the ISM, such as gas metallicity,
interstellar far-UV (FUV) radiation field and 
column density of gas and dust (e.g. \citealt{Bell07}; Bayet et al. 2012, in prep.).

Recently, large efforts have been devoted to 
measuring the CO emission in high-redshift galaxies. 
 These observations 
have mainly been carried out for higher CO rotational transitions 
 (e.g. \citealt{Tacconi10}; \citealt{Riechers10}; \citealt{Genzel10}; 
\citealt{Daddi10}; \citealt{Geach11}). Thus, in order to 
estimate molecular gas masses from these observations, a connection to 
the CO$(1-0)$ luminosity is needed, as expressed through the
$\rm CO(J\rightarrow J-1)/\rm CO(1-0)$ luminosity ratio.
The latter depends upon the excitation of the CO lines and the conditions in the ISM, 
and is therefore uncertain.
The study of several CO transitions, as well as other molecular species,
has revealed a 
wide range of ISM properties that drive large differences in the excitation
levels of CO lines in different galaxy types.
Through comparisons with PDR and LVG models, 
a broad distinction has been made
between the ISM in normal star-forming galaxies, in starburst-like galaxies,
and most recently, the ISM excited by radiation from active galactic nuclei (AGN),
suggesting large differences in gas temperature 
(e.g. \citealt{Wolfire03}; \citealt{Meijerink07}; \citealt{VanderWerf10}; \citealt{Wolfire10}; 
\citealt{VanderWerf11}; \citealt{Bayet11}).
\citet{Ivison11} show 
that large uncertainties are introduced in the study of the
CO-to-IR luminosity ratio of galaxies when including measurements which rely on an assumed 
$\rm CO(J\rightarrow J-1)/CO(1-0)$ luminosity ratio.

In recent years, efforts have also been made to 
develop a theoretical framework 
in a cosmological context to understand the relation between cold, dense gas, SF 
and other galaxy properties (e.g. 
\citealt{Pelupessy06}; \citealt{Gnedin09}; Narayanan et al. 2009, 2012; \citealt{Bournaud10}). 
In particular, a new generation of semi-analytical models of galaxy formation
have implemented improved  recipes for SF which use more physical 
descriptions of the ISM of galaxies (\citealt{Dutton09}; \citealt{Fu10};
\citealt{Cook10}; \citealt{Lagos10}). This has allowed a major step forward in the understanding  
of a wide range of galaxy properties, including gas and stellar contents and their 
scaling relations (e.g. \citealt{Lagos11}; \citealt{Kauffmann12}). Lagos et al. (2011a) 
presented simple predictions for CO emission, based on assuming a constant conversion factor 
between CO luminosity and H$_2$ mass, and 
successfully recovered the $L_{\rm CO(1-0)}/L_{\rm IR}$ ratio in normal and starburst 
galaxies from $z=0$ to $z=6$.

Despite this progress, a crucial step in the comparison between observations 
and theoretical predictions is missing: 
a physically motivated CO-H$_2$ conversion factor, $X_{\rm CO}$. 
Hydrodynamical cosmological simulations have successfully included
the formation of CO, as well as H$_2$ (e.g. \citealt{Pelupessy09}), but their 
high-computational cost does not allow a large number of galaxies
spanning a wide range of properties to be simulated to assess 
the origin of statistical relations such as that between the CO and IR luminosities. 
\citet{Obreschkow09d} presented a simple phenomenological
model to calculate the luminosities of different CO transitions, based 
on a calculation of the ISM temperature depending on the surface density of SFR
and the AGN bolometric luminosity, under the assumption of local thermodynamic equilibrium. 
However, this modelling introduces several
extra free parameters which, in most cases, are not well constrained
by observations. 

In this paper we propose a theoretical framework to statistically study the connection between 
CO emission, SF and H$_2$ mass based on a novel approach which combines a 
state-of-the-art semi-analytic model of galaxy formation with a {single gas phase} PDR model 
of the ISM which outputs the chemistry of the cold ISM. From this hybrid model 
 we estimate the CO emission in different transitions using the predicted molecular 
content, gas metallicity, UV and X-ray radiation fields 
in the ISM of galaxies, attempting to include as much of the physics determining $X_{\rm CO}$ 
as possible. {The underlying assumption is that all molecular gas is locked up in GMCs.} 
Although inferences from observations indicate that galaxies have some diffuse H$_2$  
{ in the outer parts of GMCs that is not traced by the CO emission} (e.g. \citealt{Reach94}; \citealt{Grenier05}), 
it has been suggested theoretically that this gas represents a constant 
correction of $\approx 0.3$ over a large range of media conditions \citep{Wolfire10}. We therefore 
do not attempt to model this diffuse component in this paper and focus on { the inner part of 
the PDRs exclusively, where there is CO emission}.

We show in this paper that by coupling a PDR model with the predictions of a galaxy formation model, we
are able to explain the observed CO luminosity in several CO transitions and its dependence 
on IR luminosity. The theoretical framework presented in this paper 
will help the interpretation of CO observations with the current and next 
generation of millimeter telescopes, such as 
the Atacama Large Millimeter Array\footnote{{\tt http://www.almaobservatory.org/}} (ALMA),
the Large Millimeter Telescope\footnote{{\tt http://www.lmtgtm.org/}} (LMT) and the new configuration
of the Plateu de Bureau Interferometer\footnote{{\tt http://www.iram-institute.org/EN}} (PdBI). 
These instruments will 
produce an unprecedented amount of data, helping to statistically assess
the cold gas components of the ISM in both local and high-redshift galaxies. 
 
This paper is organised as follows. In $\S 2$ we present the galaxy formation and the PDR models used 
and describe how we couple the two codes to predict the CO emission in galaxies.
$\S 3$ presents the predicted CO$(1-0)$ emission of galaxies in the local universe, and its relation 
 to other galaxy properties, and compares with available observations. 
$\S 4$ is devoted to the study of the emission of multiple CO lines in the local and high-redshift universe, 
i.e. the CO spectral line energy distribution (SLED), 
how the CO emission relates to the IR luminosity and how this depends on selected 
physical ingredients used in the model.
{In $\S 5$ we analyse the assumptions of the PDR model and how these affect the 
predictions for the CO luminosity.}
In $\S 6$, we focus on the ALMA science case for measuring the cold gas content of high-redshift 
star-forming galaxies to illustrate the predictive 
power of the model. We discuss our results and present our
conclusions in $\S 6$. In Appendix~\ref{App:COIR}, we describe how we convert the CO luminosity 
to the different units used in this paper and how we estimate it from the 
H$_2$ mass and $X_{\rm CO}$.

\section{Modelling the CO emission of galaxies}\label{modelssec}

We study the $\rm CO$ emission {from the $(1-0)$ to the $(10-9)$} rotational transitions, 
and its relation to other galaxy properties, using a modified version of the  
{\texttt{GALFORM}} semi-analytical model of galaxy formation 
described by Lagos et al. (2011a, 2011b) in combination with 
the Photon Dominated Region code, {\tt UCL$_{-}$PDR} of \citet{Bayet11}. 
In this section we describe the galaxy formation model and the physical processes included in it in $\S 2.1$, 
the {\tt UCL$_{-}$PDR} model and its main parameters in $\S 2.2$, 
give details about how we couple these two models 
to estimate the CO emission of galaxies in $\S 2.3$, and briefly describe how the CO conversion factor 
depends on galaxy properties predicted by {\tt GALFORM} in $\S 2.4$.

\subsection{The galaxy formation model} 

The {\texttt{GALFORM}} model \citep{Cole00} 
takes into account the main physical processes
that shape the formation and evolution of galaxies. These are: (i) the collapse
and
merging of dark matter (DM) halos, (ii) the shock-heating and radiative cooling
of gas inside
DM halos, leading to the formation of galactic disks, (iii) quiescent star
formation (SF) in galaxy disks, (iv) feedback
from supernovae (SNe), from AGN and from photo-ionization of the
intergalactic medium (IGM), (v) chemical
enrichment of stars and gas, and (vi) galaxy mergers driven by
dynamical friction within common DM halos, which can trigger bursts of SF 
and lead to the formation of spheroids (for a review of these
ingredients see \citealt{Baugh06}; \citealt{Benson10b}).
Galaxy luminosities are computed from the predicted star formation and
chemical enrichment histories using a stellar population synthesis model.
Dust extinction at different wavelengths is calculated
self-consistently from the gas and metal contents of
each galaxy and the predicted scale lengths of the disk and bulge components
using a radiative transfer model
(see \citealt{Cole00} and \citealt{Lacey11}). \citet{Lagos10} improved the 
treatment of SF in quiescent disks, (iii) in the above list, which allowed more 
general SF laws to be used in the model.

{\texttt{GALFORM}} uses 
the formation histories of DM halos as a starting point 
to model galaxy formation (see \citealt{Cole00}). 
In this paper we use halo merger trees extracted from the Millennium N-body
simulation \citep{Springel05}, which assumes the following  cosmological parameters: 
$\Omega_{\rm m}=\Omega_{\rm DM}+\Omega_{\rm baryons}=0.25$ (with a
baryon fraction of $0.18$), $\Omega_{\Lambda}=0.75$, $\sigma_{8}=0.9$
and $h=0.73$. The resolution of the $N$-body
simulation corresponds to a minimum halo mass of $1.72 \times 10^{10} h^{-1} M_{\odot}$. 
This is sufficient to resolve the halos that contain most of the 
H$_2$ in the universe at $z<8$ \citep{Lagos11}.

\citet{Lagos10} studied three SF laws, (i) the empirical SF law of \citet{Kennicutt98}, (ii) 
the empirical SF law of \citet{Blitz06} and (iii) the theoretical SF law of \citet{Krumholz09}. 
Here we follow Lagos et al. (2011a, hereafter L11), who adopted the empirical SF law of \citet{Blitz06} 
as their preferred model. The main successes of the L11 model include the  reproduction 
 of the optical and near-infrared 
luminosity functions (LF), the $z=0$ atomic hydrogen (HI) mass function (MF), 
the global density evolution of HI at $z<3.5$, and scaling relations between 
HI, H$_2$, stellar mass and galaxy morphology in the local Uuniverse.  

The \citet{Blitz06} empirical SF law has the form 

\begin{equation}
\Sigma_{\rm SFR} = \nu_{\rm SF} \,\rm f_{\rm mol} \, \Sigma_{\rm gas},
\label{Eq.SFR}
\end{equation}
\noindent where $\Sigma_{\rm SFR}$ and $\Sigma_{\rm gas}$ are the surface
densities of SFR and the total cold gas mass, respectively,
$\nu_{\rm SF}$ is the inverse of the SF
timescale for the molecular gas and $\rm f_{\rm mol}=\Sigma_{\rm mol}/\Sigma_{\rm gas}$ is the
molecular to total gas mass surface density ratio. The molecular and total gas
contents include the contribution from helium, while HI and H$_2$ masses only include 
hydrogen (helium accounts for $26$\% of the overall cold gas mass).
The ratio $\rm f_{\rm mol}$ is assumed to depend on
the internal hydrostatic pressure of the disk 
as $\Sigma_{\rm H_2}/\Sigma_{\rm HI}=\rm f_{\rm mol}/(f_{\rm mol}-1)=
(P_{\rm ext}/P_{0})^{\alpha}$ \citep{Blitz06}. 
The parameters values we use for $\nu_{\rm SF}$, $\rm P_{0}$ and $\alpha$ 
are the best fits to observations of spirals and dwarf galaxies, $\nu_{\rm SF}=0.5\, \rm Gyr^{-1}$,  
$\alpha=0.92$ and $\rm log(P_{0}/k_{\rm B} [\rm cm^{-3} K])=4.54$
(\citealt{Blitz06}; \citealt{Leroy08}; \citealt{Bigiel11}; \citealt{Rahman11}). 
{\citet{MacLow12} explain the relation between the $\Sigma_{\rm H_2}/\Sigma_{\rm HI}$ ratio 
 and the midplane pressure being the result of an underlying and more fundamental 
 relation between these two quantities and the local density in 
normal spiral galaxies (see also \citealt{Pelupessy09}). In this paper, however, we adopt 
the empirical relation to avoid fine-tuning of 
the parameters associated with it (see \citealt{Lagos10}).}

For starbursts the situation is less clear. Observational uncertainties,
such as the conversion factor between CO and H$_2$
in starbursts, and the intrinsic compactness of star-forming regions,
have not allowed a clear characterisation of the SF law (e.g.
\citealt{Kennicutt98}; \citealt{Genzel10}; \citealt{Combes11}). Theoretically,
it has been suggested that the SF law in starbursts is different
from that in normal star-forming galaxies: the relation between
$\Sigma_{\rm H_2}/\Sigma_{\rm HI}$ and gas pressure is expected to be different
in environments of very high gas densities typical of starbursts
(\citealt{Pelupessy06}; \citealt{Pelupessy09}; \citealt{MacLow12}), where
the ISM is predicted to always be dominated by H$_2$ independently
of the exact gas pressure. For these reasons we choose to apply the Blitz \&
Rosolowsky SF law only during quiescent SF (fuelled by the accretion of cooled gas 
onto galactic disks) and retain the original SF prescription
for starbursts (see \citealt{Cole00} and L11 for details). In the latter,
the SF timescale is taken to be proportional to the bulge dynamical timescale
above a minimum floor value (which is a model parameter) and involves the whole ISM gas content
in the starburst, giving $\rm SFR = \it M_{\rm gas}/\tau_{\rm SF}$ (see
\citealt{Granato00} and \citealt{Lacey08} for details), with  

\begin{equation}
\tau_{\rm SF}=\rm max(\tau_{min},f_{\rm dyn}\tau_{\rm dyn}). 
\label{SFlawSB}
\end{equation}

\noindent Here we adopt 
$\tau_{\rm min}=100\, \rm Myr$ and $f_{\rm dyn}=50$. This parameter choice helps to reproduce the 
observed rest-frame UV ($1500$\AA) luminosity function from $z\approx3$ to $z\approx 6$ 
(see \citealt{Baugh05}; \citealt{Lacey11}). In \citet{Lagos11} these parameters were set to 
$\tau_{\rm min}=5\, \rm Myr$ and $f_{\rm dyn}=2$, inherited from the parameter choice in 
\citet{Bower06}. However, 
the modification of these two parameters does not have any relevant influence on the results presented 
previously in Lagos et al. (2011a,b), but mainly affects the UV luminosity of very 
high redshift galaxies through the dust production during starbursts.

{In order to estimate the CO emission in starbursts, we assume 
 here that the cold gas content is fully molecular, $f_{\rm mol} = 1$.}
Note that this is similar to assuming that the Blitz \&
Rosolowsky {relation between the midplane pressure and 
the $\Sigma_{\rm H_2}/\Sigma_{\rm HI}$ ratio} holds in starbursts, given that large gas and
stellar densities lead to $f_{\rm mol} \approx 1$. Throughout the paper we will refer to galaxies 
as `starburst galaxies' if their total SFR is dominated by the starburst mode, 
$\rm SFR_{\rm starburst}>SFR_{\rm quiescent}$, 
while the rest of the galaxies will be refered to as `quiescent galaxies'.

\subsubsection{Estimating the properties of the interstellar medium of galaxies in {\texttt{GALFORM}}}\label{CalGuv}

The three properties predicted by {\tt GALFORM} which we use as inputs for the PDR model are 
(1) the ISM metallicity, $Z_{\rm gas}$, (2) the average internal FUV radiation field, $G_{\rm UV}$, 
and  (3) the average internal hard X-ray radiation field, $F_{\rm X}$. 
In this subsection we describe how these three properties are estimated and compare with observations 
in the case of $Z_{\rm gas}$. 
\begin{figure}
\begin{center}
\includegraphics[width=0.49\textwidth]{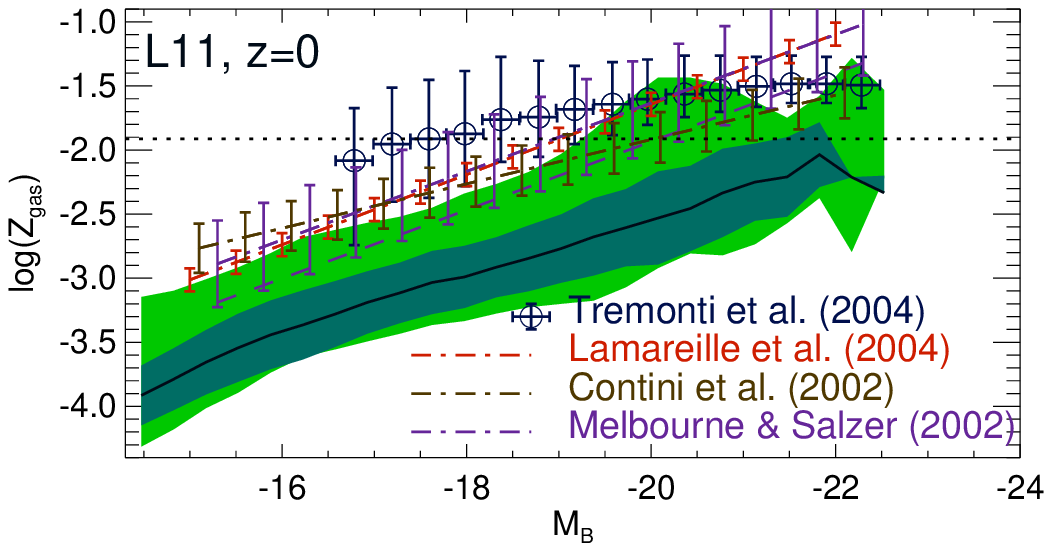}
\includegraphics[width=0.49\textwidth]{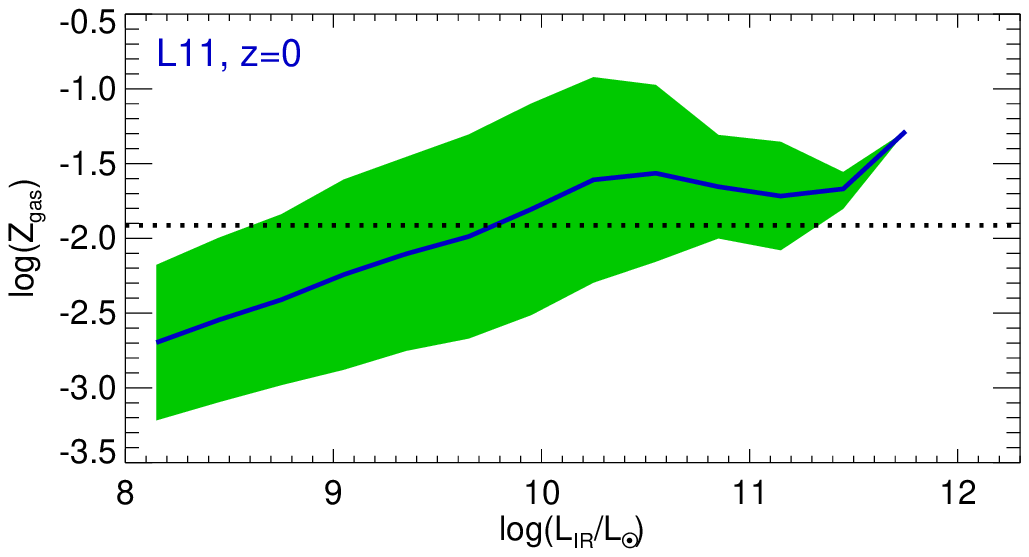}
\caption{{\it Top-panel:} Gas metallicity vs. dust extincted $B$-band absolute magnitude of 
galaxies at $z=0$ in the L11 model.
The solid line and the dark shaded area show the median and 
the $10$ to $90$ percentile range of model galaxies in the L11 model with 
an equivalent width of the H$\beta$ line $\rm EW(H\beta)>1.5$\AA.
The lighter colour 
shaded region shows the $10$ to $90$ percentile range of all model galaxies with 
SFR$>0$.  
The best fit and standard deviation of the observed gas metallicity-luminosity relations 
 are shown as dot-dashed lines with 
errorbars (\citealt{Lamareille04}; \citealt{Contini02}; \citealt{Melbourne02}).
Circles and errorbars show the median and $2\sigma$ range on the 
observational estimates of \citet{Tremonti04}, using their correction to convert the 
$g$-band luminosity-metallicity relation to the $B-$band. 
Note that the observational data 
correspond to a 
metallicity inferred from emission lines coming from the central parts of galaxies 
(i.e. emission within the fiber, which is of a diameter   
$2$~arcsec in the case of 2dFGRS in \citealt{Lamareille04}, $3$~arcsec in the case of 
SDSS in \citealt{Tremonti04}, and $2$ and $3.5$~arcsecs in the smaller surveys of 
\citealt{Contini02} and \citealt{Melbourne02}, respectively).
We also show the
\citet{Melbourne02} median relation shifted by $-0.3$~dex (dashed line) to illustrate 
the possible systematic error in the strong line method (see text for details).
{\it Bottom-panel:} Gas metallicity vs. IR luminosity 
for the L11 model. Here we include all model galaxies.
{For reference, the solar metallicity value reported by \citet{Asplund05} is shown 
as a horizontal dotted line in both panels.}}
\label{MMr}
\end{center}
\end{figure}

\begin{itemize}
\item $Z_{\rm gas}$. 
In {\texttt{GALFORM}}, $Z_{\rm gas}$ corresponds to the total mass fraction in metals in 
the ISM, and is calculated by assuming instantaneous recycling.  
$Z_{\rm gas}$ is the result of the non-linear interplay between the 
existing metal content in the ISM, the metal content of the incoming cooled gas, which originated in the 
hot halo, and the metals ejected by dying stars \citep{Cole00}. 
 
The top-panel of Fig.~\ref{MMr}  
shows the gas metallicity as a function of the $B-$band luminosity for galaxies in the model compared to 
the observational results of \citet{Contini02}, \citet{Melbourne02}, \citet{Lamareille04} and 
\citet{Tremonti04}.
A correction factor needs to be applied to the observations to convert from the inferred 
abundance of oxygen relative to hydrogen, $\rm O/H$, to $Z_{\rm gas}$. 
We use the solar
metallicity ratios reported by \citet{Asplund05}, 
$\rm O/H_{\odot}=4.57\times 10^{-4}$ and $Z_{\odot}=0.0122$. 
This choice of the value of solar abundance is 
to keep consistency with the solar abundance assumed in the {\tt UCL$_{-}$PDR} model.  
In the case of the \citet{Tremonti04} data, we applied the conversion suggested by these authors to 
derive a $B$-band luminosity-metallicity relation from their $g$-band relation. 
The luminosity-metallicity relations estimated by 
\citet{Tremonti04} and \citet{Lamareille04} 
used large area redshift surveys (the Sloan Digital Sky Survey and the 
2 Degree Field Galaxy Redshift survey, respectively). 
In the case of \citet{Contini02} and \citet{Melbourne02}, results were based on smaller 
samples of star-forming galaxies 
which were followed up in spectroscopy.
The \citet{Contini02} $B$-band luminosity-metallicity relation was derived 
from a sample of UV-selected galaxies, which includes a few higher redshift galaxies with $z>0.15$.

The observational results shown in Fig.~\ref{MMr} use abundance indicators based on 
emission lines to calculate oxygen abundances. 
To estimate oxygen abundances, \citet{Contini02}, \citet{Melbourne02} and \citet{Lamareille04} 
 use an empirical relation between oxygen abundance and the $R_{23}$ abundance ratio, 
where $R_{23}=\rm ([OII]+[OIII])/H\beta$, which is often called the `strong-line' technique. 
\citet{Tremonti04}, on the other hand, fit all prominent emission lines with a model designed to 
describe integrated galaxy spectra, which includes HII regions and diffuse ionized gas. 
\citet{Kennicutt03} 
compared the  
`strong-line' technique with abundances inferred from electron temperature measurements 
in a sample of HII regions with very high resolution spectroscopy, 
and argued that 
the empirical `strong-line' method systematically results 
in larger abundances by approximately a factor of $2$ due to uncertainties in
the nebular models used in calibration.
In order to illustrate this possible systematic error, we show as a dashed line the 
relation of \citet{Melbourne02} data shifted by $-0.3$~dex.

The L11 model predicts a lower normalization of the 
luminosity-metallicity relation than implied by observations, 
but with a similar slope. When selecting star-forming galaxies in the model 
by their H$\beta$ equivalent width, this discrepancy increases due to the 
tighter gas metallicity-luminosity relation predicted for these galaxies. 
The correction suggested by \citet{Kennicutt03} to remove the systematic introduced by 
the `strong-line' technique reduces the discrepancy between the observed and predicted median 
relations to a factor $2$, well within 
the typical dispersion of observational data (see errorbars for \citealt{Tremonti04} in 
Fig.~\ref{MMr}).
Another caveat in the comparison between observations and the model predictions 
is the fact that 
the observational data are inferred from the emission lying within a spectrograph fiber 
(typically of $3$~arcsecs or less in diameter), which typically covers 
only the central parts of the galaxy, and therefore are not global mass-weighted metallicities. 
Galaxies show metallicity gradients, with the 
central parts being more metal rich than the outskirts (e.g. \citealt{Peletier90}; \citealt{deJong96}). 
The differences 
in metallicity between centers and the outer regions of galaxies 
 can be as large as a factor $2-3$ in early-type
galaxies and $3-10$ in late-type galaxies \citep{Henry99}.
Thus, one would expect metallicities inferred from the fibers to be upper limits for the
mass-weighted ISM metallicity.
Given this caveat, the model predictions are in reasonable 
agreement with the observations.

Throughout the paper we make extensive comparisons between the  
CO and total IR luminosity (see Appendix~\ref{App:COIR} for a description of how we 
calculate the IR luminosity in {\tt GALFORM}). We plot in the bottom-panel of Fig.~\ref{MMr}, 
the gas metallicity as a function of the IR 
 luminosity. For $L_{\rm IR}<10^{10}\, L_{\odot}$, 
the gas metallicity increases with $L_{\rm IR}$, but tends to a constant value at higher 
luminosities. As we show later in the text, 
the dependence of the CO emission on IR luminosity is 
influenced by the gas metallicity ($\S 3$ and $\S 4$), as  
it alters both luminosities. The bottom panel of Fig.~\ref{MMr} will therefore help in the interpretation 
of the results later in this paper. 

\item $G_{\rm UV}$. The average internal UV radiation field, $G_{\rm UV}$, 
corresponds to a local radiation field that depends 
on the transmission of UV photons from star-forming regions and their propagation 
through the diffuse ISM.
The exact transmitted fraction of UV radiation depends on the local conditions in the ISM, such as the optical 
depth, the ratio of gas in the diffuse ISM and in GMCs, etc (see \citealt{Lacey11}). 
Whilst $G_{\rm UV}$ is a local property, we make a rough 
estimate by considering two simple approximations which are based on global galaxy 
properties. The first scaling is motivated by the close relation between UV luminosity and 
SFR \citep{Lacey11}, so that in an optically thin slab, 
the average UV flux scales approximately as $\langle I_{\rm UV}\rangle \propto \Sigma_{\rm UV}\propto \Sigma_{\rm SFR}$. 
This is expected if the UV radiation field in the wavelength range considered here ($\lambda=900-2100$\AA) 
is dominated by radiation from OB stars. 
We therefore assume that $G_{\rm UV}$ is related to the surface density of SFR by 

\begin{equation}
\frac{G_{\rm UV}}{G_{0}}=\left(\frac{\Sigma_{\rm SFR}}{\Sigma^0_{\rm SFR}}\right)^{\gamma}.
\label{para1}
\end{equation}

\noindent Here we take $\Sigma_{\rm SFR}=\rm SFR/2\, \pi\rm r^2_{50}$, where 
$\rm r_{50}$ corresponds to the half-mass radius, either of the 
disk or the bulge, depending on where the SF is taking place 
(in the disk for quiescent SF and in the bulge for starbursts). We set $\gamma=1$ so that 
$G_{\rm UV}$ increases by the same factor as $\Sigma_{\rm SFR}$. However, 
values of $\gamma=0.5-2$ do not change the predictions of the model significantly.
For the normalisation, we choose $\Sigma^0_{\rm SFR}=10^{-3}\, M_{\odot}\, \rm yr^{-1}\, kpc^{-2}$, 
so that $G_{\rm UV}=G_{0}=1.6\times 10^{-3}\, \rm erg\, cm^{-2}\, s^{-1}$ for the solar 
neighborhood \citep{Bonatto11}. 

A dependence of $G_{\rm UV}$ solely on $\Sigma_{\rm SFR}$ assumes that an increase in the local UV radiation field 
takes place if a galaxy forms stars at a higher rate per unit area, 
but does not take into account the transparency of 
the gas. 
To attempt to account for this, we consider an alternative scaling in which we 
include in a simple way the average optical depth of the ISM in the 
description of $G_{\rm UV}$.
In a slab, the transmission probability of UV photons, $\beta_{\rm UV}$, scales with the 
optical depth, $\tau_{\rm UV}$, so that 
$\beta_{\rm UV}\sim (1-e^{-\tau_{\rm UV}})/\tau_{\rm UV}$. The optical depth, on the 
other hand, depends on the gas metallicity and column density of atoms 
as $\tau_{\rm UV}\propto Z_{\rm gas}\, N_{\rm H}$. In optically thick gas ($\tau_{\rm UV}\gg 1$), 
$\beta_{\rm UV}\sim \tau^{-1}_{\rm UV}$. By assuming that the average local UV field 
depends on the average emitted UV field ($\langle I_{\rm UV}\rangle \propto \Sigma_{\rm SFR}$) times an average 
UV transmission factor, we get the scaling 

\begin{equation}
\frac{G_{\rm UV}}{G_{0}}=\left(\frac{\Sigma_{\rm SFR}/\Sigma^0_{\rm SFR}}{[Z_{\rm gas}/Z_{\odot}]\, [\Sigma_{\rm gas}/\Sigma^0_{\rm gas}]}\right)^{\gamma {\prime}}.
\label{para2}
\end{equation}

\noindent We set $\gamma{\prime}=1$, but as with Eq.~\ref{para1}, 
varying the exponent in the range $\gamma{\prime}=0.5-2$ has little impact on the model predictions (see $\S 4$).  
We use the solar neighborhood value, $\Sigma^0_{\rm gas}=10\, M_{\odot}\, \rm pc^{-2}$ \citep{Chang02}.
The parametrization of Eq.~\ref{para2} has been shown to explain the higher 
$G_{\rm UV}$ in the Small Magellanic Cloud compared to the Milky-Way, which is needed to explain the 
low molecular-to-atomic hydrogen ratios \citep{Bolatto11}.
We test these two parametrizations of $G_{\rm UV}$ against broader observations in $\S 3$ and 
$\S 4$.

\item $F_{\rm X}$. In {\texttt{GALFORM}} we model the growth and emission 
by supermassive black holes 
(SMBHs) which drive AGN in galaxies \citep{Fanidakis10}. \citet{Fanidakis10b} 
estimate the emission from accreting SMBHs over a wide wavelength range, from 
hard X-rays to radio wavelengths.
The SMBH modelling of Fanidakis et al. includes an estimate of the 
efficiency of energy production by accretion onto the black hole, 
taking into account the value of the black hole spin, which is followed through 
all the gas accretion episodes and mergers with 
other black holes. \citet{Fanidakis10b} show that 
the model can successfully explain the LF of AGN and quasars and its time evolution at different wavelengths. 
  In this work we use this SMBH modelling to 
take into account the heating of the ISM by the presence of an AGN in the galaxy, 
which has been shown to be important in the hard X-ray energy window \citep{Meijerink07}. The emission of AGN in hard X-rays 
 ($2-10$~KeV), $L_{\rm X}$, 
is calculated in Fanidakis et al. using the bolometric luminosity of the AGN and the bolometric corrections 
presented in \citet{Marconi04}. 
We estimate the average hard X-ray flux, $F_{\rm X}$, at the half-mass radius of the bulge,

\begin{equation}
F_{\rm X}=\frac{L_{\rm X}}{4\pi\, r^2_{\rm 50}}.
\label{para3}
\end{equation}
 
\end{itemize}

\subsection{The {\tt UCL$_{-}$PDR} code}

The {\tt UCL$_{-}$PDR} code attempts to fully describe the chemical and thermal evolution of 
molecular clouds under different conditions in the surrounding ISM as quantified by: 
the far-UV (FUV) radiation background, 
the cosmic ray background, the volume number density of hydrogen, 
the average dust optical depth and the gas metallicity (see Bell et al. 2006; 2007 for a 
detailed description). 
More recently, \citet{Bayet09b}
and \citet{Bayet11} explored these 
parameters for the ISM in external galaxies. 
We use the code released by \citet{Bayet11}, in which additional 
cooling mechanisms were included, such as $^{13}\rm C^{16}O$, 
$\rm ^{12}C^{18}O$, CS and OH. \citet{Bayet11} showed that these coolants 
are important when dealing with galaxies which are forming stars at high rates.

The {\tt UCL$_{-}$PDR} code is a time-dependent model which treats a cloud as 
a one-dimensional, semi-infinite slab 
illuminated from one side by FUV photons.
{Molecular gas described by the PDR code correspond to clouds having a single gas phase, 
with a single gas volume density, although gradients in temperature and chemical 
composition depend on the optical depth.} 
The radiative transfer equations are solved and the thermal 
balance between heating and cooling mechanisms is calculated 
leading to the determination of the gradients of kinetic temperature, 
chemical composition and emission line strength across the slab 
(i.e. as a function of optical extinction in the visible, $A_{\rm V}$).
The gas kinetic temperature at which this balance is achieved will be referred to  
throughout the paper as the typical kinetic temperature of molecular clouds in galaxies 
in the model, $T_{\rm K}$.

The starting point in the model is to assume that hydrogen is mostly molecular 
and that other species are atomic. The model follows the relative abundance 
of $131$ species, including atoms and molecules, using a network of more than $1,700$ chemical 
reactions (see \citealt{Bayet09}; \citealt{Bayet11}). 
The initial element abundances, dust-to-gas ratio
and H$_2$ formation rate are assumed
to scale linearly with the metallicity of the gas.

The physical mechanisms included in the {\tt UCL$_{-}$PDR} code are 
(i) H$_2$ formation on dust grain surfaces, (ii) H$_2$ photodissociation 
by FUV radiation (which we define as the integrated emission for the 
wavelength range $\lambda=900-2100$\AA), 
(iii) H$_2$ UV fluorescence, (iv) the photoelectric effect
from silicate grains and polycyclic aromatic hydrocarbons (PAHs), 
(v) C~II recombination and (vi) interaction of
low-energy cosmic-rays (CRs) with the gas, which boosts the temperature of the gas. The latter 
results in stronger CO emission from high order rotational transitions that resembles the observed 
CO emission from galaxies which host AGN. The 
CO spectral line energy distribution (commonly referred to as the CO ladder or SLED) 
therefore can be obtained for a wide range of parameters included in the {\tt UCL$_{-}$PDR} code. 

Given that the ISM of galaxies is not resolved in {\texttt{GALFORM}}, we assume 
the following fiducial properties for GMCs. We adopted a 
gas density of $n_{\rm H}=10^4\, \rm cm^{-3}$, where each model 
was run for $10^{6}~\rm yr$. {Note that $n_{\rm H}$ represents the 
total number of hydrogen nuclei.} 
The value of $n_{\rm H}$ adopted is similar to the assumption
used previously in {\texttt{GALFORM}} for GMCs
(i.e. $n_{\rm H}=7\times 10^3\, \rm cm^{-3}$;
\citealt{Granato00}; \citealt{Lacey11}), which in turn is motivated by 
the assumptions used in the {\texttt{GRASIL}} code \citep{Silvia98}, which calculates the
reprocessing of stellar radiation by dust. 
The parameters above correspond to star-forming gas which is likely to be opaque to radiation. 
Note that this dense gas phase typically has $A_{V}$ in the range $3-8$~mag. We choose 
to focus on dense gas of $A_{V}=8$~mag 
to obtain a 
$X_{\rm CO(1-0)}$ for the local neighbourhood properties consistent with 
observational results.
We expect this approximation to be accurate particularly in gas-rich galaxies, which 
to first order have a larger proportion of gas in 
this dense phase with respect to the total gas reservoir compared to more passive 
galaxies. This is simply because of the energetics and the dynamics involved in highly star-forming 
regions, 
which typically increase both density and temperature, leading to a more dense, 
opaque and fragmented medium (see Bayet 2008, 2009 for more details).
Note, however, that the assumption of a lower $A_{V}$ for lower gas surface density galaxies, 
$\Sigma_{\rm gas}<10^7 M_{\odot}\, \rm kpc^{-2}$, would not 
affect the results shown in this paper, for example simply moving galaxies along the fainter 
part of the CO luminosity function, without modifying its bright-end. 

{\citet{Wolfire03} suggested that a minimum 
density $n_{\rm H,min} \propto G_{\rm {UV}}$ is necessary
to obtain pressure balance between the warm and cold neutral media in the ISM.
All the models shown in Table~\ref{XCOs} fulfill this condition, with  
$n_{\rm H}>n_{\rm H,min}$. However, we test the effect of assuming that 
 $n_{\rm H}$ scales with the minimum density of \citet{Wolfire03}, which leads to 
$n_{\rm H} \propto G_{\rm {UV}}$. With this in mind, we ran four more models with $n_{\rm H}$ varying in such 
a way that the
$n_{\rm H}/G_{\rm {UV}}$ ratio is left invariant,
in addition to the PDR models run using $n_{\rm H}=10^4\, \rm cm^{-3}$. We analyse 
the CO luminosities prediced by this set of models in $\S 5$.}

The output of the {\tt UCL$_{-}$PDR} code 
includes the conversion factor,  $X_{\rm CO(J \rightarrow J-1)}$, between 
the intensity of a particular CO rotational transition and the column number density 
of H$_2$ molecules, 

\begin{equation}
X_{\rm CO(J \rightarrow J-1)}=\frac{N_{\rm H_2}}{I_{\rm CO(J\rightarrow J-1)}}, 
\label{XCOform}
\end{equation}  

\noindent where $N_{\rm H_2}$ is the H$_2$ column density and 
$I_{\rm CO}$ is the integrated CO line intensity (see Appendix~\ref{App:COIR}). 
This conversion factor, which is the one we are interested in here, depends on 
the conditions in the ISM.  

\begin{table*}
\begin{center}
\caption{Conversion factors from CO$(1-0)$-H$_2$ (6), CO$(3-2)$-H$_2$ (7) and CO$(7-6)$-H$_2$ (8) in units 
of $10^{20}\, \rm cm^{-2}\, (K\,km\,s^{-1})^{-1}$, and the kinetic 
temperature of the gas (9) 
for galaxies 
with different ISM conditions: (1) FUV radiation background, $G_{\rm UV}$, in units of $G_{0}= 1.6\times 10^{-3} \, 
\rm erg\, cm^{-2} s^{-1}$, (2) gas metallicity, $Z_{\rm gas}$, in units of $Z_{\odot}=0.0122$, 
(3) hard X-ray flux, $F_{\rm X}$, 
in units of $\rm erg\, s^{-1}\, cm^{-2}$ and { (4) 
total number of hydrogen nuclei, $n_{\rm H}$, in units of ${\rm cm}^{-3}$} (\citealt{Bell06}; \citealt{Asplund05}). 
%We remind the reader 
%that the conversion factors here correspond to GMCs with $n_{\rm H}=10^4\, \rm cm^{-3}$ and $A_V=8$~mag (see
%text for details).
}\label{XCOs}
\begin{tabular}{c c c c c | c c c c }
\\[3pt]
\hline
(1)& (2) &(3) &{ (4)} &{ (5)} &(6) & (7) & (8) & (9)\\
\hline
$G_{\rm UV}/G_{0}$ & $Z_{\rm gas}/Z_{\odot}$ & $F_{\rm X}/\rm erg\, s^{-1}\, cm^{-2}$ &  
 ${ n_{\rm H}}/{\rm  cm}^{-3}$ & ${ (n_{\rm H}/{\rm  cm}^{-3})/(G_{\rm UV}/G_{0})}$ & $X_{\rm CO(1-0)}$ & $X_{\rm CO(3-2)}$ & $X_{\rm CO(7-6)}$ & T$\rm _K/K$\\
\hline
      1&       0.01 &     0.01  & $10^4$ & $10^4$  &    3.585 &     2.335  &    24.091   &    39.37 \\
      1&       0.05 &     0.01  &$10^4$ & $10^4$  &       3.212 &     4.007  &    100.043   &    17.95 \\
      1&       0.1 &     0.01   & $10^4$ & $10^4$  &      3.177  &     4.168   &    218.343    &    19.44 \\
      1&       0.5 &     0.01   &$10^4$ &$10^4$   &       2.760  &     4.105   &    1281.91    &    12.33 \\
      1&       1 &     0.01  &$10^4$ &$10^4$   &       1.883 &     2.808  &    1124.70   &    10.47 \\
      1&       2 &     0.01  &$10^4$ &$10^4$   &       1.015 &     1.521  &    747.670   &    8.72 \\
      1&       0.01 &     0.1 &$10^4$ &$10^4$   &       3.620 &     1.817  &    9.311   &    51.24 \\
      1&       0.1 &     0.1 &$10^4$ &$10^4$   &       3.293 &     3.072  &    55.812   &    38.61 \\
      1&       1 &      0.1 &$10^4$ &$10^4$   &       1.519 &     2.103  &    96.234   &    19.26 \\
      1&       2 &     0.1 &$10^4$ &$10^4$   &       0.915  &     1.283  &    111.9   &    15.69 \\
      1&       0.01 &      1 &$10^4$ &$10^4$   &       6.260 &     3.688  &    63.556   &    107.78 \\
      1&       0.1 &      1 &$10^4$ &$10^4$   &       2.389 &     1.653  &    5.509   &    60.81 \\
      1&       1 &      1 &$10^4$ &$10^4$   &       2.263 &     1.853  &    56.680   &    38.82 \\
      1&       2 &      1 &$10^4$ &$10^4$   &       0.999  &     0.956   &    20.411   &    38.33 \\
      10&       0.01 &     0.01  &$10^4$ & $10^3$  &       3.856 &     2.393  &    27.359   &    38.93 \\
      10&       0.1 &     0.01  &$10^4$ &$10^3$   &       2.970 &     3.274  &    193.999   &    18.73 \\
      10&       1 &     0.01  &$10^4$ &$10^3$   &       1.139 &     1.392  &    48.991   &    10.22 \\
      10&       2 &     0.01  &$10^4$ &$10^3$   &       0.596  &     0.726   &    24.525   &    8.64 \\
      10&       0.01 &      1 &$10^4$ &$10^3$   &       6.468 &     3.759  &    66.494   &    104.35 \\
      10&       0.1 &      1 &$10^4$ &$10^3$   &       2.382 &     1.627  &    5.510   &    60.72 \\
      10&       1 &      1 &$10^4$ &$10^3$   &       2.003 &     1.518  &    35.534   &    39.13 \\
      100&       0.01 &     0.01  &$10^4$ &$10^2$   &       3.673 &     3.513  &    190.302   &    38.17 \\
      100&       0.1 &     0.01  &$10^4$ &$10^2$   &       3.174 &     3.441  &    225.733   &    18.00 \\
      100&       1 &     0.01  &$10^4$ &$10^2$   &       0.913  &     1.005  &    12.726   &    9.63 \\
      100&       0.01 &      1.0 &$10^4$ &$10^2$   &       13.73 &     2.477  &    5.428   &    98.49 \\
      100&       0.1 &      1.0 &$10^4$ &$10^2$   &       2.237 &     1.839  &    10.591   &    60.57 \\
      100&       1 &      1.0 &$10^4$ &$10^2$   &       1.861 &     1.312  &    24.277   &    38.94 \\
      1000&       0.01 &     0.01  &$10^4$ &$10$   &       4.560 &     4.061  &    295.512   &    39.35 \\
      1000&       0.1 &     0.01  &$10^4$ &$10$   &       3.059 &     3.237  &    195.177   &    16.33 \\
      1000&       1 &     0.01  &$10^4$ &$10$   &       0.809  &     0.867   &    6.771   &    8.78 \\
      1000&       2 &     0.01  &$10^4$ &$10$   &       0.374  &     0.385   &    2.094   &    7.86 \\
      1000&       0.01 &     0.1 &$10^4$ &$10$   &       2.592 &     3.841  &    102.544   &    19.58 \\
      1000&       0.1 &     0.1 &$10^4$ &$10$   &       2.149 &     3.146  &    103.455   &    19.51 \\
      1000&       1 &     0.1 &$10^4$ &$10$   &       1.519 &     2.103  &    96.234   &    19.26 \\
      1000&       0.01 &      1.0 &$10^4$ &$10$   &      15.475 &     2.664  &    5.859   &    113.55 \\
      1000&       0.1 &      1.0 &$10^4$ &$10$   &       2.722 &     1.536  &    6.020   &    59.83 \\
      1000&       1 &      1.0 &$10^4$ &$10$   &       2.263 &     1.853  &    56.680   &    38.82 \\
      { 1}   &       { 1} & { 0.01} & $ 10^3$ &$ 10^{3}$ & { 2.692} & { 7.374}  & $ 6.3\times 10^{4}$ & { 11.49}  \\
      { 100} &       { 1} & { 0.01} & $ 10^5$ &$ 10^{3}$ & { 0.793} & { 0.659}  & { 1.52}  & { 7.59}  \\
      { 1000}&       { 1} & { 0.01} & $ 10^6$ &$ 10^{3}$ & { 0.542} & { 0.358}  & { 0.421}  & { 6.16}  \\
      { 1000}&       { 1} & { 1}    & $ 10^6$ &$ 10^{3}$ & { 0.691} & { 0.43}   & { 0.513}  & { 21.18}  \\
\hline
\end{tabular}
\end{center}
\end{table*}

We present, { for} the first time, the conversion factors for different transitions
predicted by the {\tt UCL$_{-}$PDR} model. The output is listed in 
Table~\ref{XCOs} for $41$ different combinations of input properties of the ISM, { where 
$37$ models use $n_{\rm H}=10^4\, {\rm cm}^{-3}$, and $4$ have variable 
$n_{\rm H}$, chosen so that $n_{\rm H}/G_{\rm UV}$ is constant}. 
These values are intended to span the range of possibilities in the 
galaxy population as a whole, ranging from low metallicity dwarf galaxies to metal rich 
starbursts.
These models consider UV radiation field strengths of 
1, 10, 100 and 1000 times the value in our local neighbourhood ($G_{0}=1.6\times 10^{-3} \, 
\rm erg\, cm^{-2} s^{-1}$), gas metallicities ranging from $Z_{\rm gas}=0.01-2\, Z_{\odot}$ and  a flux 
(in the hard X-rays window) 
 of $0.01$, $0.1$ and 
$1\, \rm erg\, s^{-1}\, cm^{-2}$ (where 
X-rays are used as a proxy for cosmic rays; see \citealt{Papadopoulos10}; 
\citealt{Meijerink11}; \citealt{Bayet11}).
The {\tt UCL$_{-}$PDR} model inputs the cosmic rays ionization rate instead of 
hard X-rays flux. We assume a direct proportionality between the cosmic ray ionization rate and 
hard X-rays flux following the studies of \citet{Meijerink11} and \citet{Bayet11}, where
 $F_{\rm X}/F_{0}=\zeta_{\rm CR}/\zeta_{0}$, with $F_{0}=0.01\, \rm erg\, s^{-1}\, cm^{-2}$ and 
$\zeta_{0}=5\times 10^{-17}\, \rm s^{-1}$. 
We only list the CO-H$_2$ conversion parameters of three CO transitions 
in Table~\ref{XCOs}. However, the 
{\tt UCL$_{-}$PDR} model was run to output all CO transitions from $1-0$ to 
$10-9$, which we use to construct CO SLEDs in $\S 4$ and $\S 6$. { A comprehensive analysis and results of the PDR model 
listed in Table~\ref{XCOs} will be presented in 
Bayet et al. (2012, in prep.).}

From Table~\ref{XCOs} it is possible to see that the general dependence of $X_{\rm CO}$ on the 
three properties $Z_{\rm gas}$, $G_{\rm UV}$ and $F_{\rm X}$, depends on the transition considered. 
For instance, $X_{\rm CO(1-0)}$ increases as the gas metallicity decreases, but its dependence on 
$G_{\rm UV}$ and $F_{\rm X}$ depends on the gas metallicity: for solar or supersolar metallicities, 
$X_{\rm CO(1-0)}$ tends to decrease with increasing $G_{\rm UV}$ and $F_{\rm X}$, given that 
the higher temperatures increase the CO$(1-0)$ emission. However, for very subsolar gas metallicities, these 
trends are the opposite: $X_{\rm CO(1-0)}$ tends to increase with increasing $G_{\rm UV}$ and $F_{\rm X}$. 
In this case this is due to the effect of the high radiation fields and the lack of an effective CO self-shielding, 
which destroys CO molecules. In the case of higher CO transitions, for example CO$(7-6)$, $X_{\rm CO(7-6)}$  
generally increases with decreasing kinetic temperature, however this is not strictly the case  in 
every set of parameters. These general trends will help explain the relations presented 
in $\S 3-4$ and $\S 6$.

{ At very low metallicities, $Z_{\rm g}\approx 0.01 Z_{\odot}$, the CO lines become 
optically thin in some of the cases, e.g. in those models where there is a high UV and X-ray flux 
illuminating the molecular clouds. This represents a limitation of the 
PDR model given the uncertainties in the opacity effect on the CO lines. However, such galaxies are extremely 
rare in our model after selecting galaxies with $L_{\rm IR}>10^9\, L_{\odot}$, which are those we use 
to study the CO SLED at redshifts $z>0$ (e.g. only $0.05$\% of galaxies with 
$L_{\rm IR}>10^9\, L_{\odot}$ at $z=6$ have $Z_{\rm g}\le 0.01 Z_{\odot}$). This is not the case for 
very faint IR galaxies. We find that in the subsample of galaxies with 
$L_{\rm IR}>10^7\, L_{\odot}$, 
more than $10$\% of galaxies have $Z_{\rm g}\le 0.01 Z_{\odot}$ at $z>1.5$.}  

\subsection{Coupling the {\tt GALFORM} and {\tt UCL$_{-}$PDR} codes}

We use the properties $Z_{\rm gas}$, $G_{\rm UV}$ and $F_{\rm X}$ as inputs 
to the {\tt UCL$_{-}$PDR} model. For each galaxy, we calculate the $X_{\rm CO}$ conversion 
 factors for several CO transitions, and use the molecular mass of the galaxy to estimate 
the CO luminosity of these transitions (see Appendix~\ref{App:COIR}). 
We use the models from Table~\ref{XCOs} to find the $X_{\rm CO}$ conversion
 factors and the gas kinetic temperature of molecular clouds, $T_{\rm K}$, 
for each galaxy according to its ISM properties. 
Given that $Z_{\rm gas}$, $G_{\rm UV}$ and $F_{\rm X}$ are discretely sampled, we interpolate 
over the entries of Table~\ref{XCOs} on a logarithmic scale in each parameter. 
Throughout the paper we will refer to the coupled code as the {\tt GALFORM+UCL$_{-}$PDR} 
model. 

Galaxies in {\texttt{GALFORM}} can have 
SF taking place simultaneously in the disk and 
the bulge, corresponding to the quiescent and starburst SF modes, respectively. 
The gas reservoirs of these two modes are different and 
we estimate the CO luminosity of the two phases independently. 
This can be particularly important at high-redshift,  
where the galaxy merger rate is higher and where galaxies are more prone to 
have dynamically unstable disks, 
which can lead to starbursts. 
For instance, \citet{Danielson10} showed that 
two gas phases, a diffuse and a dense phase, are necessary 
to describe the CO spectral line energy distribution of the $z=2.3$ galaxy 
SMM~J2135-0102 (see also \citealt{Bothwell12}), illustrating the importance of allowing for the possibility of concurrent 
quiescent and burst episodes of SF in the modelling of CO emission.

\subsection{The dependence of the CO-H$_2$ conversion factor on galaxy properties in {\tt GALFORM}}
\begin{figure}
\begin{center}
\includegraphics[width=0.49\textwidth]{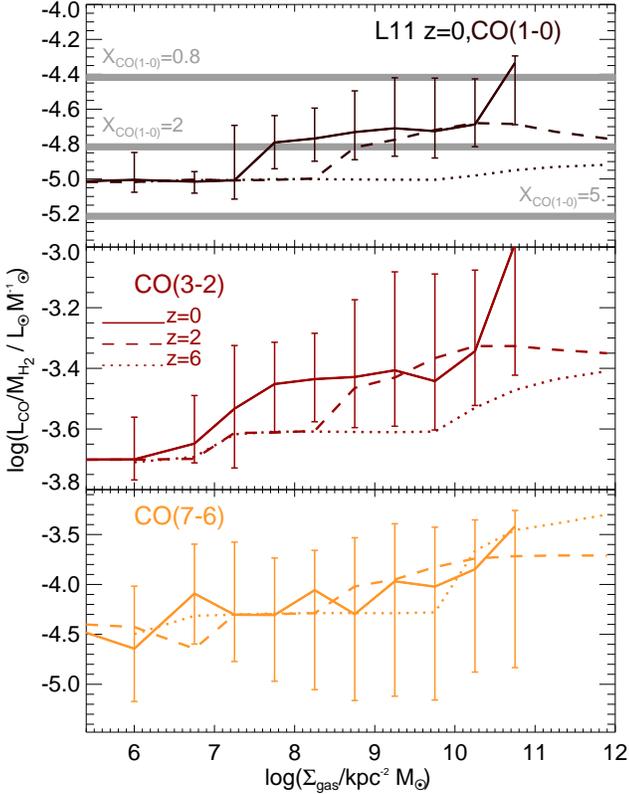}
\caption{CO luminosity-to-molecular hydrogen mass ratio as a 
function of the average gas surface density of galaxies in {\tt GALFORM} at $z=0$ (solid lines), $z=2$ 
(dashed lines) and $z=6$ (dotted lines) for the CO transitions $(1-0)$ (top panel), $(3-2)$ (middle 
panel) and $(7-6)$ (bottom panel).  
Lines show the median of the predicted distributions, and errorbars show the $10$ and $90$ percentiles, 
which are shown, for clarity, only for one redshift in each panel. 
For reference, in the top panel, the thick, horizontal lines show 
the ratios for fixed conversion factors: $X_{\rm CO(1-0)}=(0.8,2,5)\times 10^{20}\, \rm cm^{-2}\, 
(K\,km\,s^{-1})^{-1}$, corresponding to the values typically 
adopted in observational studies for 
starburst, normal star-forming and 
dwarf galaxies, respectively.}
\label{XCOsplot}
\end{center}
\end{figure}

In order to illustrate how much the $X_{\rm CO}$ conversion factor varies 
with galaxy properties, in this subsection 
we focus on the predictions of the CO luminosity-to-molecular mass ratio 
in the {\tt GALFORM+UCL$_{-}$PDR} model in the case where  
$G_{\rm UV}$ depends on $Z_{\rm gas}$, $\Sigma_{\rm SFR}$
and $\Sigma_{\rm gas}$ (see Eq.~\ref{para2}).

Fig.~\ref{XCOsplot} shows the $L_{\rm CO}/M_{\rm H_2}$ 
ratio as a function of the average ISM gas surface density, $\Sigma_{\rm gas}$,  
at three different
redshifts, and for three different CO transitions. In the case of CO$(1-0)$, we show for reference 
the standard conversion factors typically adopted in the literature for starburst, 
normal spiral and dwarf
galaxies as horizontal lines. At $z=0$ there is, on average, a positive correlation between 
$L_{\rm CO}/M_{\rm H_2}$ and $\Sigma_{\rm gas}$, with $L_{\rm CO}/M_{\rm H_2} \propto
\Sigma^{0.15}_{\rm gas}$, regardless of the CO transition. 
At higher redshifts, the relation between 
the $L_{\rm CO}/M_{\rm H_2}$ ratio and $\Sigma_{\rm gas}$ flattens, mainly due 
to the lower gas metallicities of galaxies. These trends are similar for all
the CO transitions. In the case of the CO$(1-0)$ at $z=0$, galaxies of low 
$\Sigma_{\rm gas}$ have $X_{\rm CO}$ closer to the value measured in dwarf galaxies moving 
to values closer to starburst galaxies at very high $\Sigma_{\rm gas}$. 
In terms of 
stellar mass, galaxies with $M_{\rm stellar}\approx 7\times 10^{10} M_{\odot}$, close to the Milky-Way 
stellar mass, have on average $X_{\rm CO(1-0)}\approx 2-3\times 
10^{20}\, \rm cm^{-2}\, (K\,km\,s^{-1})^{-1}$, in agreement with the measurements of the 
solar neighbourhood. 
For relatively massive galaxies, the model predicts that the $L_{\rm CO}/M_{\rm H_2}$ 
ratio evolves only weakly with redshift at 
$z\lesssim 2$, which explains the similarity between the $X_{\rm CO(1-0)}$ measured by \citet{Daddi10} 
in normal star-forming galaxies at $z\approx 1.5$ and the value for local spiral galaxies.

The coupled {\tt GALFORM+UCL$_{-}$PDR} model thus 
predicts a dependence of $X_{\rm CO}$ on galaxy properties 
which broadly agrees with observations in the local Universe and explains the few observations 
of high-redshift galaxies.

\section{The CO(1-0) emission of galaxies in the local universe}

In the local Universe, the CO$(1-0)$ emission of galaxies has been studied extensively in different 
environments with large samples of galaxies (e.g. \citealt{Keres03}; \citealt{Solomon05}
\citealt{Bothwell09}; \citealt{Saintonge11}; \citealt{Lisenfeld11}). In this section we compare 
our predictions for the CO$(1-0)$ emission of galaxies at $z=0$ and how this 
relates to their IR luminosity, with available observations.

\subsection{The CO(1-0) luminosity function}

In this subsection we focus on the CO$(1-0)$ LF and how the predictions depend
 on the assumptions and the physics of the model. 

\begin{figure}
\begin{center}
\includegraphics[width=0.5\textwidth]{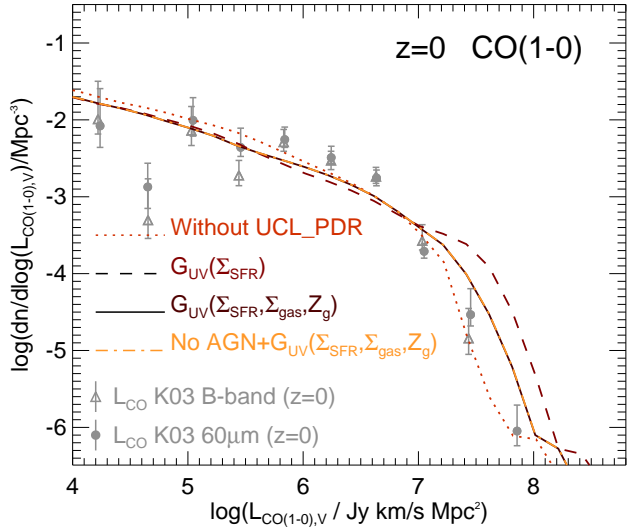}
\caption{The $z=0$ $\rm CO(1-0)$ luminosity function predicted by {\tt GALFORM+UCL$_{-}$PDR} model. 
Observational estimates of \citet{Keres03}
for samples of galaxies selected in the $B$-band (triangles) and at $60\,\mu$m (filled circles), 
are also shown. The predictions of the models are shown for 
the {\tt GALFORM+UCL$_{-}$PDR} model when (i) $G_{\rm UV}=G_{\rm UV}(\Sigma_{\rm SFR})$ (Eq.~\ref{para1}; 
dashed line), 
(ii) $G_{\rm UV}=G_{\rm UV}(\Sigma_{\rm SFR}, Z_{\rm gas}, \Sigma_{\rm gas})$ (Eq.~\ref{para2}; solid line), 
and (iii) using $G_{\rm UV}$ as in (ii) but assuming AGN do not contribute to heat the ISM (dot-dashed line). 
For reference, we also show the predictions of the {\tt GALFORM} model without
the PDR coupling, assuming 
two constant $X_{\rm CO}$ factors, 
$\rm X_{\rm CO(1-0)}=(2,0.8)\times 10^{20}\, \rm cm^{-2}\,(K\,km\,s^{-1})^{-1}$ 
for quiescent and starburst galaxies, respectively (dotted line).} 
\label{LFCOz0}
\end{center}
\end{figure}

Fig.~\ref{LFCOz0} shows the $\rm CO(1-0)$ LF at $z=0$ for the hybrid 
{\tt GALFORM+UCL$_{-}$PDR} model, 
for the two parametrizations of $G_{\rm UV}$: 
(i) $G_{\rm UV}(\Sigma_{\rm SFR})$ (Eq.~\ref{para1}) and 
(ii) $G_{\rm UV}(\Sigma_{\rm SFR}, Z_{\rm gas}, \Sigma_{\rm gas})$ (Eq.~\ref{para2}). 
We also show the latter model, (ii), without the inclusion of AGN as an ISM heating source. 
For reference, we also show the predictions of the {\tt GALFORM} model without the processing of the 
PDR model, in the simplistic case where we assume two constant conversion factors, 
$X_{\rm CO(1-0)}=2\times 10^{20}\, \rm cm^{-2}\, (K\,km\,s^{-1})^{-1}$ 
for quiescent galaxies and $X_{\rm CO(1-0)}=0.8\times 10^{20}\, \rm cm^{-2}\, (K\,km\,s^{-1})^{-1}$ for starbursts.
Observational estimates of the $\rm CO(1-0)$ LF made using 
both, $B$-band and $60\,\mu$m selected samples, are plotted 
as symbols \citep{Keres03}. 

Differences between the predictions 
of the model using different assumptions about $G_{\rm UV}$ become evident at CO luminosities brighter than the 
break in the CO LF (i.e. $\rm log(\it L_{\rm CO}/\rm Jy\, km/s\, Mpc^{2})\approx 6.7$). 
The model assuming a dependence of $G_{\rm UV}$ solely on $\Sigma_{\rm SFR}$ predicts 
a larger number density of bright galaxies due to the higher $G_{\rm UV}$
values in galaxies with large molecular 
mass and high SFRs. We show later that the kinetic gas temperatures of the {\tt GALFORM+UCL$_{-}$PDR} model, 
 when assuming
a dependence of $G_{\rm UV}$ solely on $\Sigma_{\rm SFR}$,
 are very high and also translate into unrealistic emission from higher order CO transitions. 
The values of $G_{\rm UV}$ are smaller when including the dependence on the optical depth, $\tau_{\rm UV}$, 
given 
that the increase in $\Sigma_{\rm SFR}$ is compensated by an increase in $\tau_{\rm UV}$, which brings 
$G_{\rm UV}$ down. This model predicts a LF which is closer to and in 
reasonable agreement with the observations. When AGN are not 
included as a heating mechanism, the model predictions for
the CO$(1-0)$ LF are not affected, indicating that lower CO transitions are not sensitive to the 
presence of AGN. However, as we show later (Fig.~\ref{LFCOall}), the emission in high CO
transitions is very sensitive to the presence of an AGN.
The {\tt GALFORM} model without the PDR (i.e. using two ad hoc constant values of 
$X_{\rm CO}$ for starburst and quiescent galaxies) 
gives a LF closer to the observed number
density of bright galaxies. This happens because galaxies in the bright-end of the CO LF 
mainly correspond to quiescent, gas-rich galaxies, whose $G_{\rm UV}>G_{0}$, driving 
lower $X_{\rm CO}$ in the {\tt GALFORM+UCL$_{-}$PDR} model 
compared to the value typically assumed for quiescent galaxies ($X_{\rm CO}=2\times 
10^{20}\, \rm cm^{-2}\, (K\,km\,s^{-1})^{-1}$).

The predictions of the {\tt GALFORM+UCL$_{-}$PDR} model 
using the form $G_{\rm UV}(\Sigma_{\rm SFR}, Z_{\rm gas}, \Sigma_{\rm gas})$ (Eq~\ref{para2}), give a
reasonable match to the observational data from \citet{Keres03}. We remind the reader that the model has not been tuned to
reproduce the CO LF. However, it is important to bear in mind
that the CO LF from \citet{Keres03} is not based on a blind CO survey, but instead on
galaxy samples selected using $60\,\mu \rm m$ or $B$-band
fluxes. These criteria might bias the LF 
towards galaxies with large amounts of dust or
large recent SF. 

\subsection{The CO-to-Infrared luminosity ratio} 

\begin{figure}
\begin{center}
\includegraphics[trim = 2mm 1mm 1mm 1mm,clip,width=0.49\textwidth]{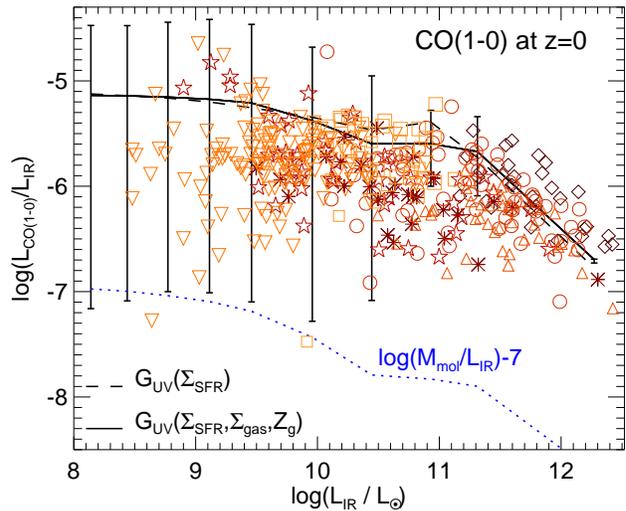}
\caption{CO$(1-0)$-to-IR luminosity ratio as a function of IR luminosity. 
Lines show the median of the {\tt GALFORM+UCL$_{-}$PDR} model when $G_{\rm UV}$ 
is estimated from Eqs.~\ref{para1} (dashed line) and \ref{para2} (solid line).
Errorbars show the 
$10$ and $90$ percentiles of galaxies in the model, and for clarity, are only shown for the 
case of $G_{\rm UV}$ estimated from Eq.~\ref{para2}. For reference, 
we also show as dotted line the predictions of the L11 model for the 
molecular-to-IR luminosity ratio, log$(M_{\rm mol}/L_{\rm IR}\, M_{\odot}/L_{\odot})$, shifted 
by an arbitrary factor of $7$~dex. Observations from 
\citet{Young91}, \citet{Yao03}, \citet{Solomon97}, \citet{Bayet09b}, \citet{Papadopoulos11}, 
\citet{Rahman11} and \citet{Lisenfeld11} are shown as symbols.}
\label{COladder2}
\end{center}
\end{figure}

Fig.~\ref{COladder2} shows the $L_{\rm CO(1-0)}/L_{\rm IR}$ ratio 
as a function of $L_{\rm IR}$. Lines show the median of the {\tt GALFORM+UCL$_{-}$PDR} model predictions when using 
$G_{\rm UV}(\Sigma_{\rm SFR})$ (Eq.~\ref{para1}; dashed line)  and 
$G_{\rm UV}(\Sigma_{\rm SFR},Z_{\rm gas},\Sigma_{\rm gas})$ (Eq.~\ref{para2}; solid line). 
Errorbars show the $10$ and $90$ percentiles of the distributions. 
We also show, for reference, the molecular mass-to-IR luminosity ratio, 
$M_{\rm mol}/L_{\rm IR}$, predicted by 
the L11 model and shifted by an arbitrary factor of $7$~dex (dotted line). 
Symbols show an observational compilation of local LIRGs and 
ULIRGs. 

The model predicts that the $L_{\rm CO(1-0)}/L_{\rm IR}$ ratio decreases as the IR luminosity 
increases. This trend is primarily driven by the dependence of the molecular mass-to-IR luminosity ratio on  
the IR luminosity which has the same form (dotted line). 
The trend of decreasing $M_{\rm mol}/L_{\rm IR}$ with increasing IR luminosity is driven 
by the gas metallicity-IR luminosity relation. Gas metallicity declines 
as the IR luminosity decreases (bottom panel of Fig.~\ref{MMr}), 
which results in lower dust-to-total gas mass ratios and therefore lower IR luminosities 
for a given SFR. The molecular mass is not 
affected by gas metallicity directly since it depends on the hydrostatic pressure of the disk 
(see $\S 2.1$). Note that this effect affects galaxies with $L_{\rm IR}<5\times 10^{10}\, L_{\odot}$, 
given that the gas metallicity-IR luminosity relation flattens above this IR luminosity 
(bottom panel of Fig.~\ref{MMr}).
The distributions of $L_{\rm CO(1-0)}/L_{\rm IR}$ predicted by the model 
extend to very low $L_{\rm CO(1-0)}/L_{\rm IR}$ ratios (as shown by the errorbars in Fig.~\ref{COladder2}). 
This is due to satellite galaxies in groups and clusters, 
which tend to have lower molecular mass-to-IR luminosity ratios, 
but that are relatively bright in IR due to their high gas metallicities (solar or 
supersolar), and therefore, large dust-to-gas mass ratios. 

At $5\times 10^{10}\, L_{\odot}<L_{\rm IR}<5\times 10^{11}\, L_{\odot}$, the 
$L_{\rm CO(1-0)}/L_{\rm IR}$-$L_{\rm IR}$ relation tends to flatten. This is due to a transition from 
galaxies dominated by quiescent SF to 
starburst galaxies, and the two different SF laws assumed in the model 
(see $\S 2.1$). The SF law determines how fast the cold gas
is converted into stars, thus playing a key role in determining 
the molecular reservoir at a given time. 
 In starburst galaxies, the SF timescale 
depends on the dynamical timescale of the bulge component with a floor
(Eq.~\ref{SFlawSB}). Starburst galaxies, which largely contribute to the number density at 
$L_{\rm IR}>5\times 10^{10}\, L_{\odot}$, have similar SF timescales
given their similar properties in stellar mass and size, 
therefore resulting in, similar molecular-to-SFR ratios, except for the brightest ones 
with $L_{\rm IR}> 5\times 10^{11}\, L_{\odot}$. This effects dominates the 
behaviour of the $L_{\rm CO(1-0)}/L_{\rm IR}$ ratio, with a second order 
contribution from variations in $X_{\rm CO}$, which tend to be small at
these high gas surface densities (see Fig.~\ref{XCOsplot}). This prediction of the model 
explains what has been observed in local and a few high-redshift galaxies: 
variations in the CO-to-IR luminosity ratio are of the same order 
as the variations of molecular mass-to-IR luminosity ratios, as inferred 
from the dust emission (\citealt{Leroy11}; \citealt{Magdis11}). 
For the brightest galaxies, 
$L_{\rm IR}> 5\times 10^{11}\, L_{\odot}$, the SF timescale decreases rapidly with increasing 
IR luminosity and consequently, the molecular mass-to-IR luminosity ratio also decreases. 

We conclude that the {\tt GALFORM+UCL$_{-}$PDR model} is able to explain
the observed CO$(1-0)$ emission of galaxies in the local Universe and its relation to the IR luminosity.

\section{The CO emission of galaxies in multiple transitions}

\begin{figure*}
\begin{center}
\includegraphics[width=1.\textwidth]{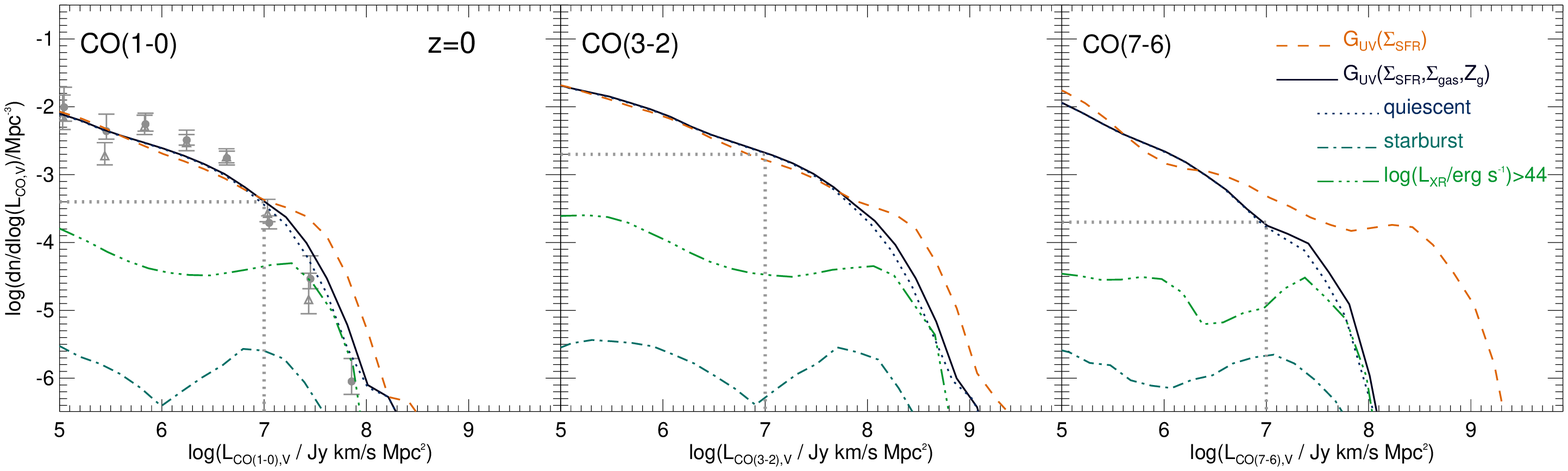}
\includegraphics[width=1.\textwidth]{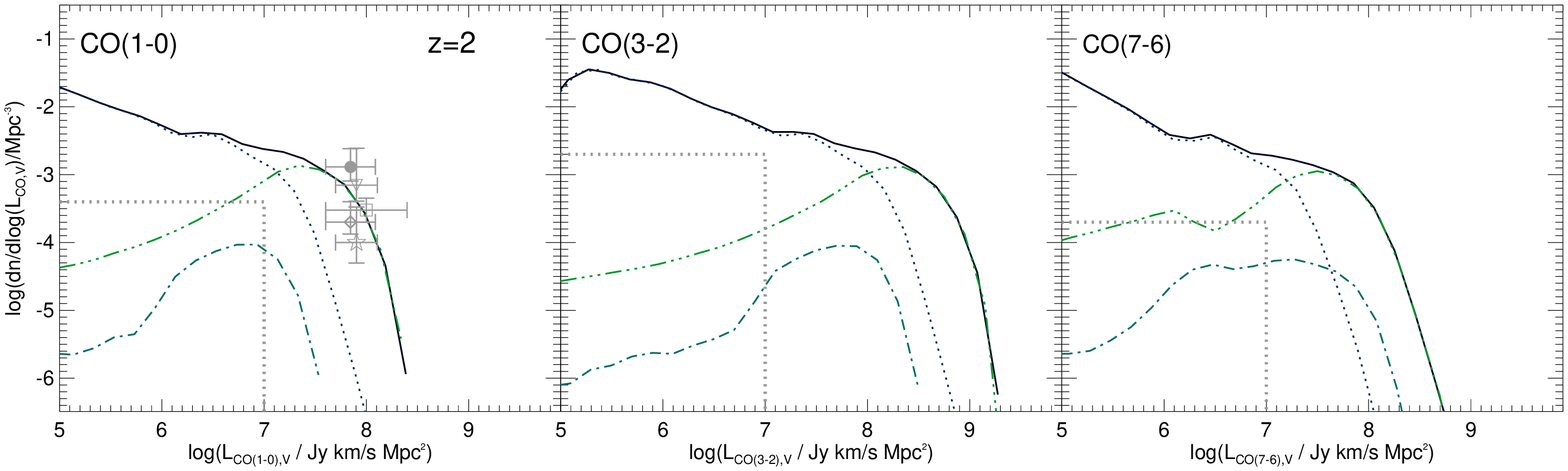}
\caption{{\it Top Row:} The $\rm CO(1-0)$ (left-hand panel), $\rm CO(3-2)$ (middle panel) and $\rm CO(7-6)$ 
(right-hand panel) luminosity functions at $z=0$
predicted by the {\tt GALFORM+UCL$_{-}$PDR} model. 
In the left-hand panel, the observational estimates of \citet{Keres03}
are shown as in Fig.~\ref{LFCOz0}.
The solid and dashed lines show the 
predictions for the two different assumptions used to estimate $G_{\rm UV}$, 
Eqs.~\ref{para1} and \ref{para2}, respectively. We use the predictions of the model 
using the $G_{\rm UV}$ parametrisation of Eq.~\ref{para2} to show the contributions to the 
LF from 
starburst galaxies (dot-dashed lines) and normal star-forming galaxies (dotted) 
without a bright AGN, and galaxies which host a bright AGN 
(X-ray luminosities $L_{\rm X}>10^{44}\, \rm erg\, s^{-1}$; triple-dot-dashed line). 
{\it Bottom Row:} The same as the top row but at $z=2$. In this set of plots we only 
show the predictions of the model using the $G_{\rm UV}$ of Eq.~\ref{para2}. { In the left-hand panel, the observational estimates of  
\citet{Aravena12} and \citet{Daddi10} are shown in symbols. The solid circle and empty diamond correspond to the 
observational data of Aravena et al. without and with correction for the overdensity of the field, respectively, 
while the empty triangle and square show the estimates of the number density when only galaxies with spectroscopically confirmed 
redshifts are considered, without and with correction for the overdensity of the field, respectively. 
The empty star corresponds to the estimate of \citet{Daddi10}.} To show the evolution in the LF, the
dotted straight lines show the number density of galaxies at a CO luminosity of
$10^7\, \rm Jy\, km/s\, Mpc^2$ at $z=0$ in both sets of rows.}
\label{LFCOall}
\end{center}
\end{figure*}

We now focus on the predictions of the {\tt GALFORM+UCL$_{-}$PDR} model 
for the CO emission of galaxies in multiple CO lines in the local and high-redshift universe. 
We focus on the CO LF of galaxies, the relation 
between the CO and IR luminosity, and the CO SLED. In contrast to the case of the CO$(1-0)$ transition, the 
available observational 
data for higher CO transitions are scarce and limited to individual objects, 
instead of homogeneous samples of galaxies. 
To carry out the fairest comparison possible at present, we 
 select model galaxies in order to sample similar IR luminosity distributions to those in the 
observational catalogues.

\subsection{The luminosity function of multiple CO lines}

\begin{figure}
\begin{center}
\includegraphics[width=0.49\textwidth]{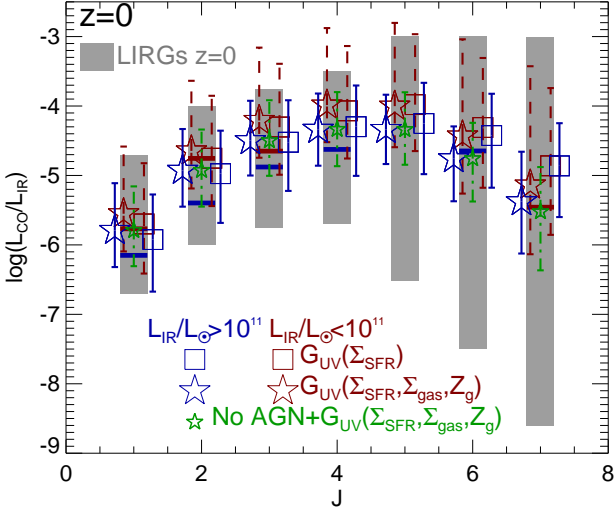}
\caption{CO$(J\rightarrow J-1)$ to IR luminosity ratio as a function of upper level J at $z=0$. 
Grey solid bars show the range of observed ratios reported by \citet{Papadopoulos11} 
for $70$ LIRGs at $z\le 0.1$. Horizontal segments show the median in the observed data 
for two IR luminosity bins, $L_{\rm IR}/L_{\odot}<10^{11}$ (dark red { with errorbars as dashed lines}) 
and $L_{\rm IR}/L_{\odot}>10^{11}$ (blue { with errorbars as solid lines}). 
We show as symbols the predictions of a sample of model galaxies randomly chosen to have the same IR 
luminosity distribution as the Papadopoulos et al. sample.
Symbols and errorbars correspond to the median and $10$ and $90$ percentiles of the  
predictions for the {\tt GALFORM+UCL$_{-}$PDR} model using the 
$G_{\rm UV}$ parametrisations of Eqs.~\ref{para1} (squares) and \ref{para2} (large stars). 
For reference we also show for the bright IR luminosity bin 
the predictions of the model using $G_{\rm UV}$ from Eq.~\ref{para2} 
when AGN are not considered as an ISM heating mechanism (small stars).}
\label{COladder1}
\end{center}
\end{figure}

The top row of Fig.~\ref{LFCOall} shows the $\rm CO(1-0)$, 
$\rm CO(3-2)$ and $\rm CO(7-6)$ LFs at $z=0$ for the 
{\tt GALFORM+UCL$_{-}$PDR} model 
using the scalings of $G_{\rm UV}(\Sigma_{\rm SFR})$ (Eq.~\ref{para1}) and 
$G_{\rm UV}(\Sigma_{\rm SFR},Z_{\rm gas},\Sigma_{\rm gas})$ (Eq.~\ref{para2}). The contributions 
from X-ray bright AGNs, with $L_{\rm X}>10^{44}\, \rm s^{-1}\, erg$, and from 
quiescent and starburst galaxies with 
$L_{\rm X}<10^{44}\, \rm s^{-1}\, erg$, 
are shown separately only for the 
model using $G_{\rm UV}(\Sigma_{\rm SFR},Z_{\rm gas},\Sigma_{\rm gas})$. 
Note that X-ray bright AGNs can correspond to both quiescent and starburst galaxies. 
The observational results 
for the $\rm CO(1-0)$ LF from \citet{Keres03} {at $z=0$ and from \citet{Aravena12} and \citet{Daddi10} at $z=2$ 
are also plotted in the top and bottom left-hand panels, respectively}. 
The model using $G_{\rm UV}(\Sigma_{\rm SFR})$ predicts a higher number density of bright galaxies 
for the three CO transitions shown in Fig.~\ref{LFCOall} due to the fact that with this assumption, 
galaxies typically have a higher value of $G_{\rm UV}$ than in the parametrisation of 
Eq.~\ref{para2}, which leads to lower values of $X_{\rm CO}$. The offset
in the bright-end between the model predictions when using 
$G_{\rm UV}(\Sigma_{\rm SFR})$ and $G_{\rm UV}(\Sigma_{\rm SFR},Z_{\rm gas},\Sigma_{\rm gas})$ 
increases for higher CO transitions, since 
$J>4$ CO transitions are more sensitive to changes in kinetic temperature, and therefore in $G_{\rm UV}$. For 
$G_{\rm UV}(\Sigma_{\rm SFR})$, galaxies are on average predicted to be  
very bright in the CO$(7-6)$ transition. As we show later, this model predicts  
an average CO$(7-6)$ luminosity brighter than observed for local 
LIRGs (see Fig.~\ref{COladder1} in $\S 4.2.1$), but still consistent with the 
observations within the errorbars. 
Quiescent galaxies in the model are responsible for shaping the faint end of the
CO LF, regardless of the CO transition. Starburst galaxies make a very small
contribution to the CO LFs at $z=0$, given their low number density.

Galaxies which host 
a X-ray bright AGN are an important contributor to the bright-end of the CO LF, along with normal 
star-forming galaxies, regardless of the
transition. In the case of the CO$(1-0)$ and CO$(3-2)$ transitions, 
this is not due to the presence of the AGN in these galaxies, but instead 
to the large molecular gas reservoir, the typically high gas metallicities, and the high SFRs, which on average 
produce higher $G_{\rm UV}$, and therefore more CO luminosity per molecular mass.
The powerful AGN is therefore a consequence of the large gas reservoirs, which fuel 
large accretion rates, 
along with a massive central black hole, and has only a secondary effect through increasing the 
kinetic temperature, that is not enough to produce a visible effect on the low CO transitions. 
However, the CO$(7-6)$ 
transition is slightly more sensitive to variations in the kinetic temperature of the gas, as 
we show in $\S 4.2$. 
The contribution from bright AGN and quiescent galaxies to the bright-end
of the CO LF at $z=0$ is very similar, regardless of the CO transition. 
This is due to the X-ray luminosity threshold chosen
to select AGN bright galaxies, $L_{\rm XR}>10^{44}\, \rm erg\, s^{-1}$, which 
takes out most of quiescent galaxies in the bright-end of the CO LF, which have hard X-ray luminosities 
 in the range $10^{43}<L_{\rm XR}/\rm erg\, s^{-1}<10^{44}$. 

The bottom row of Fig.~\ref{LFCOall} is the same as the top row but shows the LFs at $z=2$. In this case 
we only show the predictions of the {\tt GALFORM+UCL$_{-}$PDR} model in the 
$G_{\rm UV}$ approximation of  Eq.~\ref{para2}. To illustrate 
evolution between $z=0$ and $z=2$, the dotted straight 
lines show the number density of galaxies at $z=0$ with a luminosity of 
$10^7\, \rm Jy\, km/s\, Mpc^2$ in the different CO transitions. 
Bright CO 
galaxies are more common at $z=2$, which is reflected in the higher number density of galaxies with 
$L_{\rm CO,V}> 10^{7} \rm Jy\, km/s\, Mpc^{2}$ compared to $z=0$. 
Bright AGNs, which 
are more common and brighter at $z=2$ than at $z=0$, are responsible for most of the 
evolution in the bright-end of the CO LF with redshift, with a less 
important contribution from quiescent and starburst galaxies that host fainter AGN. 
In the faint-end, there is an significant 
increase in the number density of galaxies, driven by the evolution 
of quiescent galaxies. 
In general, the LF for higher CO transitions shows stronger evolution with redshift than it does 
for lower 
CO transitions, again indicating that the higher CO transitions are more sensitive to variations 
in $G_{\rm UV}$ and $F_{\rm X}$. 
From an observational point of view, measuring CO luminosity ratios, such as 
the CO$(7-6)$-to-CO$(1-0)$ ratio, is promising for 
constraining the average physical state of the molecular gas. 
However, in terms of 
estimating the total molecular mass in galaxies, lower CO transitions are more
useful, given their lower sensitivity to changes in the conditions in the
ISM in galaxies. { Our predictions for the CO$(1-0)$ at $z=2$ agree very well with 
the observed number density of bright CO$(1-0)$ galaxies reported by 
\citet{Daddi10} and \citet{Aravena12}.
However, the uncertainty in the inferred space density displayed by the 
observations at $z=2$ is large, suggesting that further observations, desirably from CO blind surveys,  
are necessary to put better constraints in the CO LF.}

Our predictions for the LF show that 
intermediate CO transitions are brighter in units of the velocity-integrated CO 
luminosity than lower and higher order CO 
transitions. This trend is similar to the predictions of \citet{Obreschkow09d}, who used a
completely different approach, which relied on estimating a gas temperature based on the SFR
surface density or AGN bolometric luminosity under local thermodynamic
equilibrium (i.e. a single gas phase). However, Obreschkow et al. predict a 
significant decrease in the number density of faint CO galaxies as the 
upper level $\rm J$ increases, behaviour that is not seen in our model. 
A possible explanation for this is that their model assumes local thermodynamic
equilibrium, for which high order CO transitions would be thermalised. This, in addition to 
the parameters Obreschkow et al. use to estimate $L_{\rm CO}$, can 
lead to much lower CO emission in high-J transitions compared to that in our model. 
For example at $T_{\rm K}=10$~K, and using the equations and parameters given in 
Obreschkow et al., a ratio of $L_{\rm CO(7-6),V}/L_{\rm
CO(1-0),V}\approx 10^{-3}$ is obtained, while our model predicts 
$L_{\rm CO(7-6),V}/L_{\rm
CO(1-0),V}\approx 0.1$ for the same temperature. Our approach does not 
require any of these assumptions given that the PDR model is designed to 
represent much more accurately the excitation state of 
GMCs.

In general, the {\tt GALFORM+UCL$_{-}$PDR} model predicts a higher number
density of bright galaxies at high-redshifts, a trend which is slightly more pronounced for the higher CO
transitions. For low CO transitions, the main driver of this effect is the higher
number density of galaxies with large molecular gas reservoirs at high redshift (see
\citealt{Lagos11}). For high CO transitions what makes the effect stronger is the higher
average kinetic temperatures of the gas in molecular clouds at high redshifts
(see Fig.~\ref{TgasEvo}). 

\begin{figure}
\begin{center}
\includegraphics[width=0.49\textwidth]{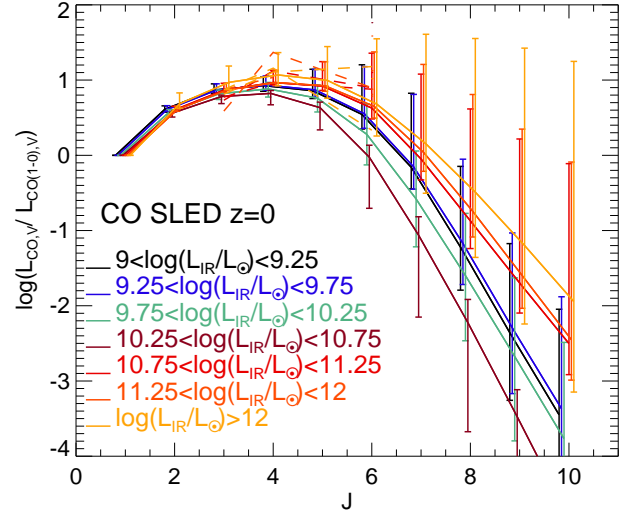}
\includegraphics[width=0.49\textwidth]{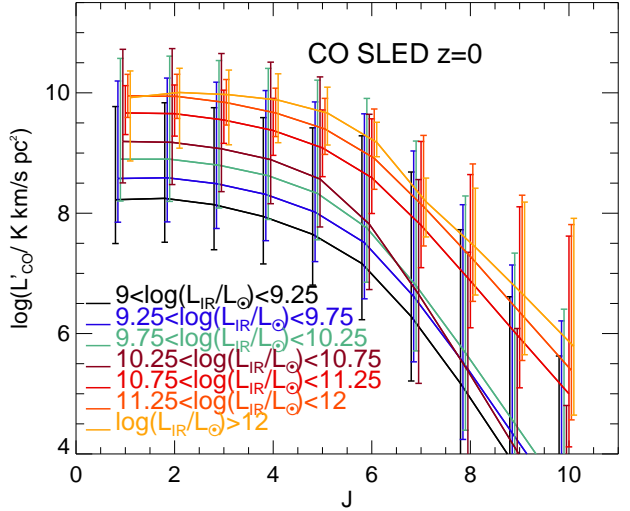}
\caption{ {\it Top panel:} The velocity-integrated luminosity normalized to the CO$(1-0)$ luminosity 
as a function of the upper quantum level of the CO rotational transition, $J$, 
predicted by the {\tt GALFORM+UCL$_{-}$PDR} model at $z=0$ for galaxies 
with IR luminosities in different ranges, as labelled, and 
using the $G_{\rm UV}$ parametrisation of Eqs.~\ref{para2}. Solid lines and errorbars show the
medians and $10$ and $90$ percentile ranges of the distributions. 
Dashed lines show individual galaxies from the \citet{Papadopoulos11} observational sample, following 
the same colour code as the model galaxies.
{\it Bottom panel:} Brightness temperature luminosity, $L^{\prime}_{\rm CO}$, 
as a function of $J$ for galaxies with IR luminosities in different ranges, as labelled.
{ Note that $L^{\prime}_{\rm CO(1-0)}$ monotonically increases with IR luminosity}.}
\label{COladderz02}
\end{center}
\end{figure}

\subsection{The CO-to-IR luminosity ratio and the CO SLED} 

In this section we study the CO-to-IR
luminosity ratio for multiple CO lines and compare to observational
data in the local and high redshift Universe.

\subsubsection{The CO-to-IR luminosity ratio and CO SLED in the local universe}

Observations have shown that emission from multiple CO transitions can help to constrain the 
state of the ISM in galaxies through the comparison with the predictions of 
PDR and LVG models. 
In the local Universe, around $100$ galaxies have been observed in more than one CO transition.
However, the 
one caveat to bear in mind is the selection of these samples, as they are built from studies 
of individual galaxies and are, therefore, 
inevitably biased towards bright galaxies. 
To try to match the composition of the observational sample, we select galaxies from the model with the same 
distribution of IR luminosities as in the observed samples.
In this section we present this comparison, which allows us to study 
whether or not the model predicts galaxies that reproduce the observed CO ladder.
 
We compare the model predictions with the observational catalogue 
presented in \citet{Papadopoulos11}. This sample comprises $70$ LIRGs and ULIRGs 
at $z\le 0.1$ which have the emission of several CO transitions measured, as well as other 
molecular species. The IR luminosities  of these galaxies cover the range 
$10^{10}-5\times 10^{12}\, L_{\odot}$. 
We randomly select galaxies in the model at $z=0$ to have the same distribution of IR 
luminosities as the sample of \citet{Papadopoulos11}.

Fig.~\ref{COladder1} shows the predicted CO-to-IR luminosity ratio 
for different transitions 
compared to the observational data, in two bins of IR luminosity, 
$10^{10}<L_{\rm IR}/L_{\odot}<10^{11}$ and $10^{11}<L_{\rm IR}/L_{\odot}<5\times 10^{12}$. 
In the case of the observations, gray bands show  
the whole range of observed CO-to-IR luminosity ratios, 
while the horizontal segments show the medians of the bright (blue) and faint (dark red) 
IR luminosity bins.
In the case of the model, we show the medians and $10$ and $90$ 
percentiles of the distributions as symbols and errorbars, respectively, where 
dark red and blue symbols corresponds to the low and high luminosity bins, respectively.
Model predictions are presented for the two parametrisations of 
$G_{\rm UV}$ discussed in $\S 3.1.1$ (see Eqs.~\ref{para1}-\ref{para2}).
We also show for the bright IR
luminosity bin, the CO-to-IR 
luminosity ratios for the model using $G_{\rm UV}(\Sigma_{\rm SFR},Z_{\rm gas},\Sigma_{\rm gas})$ 
(Eq.~\ref{para2}) and
assuming no heating of the ISM by AGN (small stars).  
The model predicts $L_{\rm CO}/L_{\rm IR}$ ratios which are well within the observed 
ranges. For higher CO transitions, the model predicts broader distributions 
of the $L_{\rm CO}/L_{\rm IR}$ ratio 
than at low CO transitions, independent of the $G_{\rm UV}$ 
parametrisation. The model also predicts that galaxies 
in the bright IR luminosity bin have slightly lower $L_{\rm CO}/L_{\rm IR}$ ratios compared to 
galaxies in the faint IR luminosity bin.

The two parametrisations of $G_{\rm UV}$ predict $L_{\rm CO}/L_{\rm IR}$ ratios 
that are only slightly offset, except for the highest CO transitions, $\rm J>5$, where the 
predictions differ by up to $\approx0.5$dex. 
At $\rm J>5$, the model in which $G_{\rm UV}$ depends on the average UV optical depth, 
$G_{\rm UV}(\Sigma_{\rm SFR}, Z_{\rm gas},\Sigma_{\rm gas})$, predicts on average  
$L_{\rm CO}/L_{\rm IR}$ ratios in better agreement with the observations than those predicted 
by the $G_{\rm UV}$ depending solely on $\Sigma_{\rm SFR}$.
This is due 
to the fact that galaxies with very high $\Sigma_{\rm SFR}$, which drives high UV
 production, also tend to have 
high $Z_{\rm gas}\, \Sigma_{\rm gas}$, decreasing the UV ionizing background if
the average UV optical depth is considered.  
For lower CO transitions, the difference between the predictions of the model when $G_{\rm UV}$ is estimated 
as in Eqs.~\ref{para1} and \ref{para2}, becomes more evident for galaxies 
that are bright in CO, which affects the bright end of the CO 
luminosity function, as discussed in $\S 3.1$. 
Galaxies in the fainter 
IR luminosity bin correspond primarily to normal star-forming galaxies with 
$G_{\rm UV}/G_{0}\approx 1-10$, while galaxies in the brighter IR luminosity bin are a 
mixture of normal star-forming and starburst galaxies. The range of  
$G_{\rm UV}$ in these galaxies varies significantly. Starburst galaxies usually 
have larger $G_{\rm UV}$ in the range $G_{\rm UV}/G_{0}\approx10-10^3$. 
Faint IR galaxies in the model, with $L_{\rm
IR}<10^{9}\, L_{\odot}$, can also
correspond to passive galaxies, whose UV ionizing background is very small,
$G_{\rm UV}/G_{0}\approx 0.01-1$. 

The variation in $G_{\rm UV}$ within the faint and bright IR
luminosity bins has a direct consequence on the range of gas kinetic temperatures displayed by
galaxies in each bin. Galaxies in the faint IR bin have $T_{\rm K}\approx
10-20\, \rm K$, while galaxies in the bright IR bin have $T_{\rm K}\approx
10-60\, \rm K$. The presence of an AGN also has an effect on the kinetic
temperature of the gas, and therefore on the CO emission of galaxies, 
as seen from the small stars in Fig.~\ref{COladder1}. When assuming that the AGN
does not heat the ISM of galaxies, galaxies appear to have lower CO-to-IR 
ratios for 
$J>6$ transitions by a factor $\approx 1.7$, while 
lower transitions are largely unaffected. This indicates again
that high CO transitions are useful to constrain the effect of AGN in heating
the ISM.

Fig.~\ref{COladderz02} shows the CO SLED 
in units of velocity-integrated CO luminosity, $L_{\rm CO,V}$ (top panel), and 
brightness temperature luminosity, $L^{\prime}_{\rm CO}$ (bottom panel), 
for $z=0$ galaxies with IR luminosities in different luminosity bins, as labelled.
Individual galaxies from the 
\citet{Papadopoulos11} observational sample of LIRGs are 
shown as dashed lines in the top panel of Fig.~\ref{COladderz02}.   
A typical way to show the CO SLED is velocity-integrated luminosity 
normalised by $L_{\rm CO(1-0),V}$, 
given that this way the SLED shows a peak,  which 
indicates the degree of excitation: the higher the J of the peak, the higher the gas kinetic temperature of 
molecular clouds, which typically indicates more SF and/or AGN activity (e.g. \citealt{Weiss07}). 
When the CO SLED is shown in brightness temperature luminosity there is no clear peak 
(bottom panel of Fig.~\ref{COladderz02}) . This happens 
because $L_{\rm CO,V}$ and $L^{\prime}_{\rm CO}$ have different 
dependencies on $J$ (see Appendix~\ref{App:COIR}). 
The {\tt GALFORM+UCL$_{-}$PDR} model predicts a peak in the CO SLED at $\rm J=4$ for 
galaxies with $L_{\rm IR}\lesssim 10^{11}\, L_{\odot}$ and at $\rm J=5$ 
for galaxies with $L_{\rm IR}\gtrsim 10^{11}\, L_{\odot}$, due to the starburst nature of 
the latter. 
We find that the lowest IR luminosity bin, $10^{9}<L_{\rm IR}/L_{\odot}<1.7\times 10^{9}$, 
shows a peak at higher J values, closer to 
starburst galaxies. This is due to the lower gas metallicities of these galaxies which increases 
$G_{\rm UV}$ and $T_{\rm K}$. 
Our predictions agree with the observed peaks of LIRGs (dashed lines). However, we remind the 
reader that the LIRG catalogue of \citet{Papadopoulos11} is not a statistically complete sample.  
Further observations are needed 
to construct volume-limited samples of galaxies with CO measurements in order to 
better constrain the physics of the ISM.

\subsubsection{Redshift evolution of the CO-to-IR luminosity ratio and the 
CO SLED at high redshift}

\begin{figure*}
\begin{center}
\includegraphics[trim = 2mm 1mm 1mm 1mm,clip,width=1.0\textwidth]{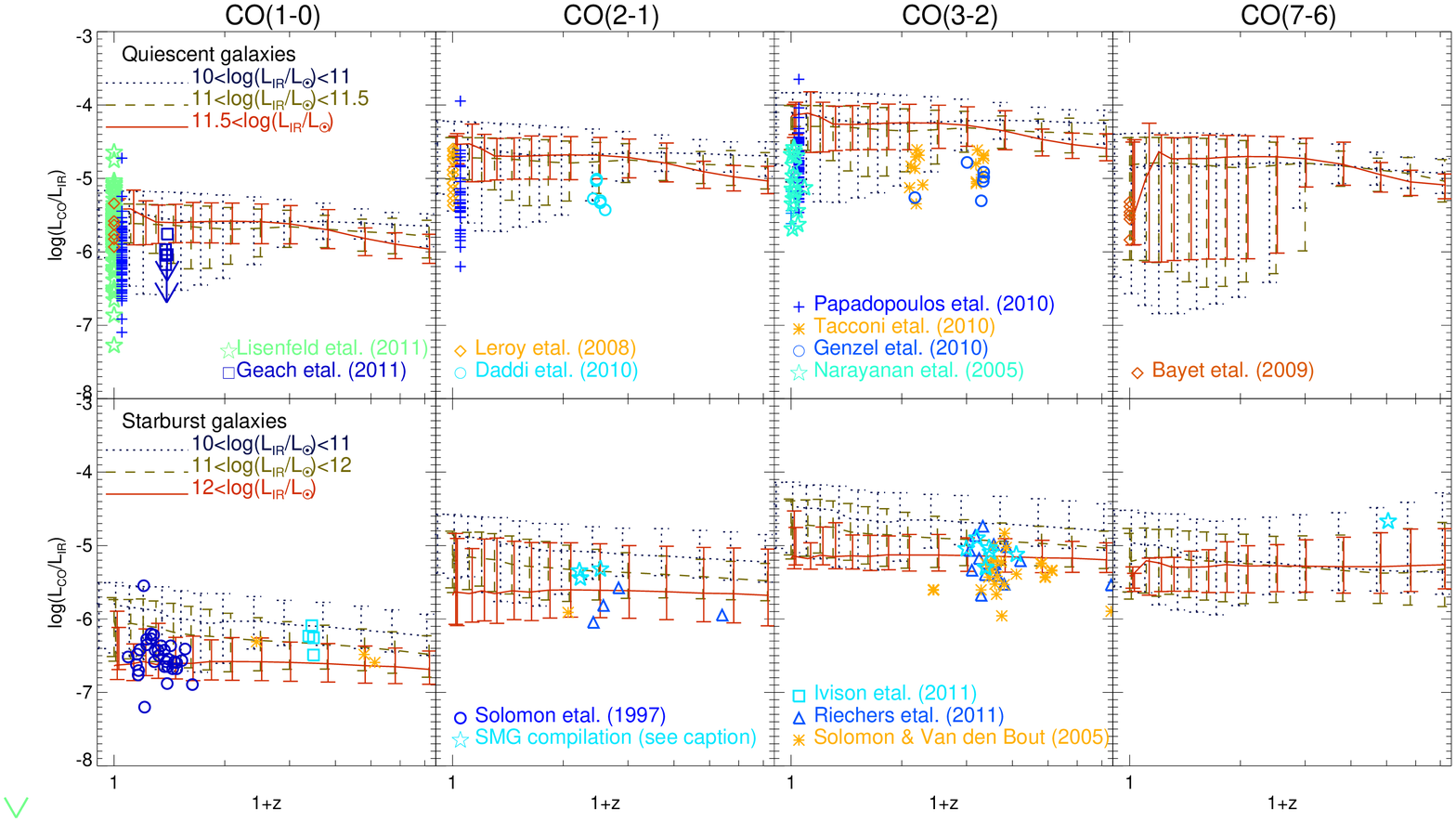}
\caption{{\it Top Row:} CO-to-IR luminosity ratio as a 
function of redshift for quiescent
galaxies in the {\tt GALFORM+UCL$_{-}$PDR} model and for four CO transitions, 
CO$(1-0)$ (left-hand panel), CO$(2-1)$ (middle-left panel), CO$(3-2)$ (middle-right panel) and CO$(7-6)$ 
(right-hand panel), for IR luminosities in different ranges, as shown in the label. 
Lines and errorbars represent the predicted median $L_{\rm CO}/L_{\rm IR}$ and
the $10$ and $90$ percentiles of the distributions.
Also shown are the following
observational data: local normal spiral galaxies from \citet{Leroy08} (diamonds), 
\citet{Lisenfeld11} (stars), \citet{Bayet09b} (diamonds);
 LIRGs from \citet{Narayanan05} (stars) and \citet{Papadopoulos11} (crosses);  
star-forming galaxies at $z\approx 1.2$ and $z\approx 2.3$ from
\citet{Tacconi10} (asterisks) and \citet{Genzel10} (circles); normal star-forming galaxies, 
BzK selected, from \citet{Daddi10} (circles); normal star-forming galaxies 
at intermediate redshifts from \citet{Geach11} (squares). Arrows indicate 
upper limits of IR sources with no detections of CO. {\it Bottom Row:} 
the same as the top row, but for starburst galaxies. 
Also shown are the observational results for ULIRGs from \citet{Solomon97} (circles), 
SMG from \citet{Solomon05} (asterisks; which compiled the SMG data from 
Frayer et al. 1998,1999, \citealt{Neri03},
\citealt{Sheth04} , \citealt{Greve05}), \citet{Ivison11} (squares) and 
a compilation including \citet{Tacconi06}, \citet{Casey09}, \citet{Bothwell10} and
\citet{Engel10} (stars), and QSOs from \citet{Riechers10} (triangles). We remind the reader 
that most of the observational data plotted here do not directly measure total IR luminosity, but infer it 
from an estimated SFR or from mid-IR or the submillimeter emission (see text for details).}
\label{LFCOIR}
\end{center}
\end{figure*}

\begin{figure}
\begin{center}
\includegraphics[width=0.49\textwidth]{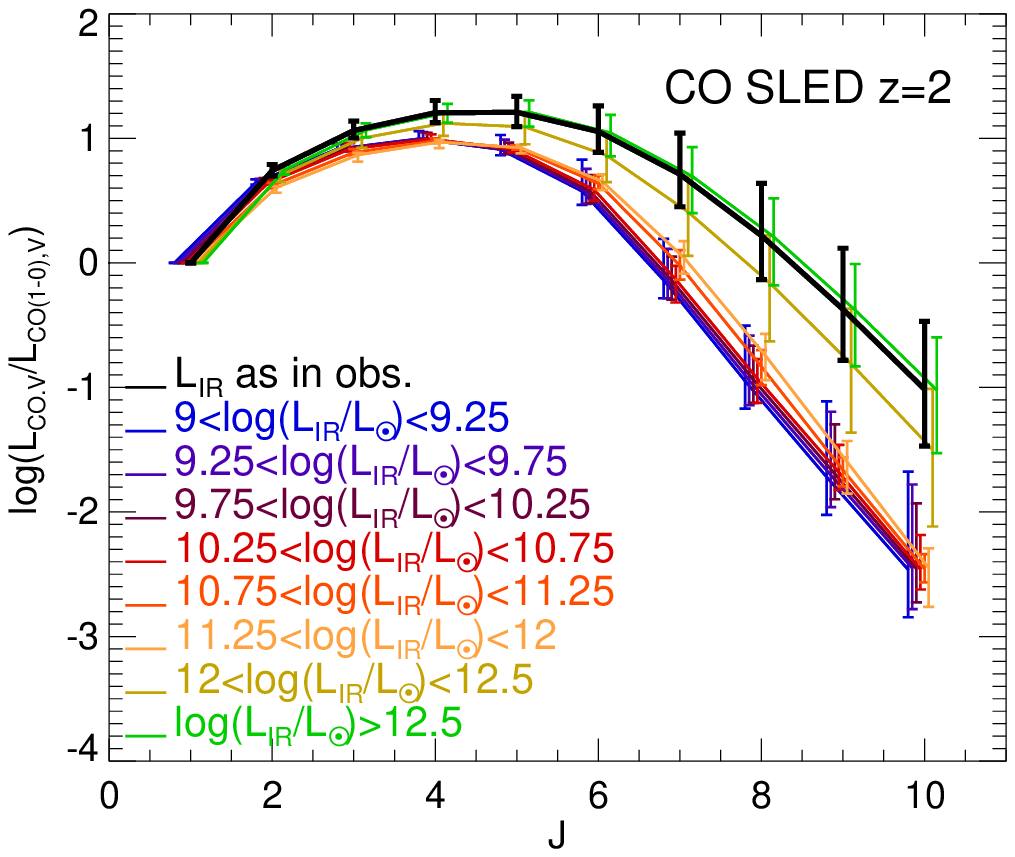}
\includegraphics[width=0.49\textwidth]{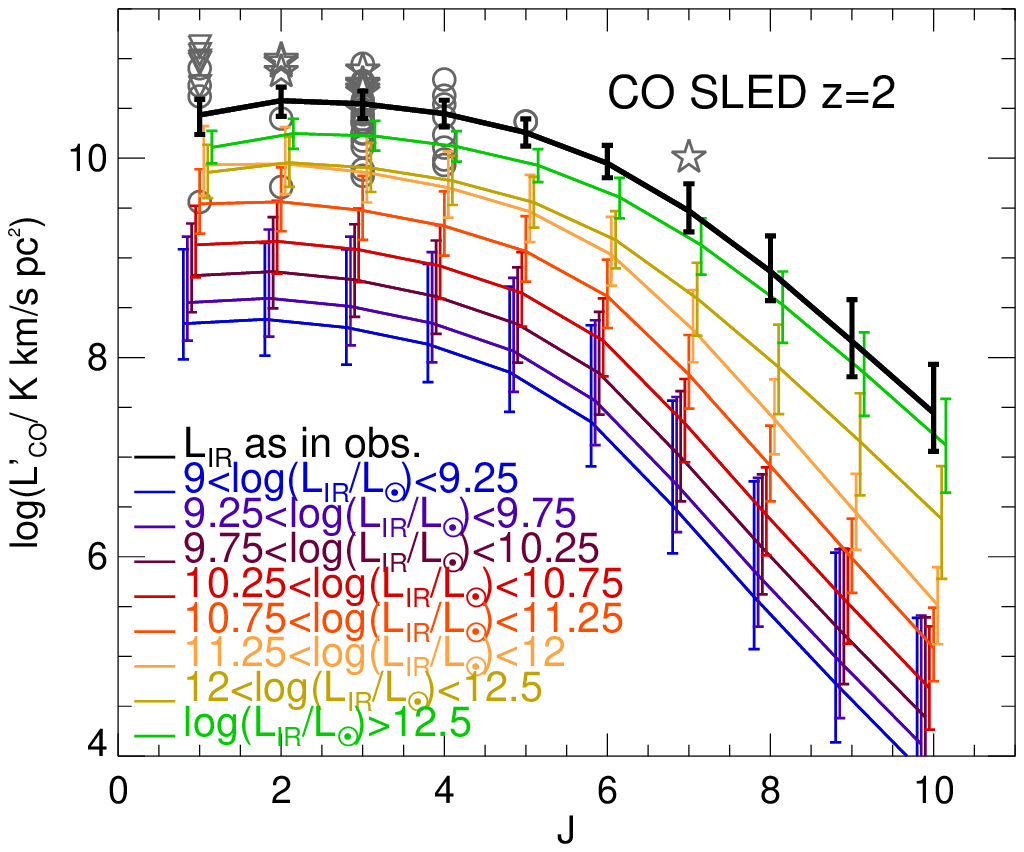}
\caption{As in Fig.~\ref{COladderz02}, but at $z=2$. 
In the bottom panel we also show individual observations of SMGs from Frayer et al. (1998,1999), \citet{Neri03}, 
\citet{Sheth04}, \citet{Greve05}, \citet{Tacconi06}, \citet{Casey09}, \citet{Bothwell10}, 
\citet{Engel10} and \citet{Ivison11}, which have a median redshift of $z\approx 2$. Symbols 
for these different sets of data are as in the bottom row of Fig.~\ref{LFCOIR}.
In order to infer a typical CO SLED of SMGs, we have scaled their CO luminosities to a common 
IR luminosity, assuming that 
the observed $L^{\prime}_{\rm CO}/L_{\rm IR}$ is conserved (see text for details).
We choose to scale the CO luminosities to the median IR luminosity of the sample,  
$\langle L_{\rm IR}\rangle \approx 8\times 10^{12}\, L_{\odot}$.
We also show as a black solid line the predictions of the {\tt GALFORM+UCL$_{-}$PDR} model for a sample of model galaxies 
selected to have the same IR luminosity distribution as the compilation of observed SMGs, whose  
CO luminosities have been scaled in the same way as was done in the observational sample. 
{ Note that $L^{\prime}_{\rm CO(1-0)}$ monotonically increases with IR luminosity}}
\label{COladderz2}
\end{center}
\end{figure}

At high redshifts, observational data on the CO emission from galaxies is based on
studies of individual galaxies, typically  LIRGs, ULIRGs, QSOs and SMGs. This has
allowed the characterisation of the CO-to-IR luminosity ratio for bright normal
star-forming and starburst galaxies. In this section we compare these
observations with the predictions of the {\tt GALFORM+UCL$_{-}$PDR} model.

Fig.~\ref{LFCOIR} shows the redshift evolution of the $L_{\rm CO}/L_{\rm IR}$ 
luminosity ratio for model galaxies for four different CO transitions, in
different bins of IR luminosity.
Quiescent and starburst galaxies are shown separately in the top and
bottom rows, respectively.
We also show a large 
compilation of observational results of local and high redshift normal
star-forming galaxies, local LIRGs, 
 local ULIRGs, high redshift colour selected galaxies, SMGs,  
and local and high redshift QSOs, and plot them in the panel
corresponding to the CO transition that was studied in each case. We plot
observed CO-to-IR luminosity ratios in the top rows of Fig.~\ref{LFCOIR} 
if observed galaxies correspond to normal star-forming galaxies or LIRGs,
or in the bottom row if they are classified as starburst galaxies (ULIRGs,
SMGs or QSOs). We warn the reader that most of the observational data 
do not directly measure total IR luminosity, but infer it from either the emission in 
 mid-IR or sub-millimeter bands, such as $24\mu$m or $850\mu$m, or an observationally inferred SFR. 
Thus, the comparison between the model predictions and the observations in 
Fig.~\ref{LFCOIR} has to be done with care. 

There is a weak trend of lower $L_{\rm CO}/L_{\rm IR}$ ratios as $L_{\rm IR}$
increases in both galaxy types 
as was shown at $z=0$ in Fig.~\ref{COladder2}.
In the case of quiescent galaxies, this trend 
is driven by the gas metallicity: IR faint galaxies have lower
metallicities, which, on average, decrease the dust opacity and the corresponding
IR luminosity, producing higher $L_{\rm CO}/L_{\rm IR}$ ratios. 
In the case of starburst galaxies, the main driver of 
the decreasing $L_{\rm CO}/L_{\rm IR}$ ratio with increasing $L_{\rm IR}$ is
the accompanying decrease in molecular mass for a given SFR due to the dependence of the SF
law on the dynamical timescale of the bulge in starbursts. 
This reduces the SF timescale in the most massive and brightest galaxies.

Observations shown in the panels corresponding to quiescent galaxies at intermediate and high redshifts 
show galaxies selected through different methods: 
 \citet{Geach11} measured the CO$(1-0)$ emission in a 24$\mu$m-selected sample at 
$z\approx 0.4$ of galaxies infalling into a rich galaxy cluster, 
\citet{Daddi10} measured the CO$(2-1)$ emission in a colour-selected sample 
of galaxies (BzK; see $\S 6.1$), and \citet{Tacconi10} and \citet{Genzel10} measured 
CO$(3-2)$ in a sample of normal star-forming galaxies located on the 
star forming sequence of the SFR$-M_{\rm stellar}$ plane. In the case of \citet{Daddi10}, 
\citet{Tacconi10} and \citet{Genzel10}, IR luminosities 
are inferred from the SFR, which in turn is estimated from the rest-frame UV and mid-IR emission with 
an uncertainty of a factor $\approx 2$. The conversion between SFR and IR luminosity used in these works 
corresponds to  
the local Universe relation calibrated for solar metallicity. 
High-redshift galaxies tend to have lower metallicities  
(e.g. \citealt{Mannucci10}; \citealt{Lara-Lopez10}), for which the use of the local Universe relation 
could possibly lead to an overestimate of the IR luminosity, and therefore, an underestimate of the 
$L_{\rm CO}/L_{\rm IR}$ ratio. Given this caveat, the apparent discrepancy of $\approx 0.3$~dex 
between the model predictions and 
the high redshift observations does not seem to be critical. 
Accurate IR luminosity measurements for high redshift galaxies are needed to better assess how the  
model predictions compare with the observations.

In the case of starbursts, observations correspond to  
the brightest galaxies observed in the local and high redshift Universe. 
IR luminosities for these galaxies are usually inferred from 
far-IR or sub-mm bands, e.g. $850\, \mu$m, and they are predicted to have gas metallicities close to 
solar, for which uncertainties in the IR luminosity are expected to be less important 
than in normal star-forming galaxies. 
These bright galaxies should be compared to the model predictions for the brightest IR galaxies. 
The model predicts a mean $L_{\rm CO}/L_{\rm IR}$ ratio and its evolution in good agreement with 
observations. These galaxies in the model are predicted to have gas kinetic
temperatures of $\approx 50$~K (see red lines in Fig.~\ref{TgasEvo}).

Fig.~\ref{COladderz2} is similar to Fig.~\ref{COladderz02}, but shows the CO SLED at $z=2$. 
The redshift is chosen to match the median of 
the SMG observational compilation also shown in Fig.~\ref{COladderz2}. This catalogue comprises $50$ SMGs observed in 
various CO transitions with IR luminosities in the range 
$L_{\rm IR}\approx 10^{12}-4\times 10^{13}\, L_{\odot}$. 
In order to infer a typical CO SLED of SMGs, we scale the CO luminosities in the SMG observational catalogue 
to the median IR luminosity of the sample, 
$\langle L_{\rm IR}\rangle \approx 8\times 10^{12}\, L_{\odot}$. We do this by assuming 
that the $L^{\prime}_{\rm CO}/L_{\rm IR}$ ratio for a given source is conserved. Thus, 
the $L^{\prime}_{\rm CO}$ plotted in Fig.~\ref{COladderz2} corresponds to the observed CO 
luminosity scaled by a factor $L_{\rm IR}/\langle L_{\rm IR}\rangle$.
These scaled observations are shown as symbols 
in the bottom panel of Fig.~\ref{COladderz2}. With the aims of performing a fair
 comparison to the model predictions, we select galaxies in the model to have the same 
IR luminosity distribution as the observational sample and then 
scale their CO luminosities following the same procedure as with the observations. 
This is shown as the black solid line in Fig.~\ref{COladderz2}. 
The CO lines with the best statistics in the
observational sample
are the CO$(1-0)$, CO$(2-1)$, CO$(3-2)$ and CO$(4-3)$. The latter three 
 correspond to the ones the model matches the best. In the case of the CO$(1-0)$, there 
is a slight discrepancy between the model and the observations, but still consistent 
with the dispersion predicted by the model.
At higher-J values, there { are} only two observations, one in agreement 
and the other one slightly above the model predictions. 
However, the low number statistics prevents us { from}  
determining how representative these points are of the general SMG population. 
The model predicts the peak of the CO SLED occurs, on average, at $J=5$ for these very 
luminous IR galaxies. 

For the general galaxy population, the model predicts that  
the brightest IR galaxies have slightly flatter CO SLEDs than
fainter IR counterparts. Differences in the CO SLEDs of faint- and bright-IR galaxies at $z=2$
are predicted to be smaller than for $z=0$ galaxies.
In other words, at a fixed IR luminosity, high redshift
galaxies tend to have shallower CO SLEDs compared to their $z=0$ counterparts. This is due to a tendency of
increasing average gas kinetic temperature in molecular clouds with increasing redshift
(see Fig.~\ref{TgasEvo}), driven by the systematically lower metallicities and 
higher SFR surface densities of high redshift galaxies.

\subsubsection{Kinetic temperature evolution}
\begin{figure}
\begin{center}
\includegraphics[width=0.5\textwidth]{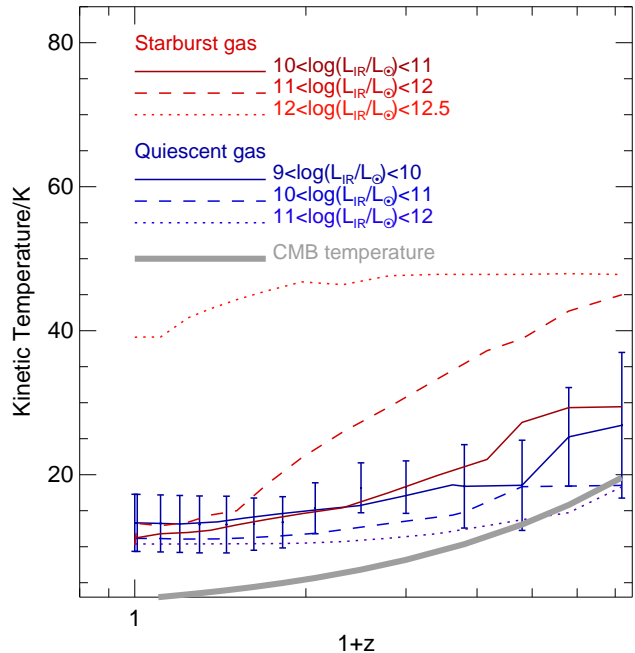}
\caption{Kinetic temperature of the gas in molecular clouds for quiescent (blue curves) and starburst 
 (red curves) gas phases in the ISM of galaxies as a function of redshift and for various bins of IR luminosity, 
as labelled. 
 Lines show the median of the predicted $T_{\rm K}$, while errorbars correspond to the $10$ and $90$ 
percentile range. For clarity errorbars are only shown in one IR luminosity bin for the quiescent gas phase. 
For reference, the Cosmic Microwave Background temperature is shown as a thick grey solid line. 
Note that galaxies can have SF taking place simultaneously in the galactic disk and the bulge, for which 
they would have two gas phases, a quiescent and a starburst gas phase, respectively. { 
For the same line style, starbursts have higher $T_{\rm K}$ than quiescent gas in galaxies.}}
\label{TgasEvo}
\end{center}
\end{figure}

The variation of the $L_{\rm CO}/L_{\rm IR}$
luminosity ratio with IR luminosity differs between quiescent and starburst
galaxies, and is related to variations in the gas kinetic temperature, which depend on IR luminosity
and redshift. Fig.~\ref{TgasEvo} shows the gas kinetic temperature of molecular clouds
for quiescent (blue lines) and starburst (red lines) gas phases for different IR
luminosity bins, as a function of redshift. 
For reference, we also show the evolution 
of the temperature of the Cosmic
Microwave Background (CMB) with redshift. Note that some
galaxies undergoing quiescent SF appear to have kinetic temperatures below the CMB temperature at
$z\gtrsim 4.5$. For these galaxies, extra heating from the CMB needs to be
included in the {\tt UCL$_{-}$PDR} model to describe the thermal and chemical
state of these galaxies. This represents a limitation of the current {\tt
GALFORM+UCL$_{-}$PDR} model. However, this only becomes relevant at very
high redshift and for quiescent galaxies. 

The relation between the kinetic temperature of the gas and IR luminosity is primarily 
dominated by gas metallicity in quiescent galaxies and by the UV radiation field in starburst galaxies. 
In the case of quiescent galaxies, the gas metallicity increases as the IR luminosity increases, 
and therefore the gas cools more efficiently, decreasing $T_{\rm K}$ as $L_{\rm IR}$ 
increases. This is true only 
in this quiescent regime given that $G_{\rm UV}$ only varies around 
$1-10\times G_{0}$. In the case of starburst galaxies, as the IR luminosity increases so does the 
UV radiation field, $G_{\rm UV}$, which boosts the kinetic temperature of the gas, driving a 
positive relation between $T_{\rm K}$ and $L_{\rm IR}$. 

In general, starburst galaxies tend to have higher $T_{\rm K}$ 
than quiescent galaxies. 
The {\tt GALFORM+UCL$_{-}$PDR} model
predicts that both cool and a warm ISM phases should be present in the high redshift universe, 
particularly in relatively bright galaxies (but not exclusively in the brightest ones), given that 
a large fraction of galaxies in the model at high redshift have SF taking place simultaneously 
in both the disk and bulge components.

\section{Assessing the robustness of the model predictions}\label{Sec:Assumptions}

{ We analyse in this section how the predictions of the coupled model presented in 
$\S 3$~and~$\S 4$ depend on the assumptions made in the PDR model. We focus on (i) the 
effect of metallicity, and (ii) the effect of a varying  
hydrogen number density (as opposed to the fixed density adopted previously). For a detailed 
analysis on how other assumptions in the PDR modelling affect the results, e.g. the assumed geometry, 
see \citet{Rollig07}.

We have shown that gas metallicity has an important effect on the predicted CO luminosity and 
SLED, particularly for relatively IR-faint galaxies. These variations 
with metallicities have been extensively analysed in PDR and LVG models, such as in 
\citet{Wolfire03}, \citet{Bell06} and \citet{Weiss07}. However, 
%in general 
comparisons between observations and PDR or LVG models with the aim of inferring 
average GMC properties tend to ignore the metallicity effect 
by assuming that the metallicity is fixed at solar or super-solar 
(e.g. \citealt{Hitschfeld08}; \citealt{Danielson10}; \citealt{Nagy12}). In order to assess 
how much our predictions change if we ignore changes in metallicity we perform the same 
calculations as in $\S 4$ but ignore the metallicity information 
in {\tt GALFORM}. We therefore select the subset of the PDR models 
shown in Table~\ref{XCOs} that have $Z_{\rm g}=1\, Z_{\odot}$ and calculate the $X_{\rm CO}$ for each galaxy from that subset, 
regardless of its actual metallicity.  

\begin{figure}
\begin{center}
\includegraphics[trim = 1mm 0mm 0mm 1mm,clip,width=0.5\textwidth]{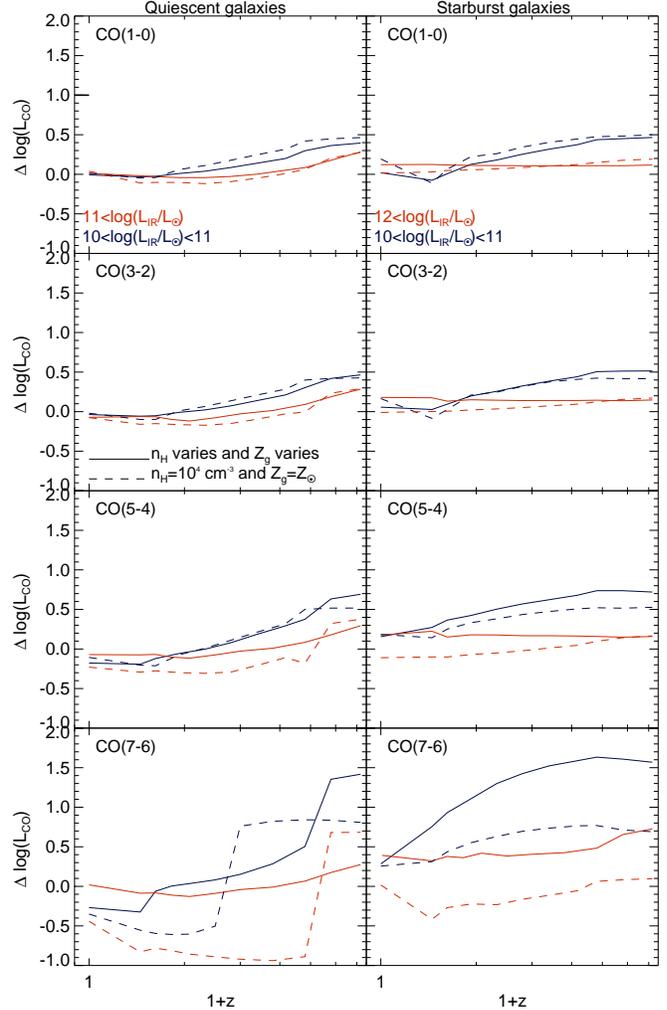}
\caption{{ Logarithm of the median ratio between the predicted CO luminosity in the PDR model variants and 
the standard {\tt GALFORM+UCL$_{-}$PDR} model as a function 
of redshift 
for two alternative PDR models. 
Quiescent and starburst galaxies are shown in the left-hand and right-hand columns, respectively, 
and for four CO transitions, 
CO$(1-0)$ (top panels), CO$(3-2)$ (middle-top panels), CO$(5-4$) (middle-bottom panels) and CO$(7-6)$ 
(bottom panels), and two IR luminosities ranges (blue and red, as labelled), which are different in the right 
and left-hand columns to enhance differences.
The assumptions made in these two alternative PDR models are: (i) constant $n_{\rm H}=10^4\, {\rm cm}^{-3}$ and metallicity, 
$Z_{\rm g}=Z_{\odot}$ (dashed lines), and (ii) varying $Z_{\rm g}$ and $n_{\rm H}$ 
(solid lines).}}
\label{CompareZ}
\end{center}
\end{figure}

Fig.~\ref{CompareZ} shows the ratio between the predicted CO luminosity 
in the standard {\tt GALFORM+UCL$_{-}$PDR} model and the variant with a fixed gas metallicity, for $4$ CO transitions 
and two IR luminosity ranges. 
Low CO transitions ($J\le 4$) are only slightly affected by this change in the PDR models, 
with differences of less than a factor $3$.
IR bright galaxies show the 
least variation in CO luminosity with respect to the standard model  
due to their already high gas metallicities, which tend to be close to solar or super-solar. 
The CO luminosities of fainter galaxies in the IR are more affected given that they show larger gas metallicity differences, 
as Fig.~\ref{MMr} shows. As we move to higher CO transitions, differences with respect to the standard model increase 
 to  up to a factor $\approx 10$. 
This is driven by the generally larger variation of the population level of 
high CO transitions with cloud properties, as we described in $\S 4$. 
These results indicate that to assume a 
gas metallicity for observed galaxies might lead to a misinterpretation of the data, 
particularly when analysing high CO transitions. This effect has also been previously seen in 
detailed ISM hydro-dynamical simulations, such as in \citet{Feldman12}. 

Throughout the paper we have so far assumed that GMCs are characterised by a constant hydrogen 
density of $n_{\rm H}=10^4\, {\rm cm}^{-3}$, 
which, we have shown, allows us to explain the observed CO luminosities of local and high-redshift galaxies. 
However, it is interesting to study the impact of allowing the hydrogen number density, $n_{\rm H}$, to vary. 
This is because simulations and theoretical models suggest that a 
minimum density of hydrogen in GMCs is required to assure pressure equilibrium between a thermally 
supported warm medium and a turbulence supported cold neutral medium  
  (\citealt{Wolfire03}; \citealt{Krumholz09b}). This minimum density 
depends on the UV flux, hard X-ray flux and metallicity as described by \citet{Wolfire03},

\begin{equation} 
n_{\rm H,min}\propto \frac{G_{\rm UV}}{1+3.1(G_{\rm UV}Z_{\rm g} F_{\rm X})^{0.37}},
\end{equation}

\noindent where 
$n_{\rm H,min}$, $G_{\rm UV}$, $Z_{\rm g}$ and  $F_{\rm X}$ are in units of 
${\rm cm^{-3}}$, $G_{0}$, $Z_{\odot}$ and $10 \rm erg\, s^{-1}\, cm^{-2}$ (following our conversion between 
$F_{\rm X}$ and $\zeta_{0}$ described in $\S 2.2$). 
We explore the effect of assuming that $n_{\rm H}\propto n_{\rm H,min}\propto G_{\rm UV}$ 
on the predictions presented in $\S 3$ and $\S 4$. 
For this we 
select a subset of PDR models from Table~\ref{XCOs}, so that 
$n_{\rm H}= 10^3 {\rm cm}^{-3}\, (G_{\rm UV}/G_{\rm 0})$. 
We repeat the analysis of $\S 4$ using this subset of 
PDR models. The ratio between the predicted CO luminosity
in the standard {\tt GALFORM+UCL$_{-}$PDR} and the model using a variable $n_{\rm H}$ 
is shown in Fig.~\ref{CompareZ} as solid lines.}

\begin{figure}
\begin{center}
\includegraphics[width=0.5\textwidth]{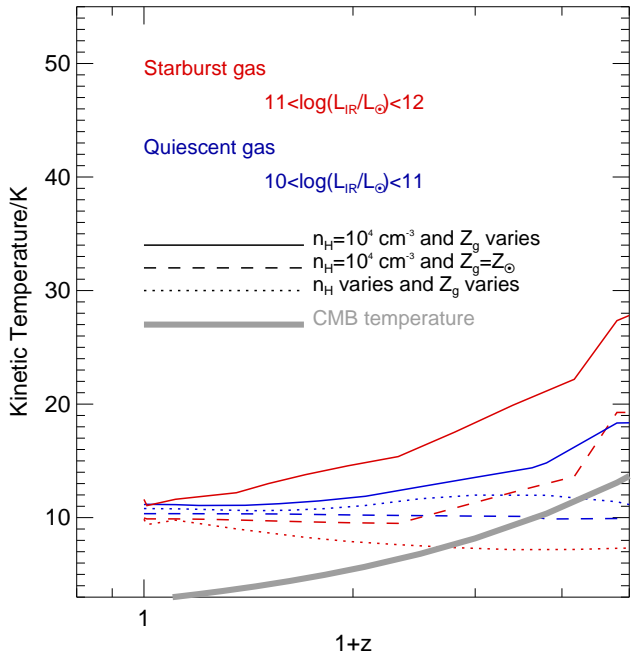}
\caption{Kinetic temperature of the gas in molecular clouds for quiescent (blue curves) and starburst 
 (red curves) gas phases in the ISM of galaxies as a function of redshift and for IR luminosities in the range 
$10^{10}-10^{11}\, L_{\odot}$ for quiescent gas and $10^{11}-10^{12}\, L_{\odot}$ for starburst gas.
 Lines show the median of the predicted $T_{\rm K}$, for the two models of Fig.~\ref{CompareZ}, in addition to the 
standard {\tt GALFORM+UCL$_{-}$PDR} model, which assumes  $n_{\rm H}=10^4\, {\rm cm}^{-3}$ and varying 
metallicity, as labelled. For reference, the Cosmic Microwave Background temperature is shown as a thick grey solid line. 
{ For the same line style, starbursts have higher $T_{\rm K}$ than quiescent gas in galaxies.}}
\label{TgasEvo2}
\end{center}
\end{figure}

{ Allowing $n_{\rm H}$ to vary has a small impact on the predicted CO luminosities at low CO transitions 
typically less than a factor of $3$, and gives slightly smaller differences with respect to the 
standard {\tt GALFORM+UCL$_{-}$PDR} model than the model which uses a $Z_{\rm g}=Z_{\odot}$ and 
 $n_{\rm H}=10^4\, {\rm cm}^{-3}$ (dashed lies in Fig.~\ref{CompareZ}). This suggests 
that including the gas metallicity information from {\tt GALFORM} in the PDR has a similar 
impact on the predicted CO luminosities than the assumption of a constant $n_{\rm H}=10^4\, {\rm cm}^{-3}$. 
This again supports our interpretation 
of the major role that metallicity plays in determining $X_{\rm CO}$ (Bayet et al. 2012, in prep.). 
When moving to high CO transitions, 
 deviations from the CO luminosities predicted by the model with variable $n_{\rm H}$ become more important. 
There is a tendency to produce fainter (brighter) CO emission from high 
CO transitions at low (high) redshifts with 
respect to the standard model. This is because in the standard model there is a clear 
increase of temperature with redshift for both, quiescent and starburst gas, that is not as obvious in the case of the quiescent 
gas in the subset of 
PDR models in which $n_{\rm H}$ is varied (Fig.~\ref{TgasEvo2}).}

{ We have argued that 
 metallicity plays an important role in determining the average transmission of UV photons in galaxies and 
that therefore affects the incident UV flux. This translates 
into low metallicity galaxies having higher temperatures. Fig.~\ref{TgasEvo2} shows the kinetic temperature evolution 
for the standard {\tt GALFORM+UCL$_{-}$PDR} model and the two subsets of PDR models we describe above for two ranges 
of IR luminosities. 
In the subset of PDR models where the metallicity is fixed, there is no visible evolution of $T_{\rm K}$ with 
redshift for quiescent gas. 
This supports our conclusion that the ISM metallicity evolution is a main driver of the increasing $T_{\rm K}$ 
with redshift in quiescent gas in the standard {\tt GALFORM+UCL$_{-}$PDR} model ($\S 4.2.3$). In the PDR subset of models where $n_{\rm H}$ 
varies, the trend between $T_{\rm K}$ and $z$ for the quiescent gas is only weakly recoveredi at $z<2.5$.
 In the case of starburst gas, the main driver of the trend of increasing $T_{\rm K}$ with redshift, is the increasing $G_{\rm UV}$ 
with redshift, that is only weakly affected by metallicity (see $\S 4.2.3$). Thus, the subset of PDRs with 
fixed metallicity recovers the trend of the standard {\tt GALFORM+UCL$_{-}$PDR} model. 
In the case of the subset of PDRs with variable $n_{\rm H}$, the trend is lost 
due to the more efficient cooling in the PDR due to the higher $n_{\rm H}$.

The small deviations in the emission of 
low CO transitions introduced by different assumptions in the PDR modelling 
 suggest that the predictions presented in this paper for these transitions are robust under these changes.
However, high CO transitions are more sensitive to the assumptions in the PDR modelling. The small number of observations 
available in these transitions does not so far allow us to distinguish these different possibilities. 
More data on these high CO transitions are needed, particularly if they cover a wide redshift range. 
Homogeneity in the observed samples, even though it is desired, it is not essential given that our 
modelling permits the prediction of a plethora of galaxy properties, which allows us to select galaxies to 
have similar properties to the observed ones.}

\section{Predictive power of the {\tt GALFORM+UCL$_{-}$PDR} model}

We have shown that the predictions of the {\tt GALFORM+UCL$_{-}$PDR} model for CO 
emission are in good agreement with 
observations of galaxies in the local and high redshift Universe.
Consequently, this coupled model is a powerful theoretical tool 
to study the observability of molecular lines in different types of galaxies and 
can therefore contribute to the 
development of science cases for the new generation of millimeter telescopes. In this section we 
focus on star-forming galaxies at high redshift, selected through two different 
techniques based on broad band colours: (i) BzK colour selection \citep{Daddi04}, which can be used 
to select star-forming 
galaxies in the redshift range $1.4\lesssim z\lesssim 2.5$, and (ii) the Lyman-break technique 
which is used to select star-forming galaxies at $z\sim3-10$. 
In this section, 
we study the observability of these star-forming galaxies with ALMA. We use 
AB magnitudes throughout this section.

\subsection{BzK galaxies}

\begin{table*}
\begin{center}
\caption{Properties of the four star-forming BzK galaxies at $z=2$: 
row (1) observer frame, extincted K-band absolute magnitude,  
(2) BzK colour index, (3) molecular gas mass, (4) stellar mass, (5) SFR, (6) gas
metallicity, (7) molecular gas half-mass radius, (8) line-of-sight CO velocity width, 
(9) velocity-integrated CO line flux of the CO$(1-0)$, the (10) CO$(3-2)$ and 
the (11) CO$(6-5)$ emission lines, 
in (12), (13) and (14) the integration time to get a 
detection at the level indicated in the
parenthesis in the band~1, band~3 and band~6 of ALMA for the CO$(1-0)$, CO$(3-2)$ and CO$(6-5)$ emission lines, 
respectively. We used the full ALMA configuration ($50$ antennae) and 
average water vapor conditions ($\approx 1.2$-$1.5$mm of column density) to simulate the observations 
of these galaxies. We also list the central frequency, $\nu_{\rm c}$, bandwidth, $\Delta \nu$, 
and angular resolution, $R$,  
used to simulate the observations.}\label{BzKprops}
\begin{tabular}{l c c c c }
\\[3pt]
\hline
Properties $z=2$ sBzK                         & BzK+gal1 & BzK+gal2 & BzK+gal3 & BzK+gal4 \\
\hline 
(1) $M_K-5\rm log(h)$                    & -22.9 & -21.3   & -21.3 & -21.5  \\
(2) BzK index                                  & 0.23  & 0.95    & 1.2   & 1.52 \\
(3) log($M_{\rm mol}/M_{\odot}$)               & 9.6    & 9.9  & 9.3   & 10.6   \\
(4) log($M_{\rm stellar}/M_{\odot}$)           & 10.6   & 10.6 & 10.2 & 10.7  \\
(5) $\rm SFR/M_{\odot}\, yr^{-1}$               & 3.7     & 3.72  & 1.0   & 19.3    \\
(6) $Z_{\rm gas}/Z_{\odot}$                                & 0.63        & 1.15      & 1.33       & 1.4       \\
(7) $r^{\rm mol}_{\rm 50}/\rm kpc$             & 0.8     & 2.0 & 2.8   & 1.9  \\
(8) $\sigma^{\rm los}_{\rm CO}/\rm km\, s^{-1}$          & 119   & 199 & 177  & 273  \\
(9) $S_{\rm CO(1-0),V}/\rm mJy\, km\, s^{-1}$  & 16.95 & 8.7 & 2.4 & 46.38 \\ 
(10) $S_{\rm CO(3-2),V}/\rm mJy\, km\, s^{-1}$  & 101.9 & 52.8   & 14.5  & 279.9  \\
(11) $S_{\rm CO(6-5),V}/\rm mJy\, km\, s^{-1}$  & 12.7 & 8.1   & 2.2  & 42.8  \\
(12) $\tau_{\rm int, band 1}$(1-0) (1$\sigma$ noise)  & 3h (0.03mJy) & 5.3d (0.0043mJy) & 56d (0.00135mJy) & 8.6h (0.017mJy) \\
(13) $\tau_{\rm int, band 3}$(3-2) (1$\sigma$ noise)  & 21.6m (0.17mJy) & 14.9h (0.027mJy) & 4.3d (0.01mJy) & 1h (0.1mJy) \\
(14) $\tau_{\rm int, band 6}$(6-5) (1$\sigma$ noise)  & 4.8h (0.67Jy) & 5.7d (0.004mJy) & 53d (0.0013mJy) & 8.7h (0.016mJy) \\
\hline 
band configuration                            & $\nu_{\rm c}$ & $\Delta \nu$ & $R/$arcsec \\
\hline
(15) band~1                              & 38.3~GHz & 0.05~GHz & 4.8 \\         
(16) band~3                              & 115.27~GHz & 0.11~GHz & 1.48 \\
(17) band~6                              & 230.49~GHz & 0.2~GHz & 0.67 \\
\hline
\end{tabular}
\end{center}
\end{table*}

\begin{figure*}
\begin{center}
\includegraphics[width=0.23\textwidth]{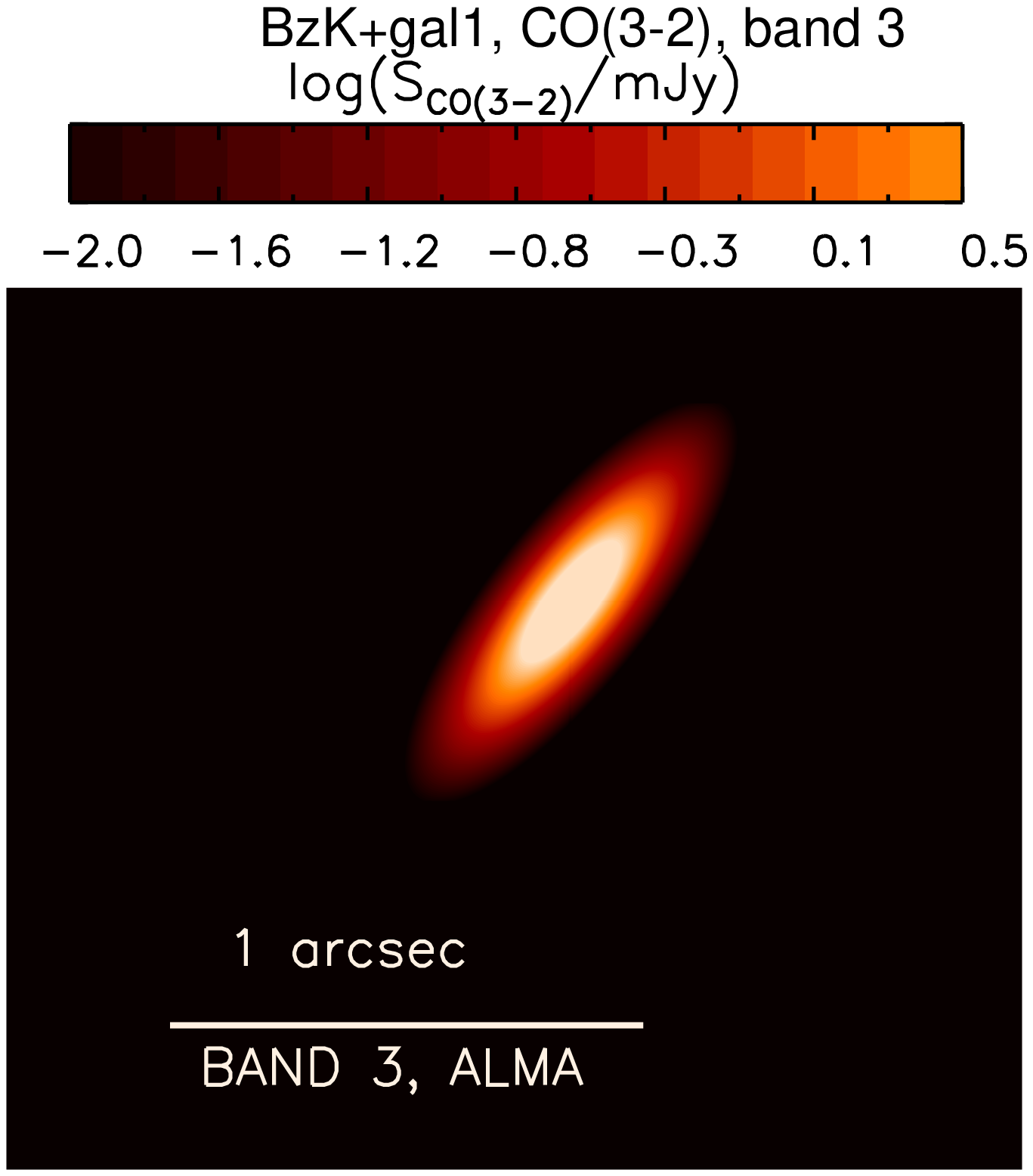}
\includegraphics[width=0.23\textwidth]{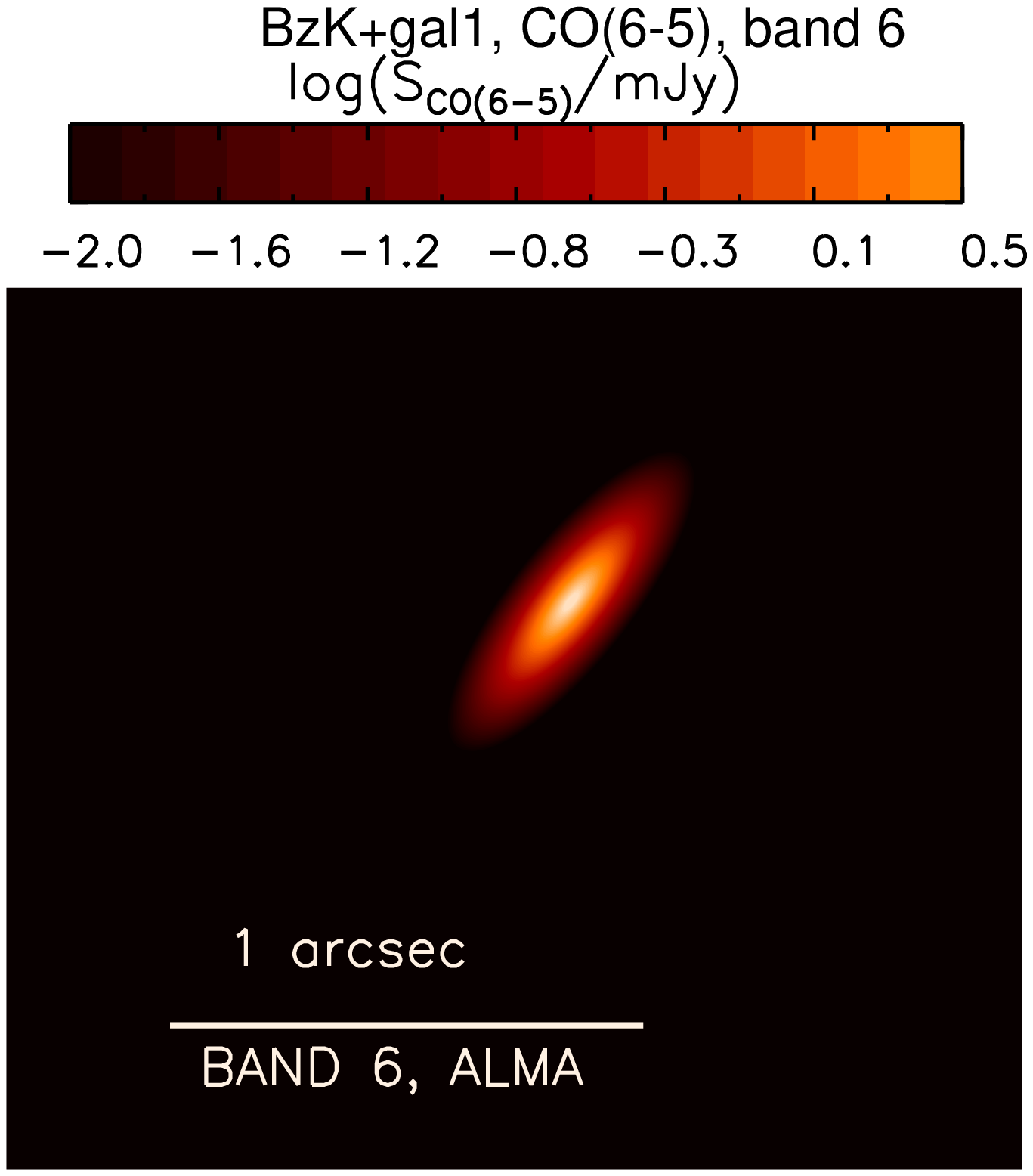}
\includegraphics[width=0.23\textwidth]{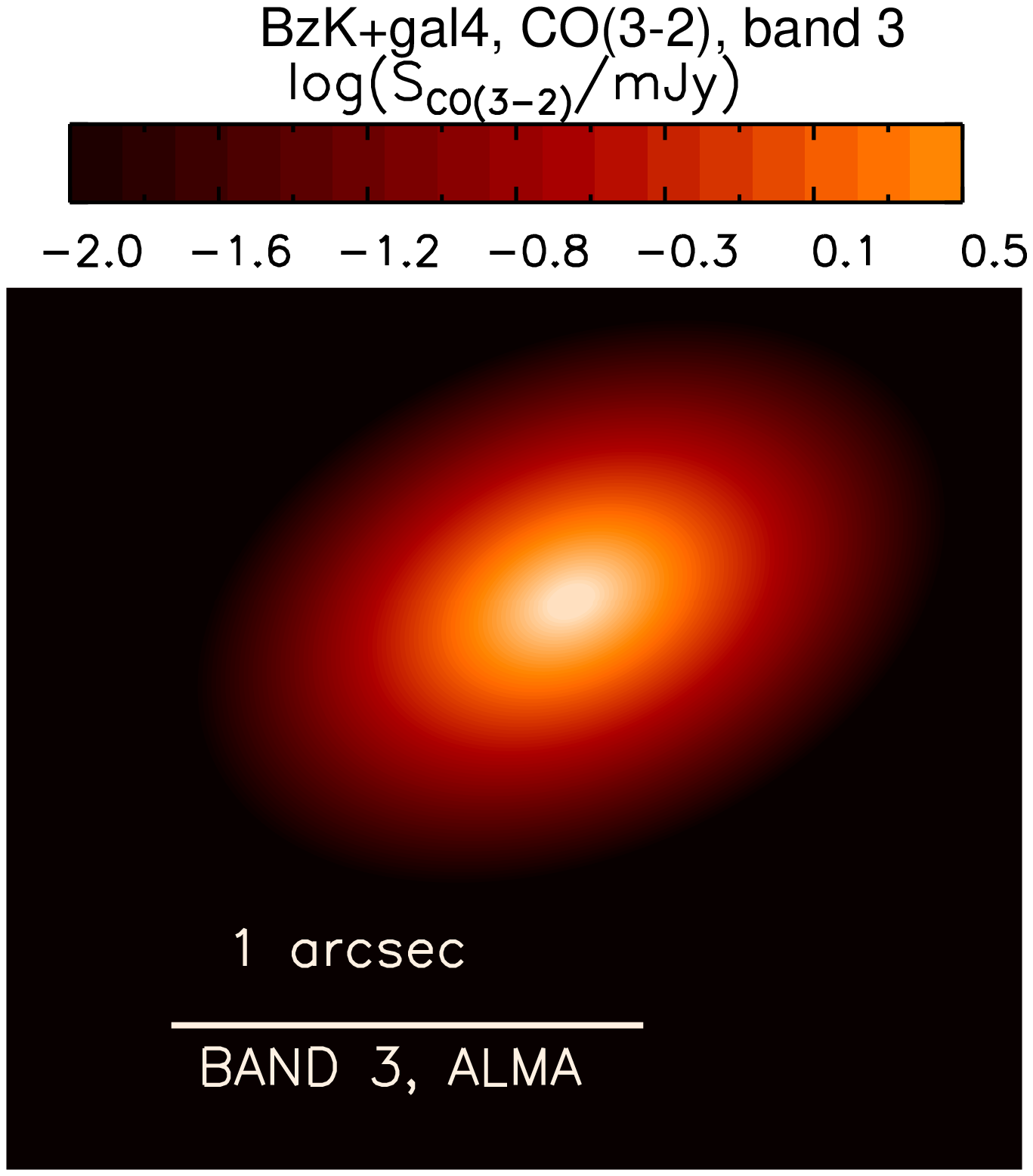}
\vspace{0.2cm}
\includegraphics[width=0.23\textwidth]{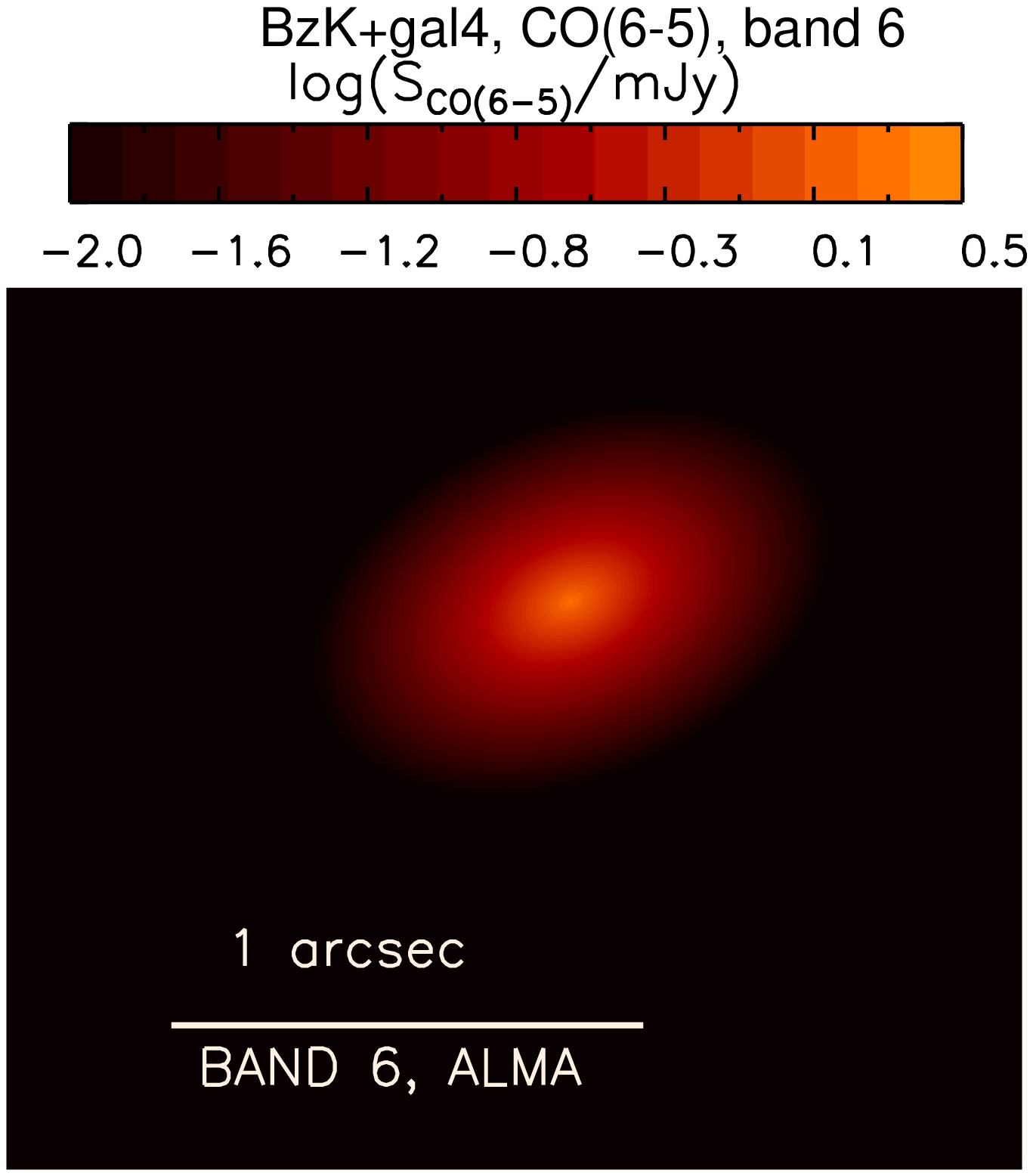}
\includegraphics[width=0.23\textwidth]{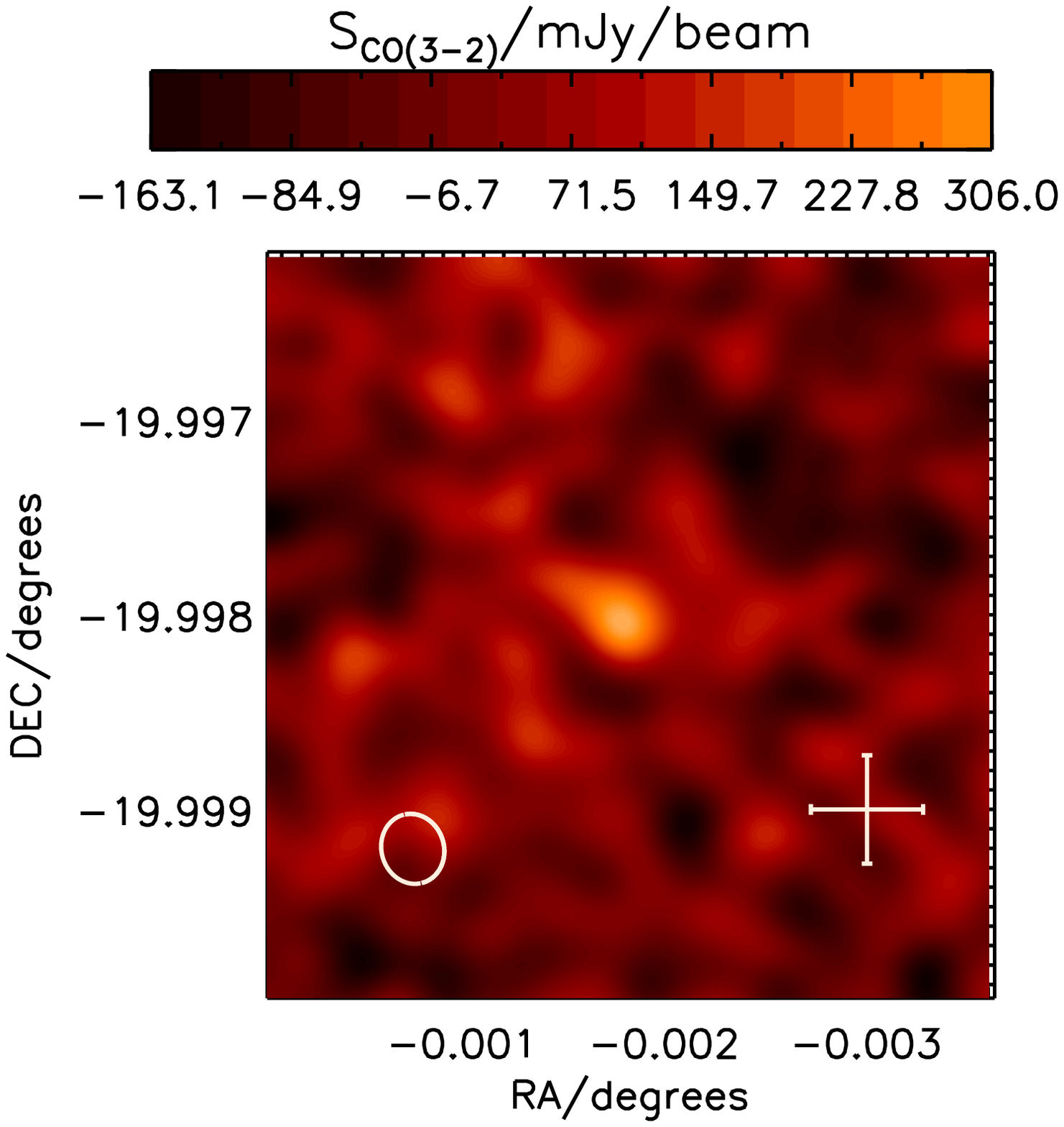}
\includegraphics[width=0.23\textwidth]{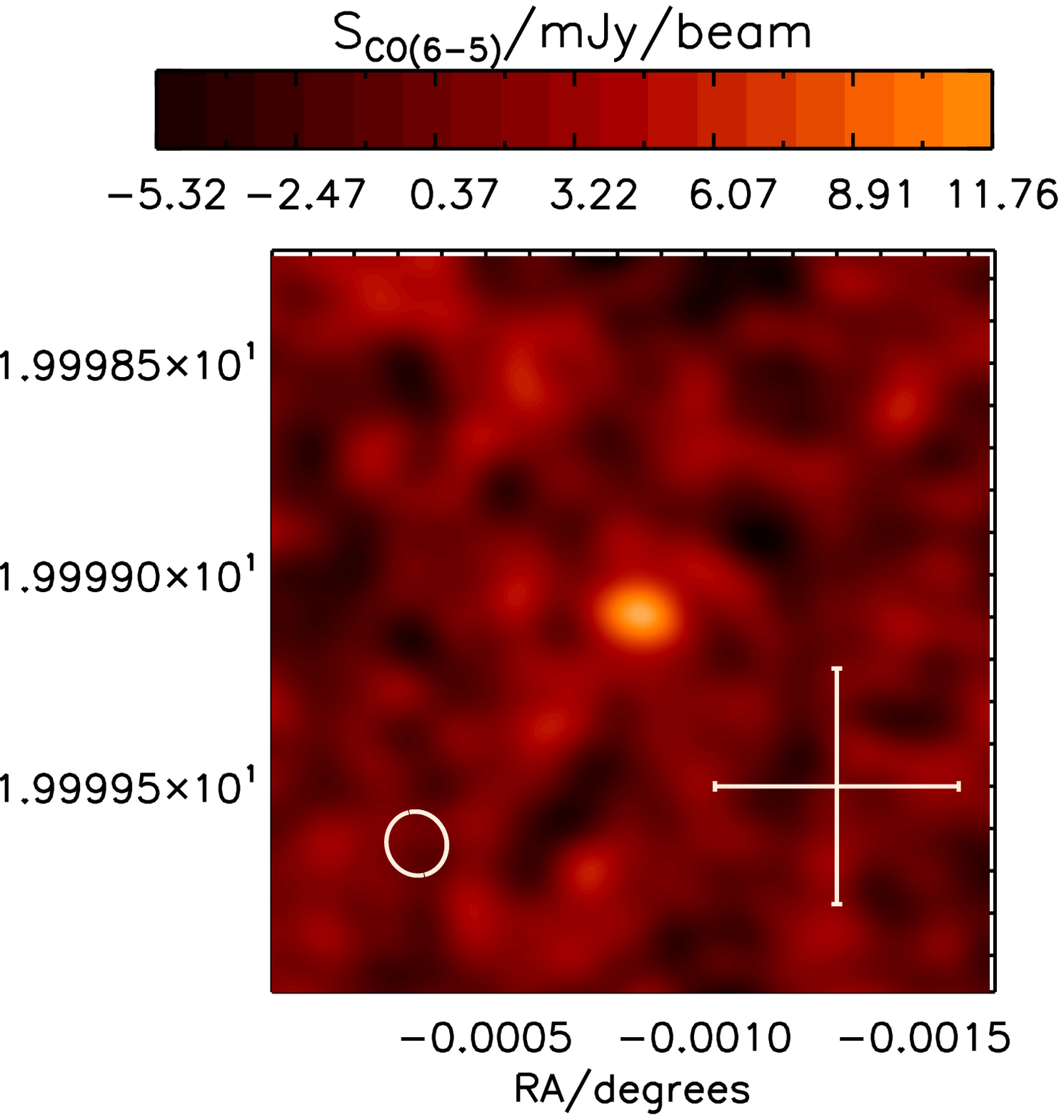}
\includegraphics[width=0.23\textwidth]{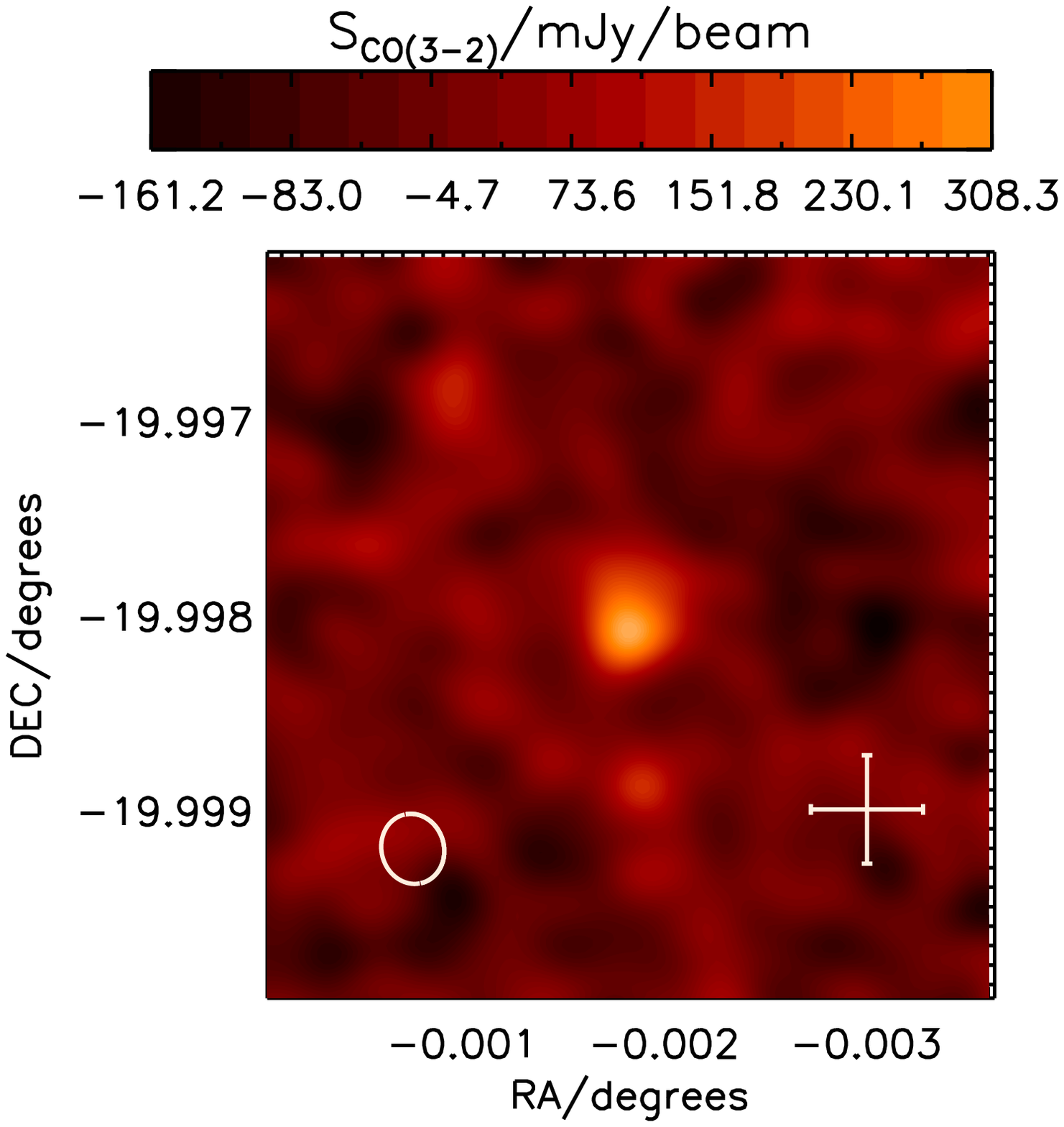}
\includegraphics[width=0.23\textwidth]{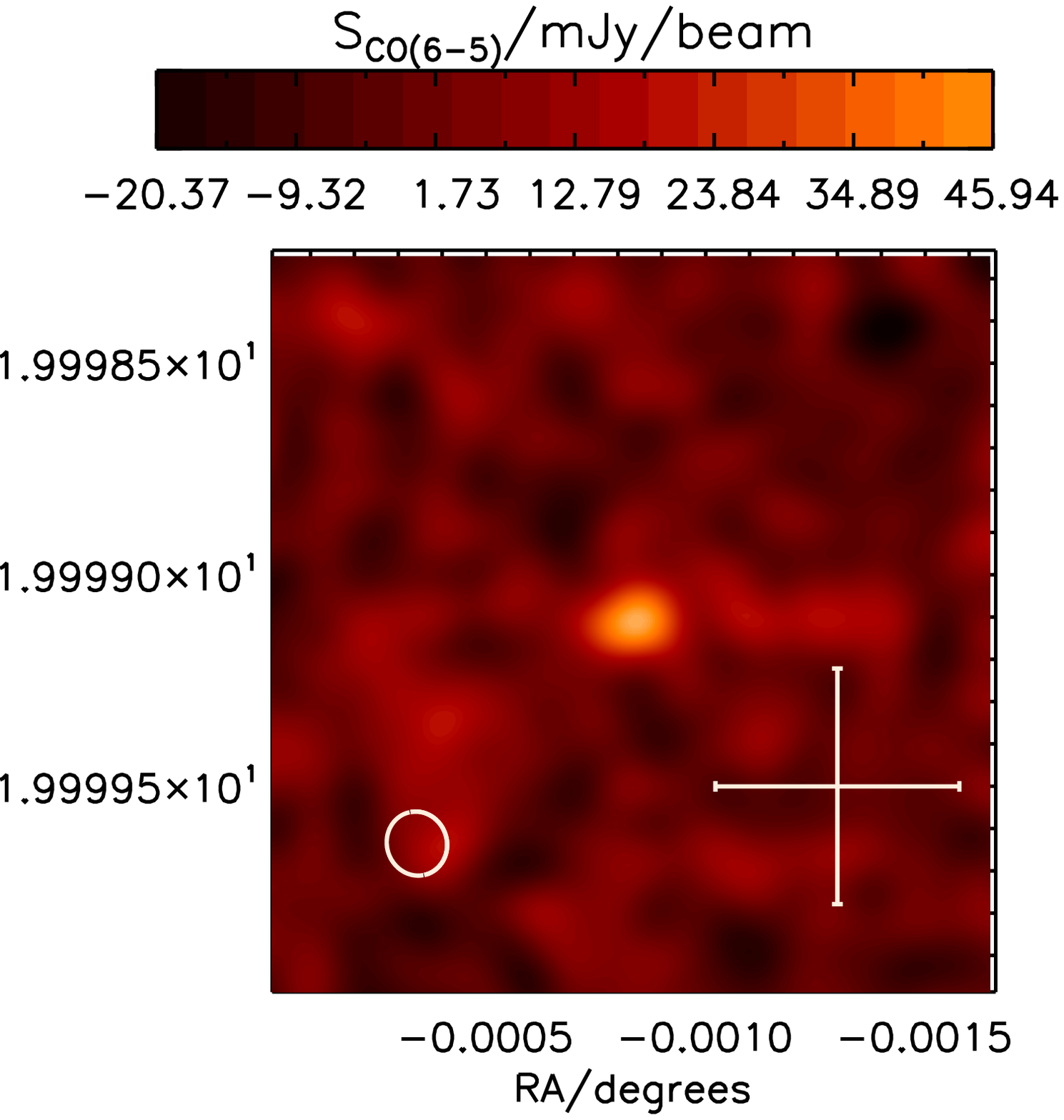}
\caption{Two  star-forming BzK galaxies at $z=2$, BzK+gal1 
(left-hand and middle-left panels) and 
 BzK+gal4 
(right-hand and middle-right panels).
{\it Top panels}: the CO$(3-2)$ (left panel) and CO$(6-5)$ (right panel) flux maps in logarithmic units of 
$\rm mJy/pixel$. Horizontal lines show 
1~arcsec scale and the band in which the CO transition would be observed in ALMA 
at this redshift is labelled (band~3 and 6, 
for the CO$(3-2)$ and CO$(6-5)$, respectively).  
{\it Bottom panels:} the simulated observations of the CO$(3-2)$ and CO$(6-5)$ 
flux maps in declination vs. right ascension. Flux is in units of $\rm mJy/beam$. 
Maps corresponds to the outputs of the {\tt CASA} software, 
after convolving the original map with the primary beam and 
including an atmospheric model for the background noise, in the full ALMA configuration 
($50$ antennae). Ellipses at the bottom-left corner 
indicate the beam size and shape and the cross shows 
1$\times$1~arcsec$^2$. The flux scale is shown at the top of each panel.
Some relevant properties of these galaxies are listed in Table~\ref{BzKprops}, along with the integration time
 used to generate the {\tt CASA} maps.}
\label{Proj3b1}
\end{center}
\end{figure*}

The BzK colour selection has shown to be efficient at selecting galaxies 
around $z\sim 2$ \citet{Daddi04}. 
BzK-selected galaxies have been used to study 
the build-up of the stellar mass, the SF history of the Universe and 
properties of star-forming and passive galaxies at the peak of SF activity 
(e.g. Daddi et al. 2005, 2007; \citealt{Lin11}). 
The BzK criterion is based on observer frame magnitudes in 
the $B$, $z$ and $K$ bands. Star-forming galaxies, also referred to as sBzK, 
are selected as those whose BzK colour index 
BzK$=(z-K)_{\rm AB}-(B-z)_{\rm AB}>-0.2$.

The model predicts that the BzK criterion is a very efficient way to select star-forming galaxies; 
with only a small contamination of 
$\approx 10$\% of galaxies outside the redshift range $1.4<z<2.5$. Furthermore,   
for a given limit $m_{K}$, there is a positive correlation 
between the BzK colour index and the SFR of galaxies, 
although with a large dispersion 
(\citealt{Merson12}). 
Large BzK colour indices (BzK$>1.5$) correspond almost exclusively to  
highly star-forming galaxies, SFR$\gtrsim 10\, M_{\odot}\, \rm yr^{-1}$.

We take the galaxy population predicted by the 
{\tt GALFORM} model at $z=2$ and select a sample 
of sBzK galaxies based on their BzK colour indices and apparent K-band magnitudes: 
BzK$>-0.2$ and $m_{\rm K}<24$. The latter cut
 corresponds roughly to the deepest K-band surveys to date (\citealt{Bielby11}). 
We randomly select four galaxies from this $z=2$ BzK sample in bins of BzK colour index,  and 
 list selected properties in Table~\ref{BzKprops}. 
In Table~\ref{BzKprops}, 
the CO line velocity width, $\sigma^{\rm los}_{\rm CO}$, corresponds to the line-of-sight circular
velocity and the velocity dispersion in the case of the disk
and the bulge components, respectively (considering a random inclination).
We choose to focus on the
CO$(1-0)$, CO$(3-2)$ and CO$(6-5)$ emission of these galaxies, which fall into band~1, 3 and 6 of ALMA,
respectively, but note that other CO lines also fall into the ALMA bands at this redshift.

We used the software {\tt CASA}\footnote{Specifically, we use the {\tt ALMA~OST} software developed 
by the ALMA regional centre in the UK, {\tt http://almaost.jb.man.ac.uk/}.}, 
which is part of the observational tools associated with ALMA, 
to simulate observations of model galaxies, by including instrumental and atmospheric effects, 
such as the convolution with the primary beam and the sky noise. 
We calculate the integration times\footnote{Integration
times were calculated using the ALMA sensitivity
calculator {\tt https://almascience.nrao.edu/call-for-proposals/ sensitivity-calculator.}}
necessary to obtain a theoretical root mean square sensitivity of
at least $5$ times lower than the peak CO flux with the 
 full ALMA configuration ($50$ antennae) under average 
water vapor conditions ($\approx 1.2-1.5 \,\rm mm$ of column density). 
The peak CO flux corresponds to $s_{\nu}=S_{\rm CO,V}/\sigma_{\rm los}$. 
Input parameters used 
in the {\tt CASA} software and integration times are listed in Table~\ref{BzKprops} 
for both CO emission lines
considered. 

\begin{table*}
\begin{center}
\caption{Properties of the four Lyman-break galaxies at $z=3$ plotted in Fig~\ref{Proj31}: 
row (1) rest-frame, extincted UV magnitude, columns (2)-(7) are as in Table~\ref{BzKprops}, 
 (8) velocity-integrated CO line flux of the CO$(3-2)$, the (9) CO$(5-4)$, 
and (10) the CO$(6-5)$ emission lines,  
(11), (12) and (13) show the integration times to get a 
detection at the sensitivity indicated in the
parentheses in band~3, 4 and 5 of ALMA for the CO$(3-2)$, CO$(5-4)$ and CO$(6-5)$ emission lines, 
respectively. We also list $\nu_{\rm c}$, $\Delta \nu$ and $R$ 
used to simulate observations.}\label{LyBprops}
\begin{tabular}{l c c c c }
\\[3pt]
\hline
Properties $z=3$ LBGs                         & LBG+gal1 & LBG+gal2& LBG+gal3& LBG+gal4 \\
\hline 
(1) $M_{\rm rest}(1500$\AA$)-5\rm log(h)$                    & -18.4      & -19.25    & -20.46    & -20.77     \\
(2) log($M_{\rm mol}/M_{\odot}$)               & 9.6    & 9.7  & 9.9   & 9.93   \\
(3) log($M_{\rm stellar}/M_{\odot}$)          & 8.3   & 9 & 9.6 & 9.8  \\
(4) $\rm SFR/M_{\odot}\, yr^{-1}$              & 1.8     & 2.1  & 9.5   & 25.2    \\
(5) $Z_{\rm gas}/Z_{\odot}$                    & 0.05        & 0.07      & 0.17       & 0.15        \\
(6) $r^{\rm mol}_{\rm 50}/\rm kpc$             & 1.8     & 1.1 & 2.2   & 2.35  \\
(7) $\sigma^{\rm los}_{\rm CO}/\rm km\, s^{-1}$        & 73   & 58 & 66  & 103  \\
(8) $S_{\rm CO(3-2),V}/\rm mJy\, km\, s^{-1}$          & 13.76 & 18.42   & 58.04  & 63  \\
(9) $S_{\rm CO(5-4),V}/\rm mJy\, km\, s^{-1}$          & 9.7 & 12.97   & 41.2  & 72.6  \\
(10) $S_{\rm CO(6-5),V}/\rm mJy\, km\, s^{-1}$          & 2.1 & 2.8   & 11.6  & 39.3  \\
(11) $\tau_{\rm int, band 3}$(3-2) ($1\sigma$ noise)   & 2.8h (0.038mJy) & 1h (0.063mJy) & 7.7m (0.18mJy) & 7m (0.19mJy) \\
(12) $\tau_{\rm int, band 4}$(5-4) ($1\sigma$ noise)   & 8.7h (0.026mJy) & 3.1h(0.045mJy) & 22.8m (0.127mJy) & 8.1m (0.214mJy) \\
(13)  $\tau_{\rm int, band 5}$(6-5) ($1\sigma$ noise)   & 53d (0.0028mJy) & 42d (0.0032mJy) & 2.9d (0.012mJy) & 9.4h (0.038mJy) \\
\hline
band configuration                            & $\nu_{\rm c}$ & $\Delta \nu$ & $R/$arcsec \\
\hline
(14) band~3                              & 86.45~GHz & 0.1~GHz & 1.48 \\
(15) band~4                              & 143.7~GHz & 0.09~GHz & 1.07 \\
(16) band~5                              & 172.5~GHz & 0.11~GHz & 0.89 \\
\hline
\end{tabular}
\end{center}
\end{table*}

\begin{figure*}
\begin{center}
\includegraphics[width=0.23\textwidth]{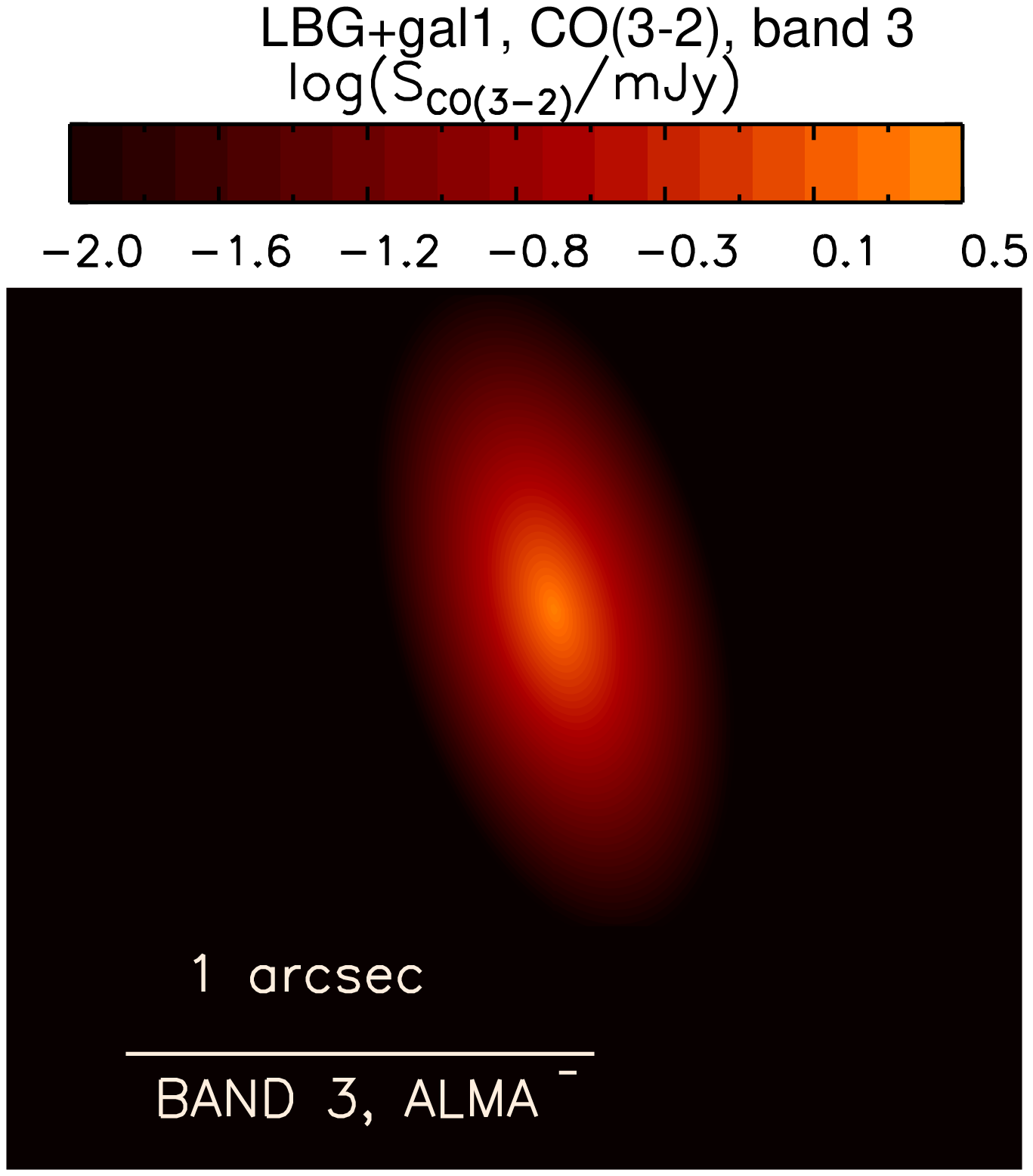}
\includegraphics[width=0.23\textwidth]{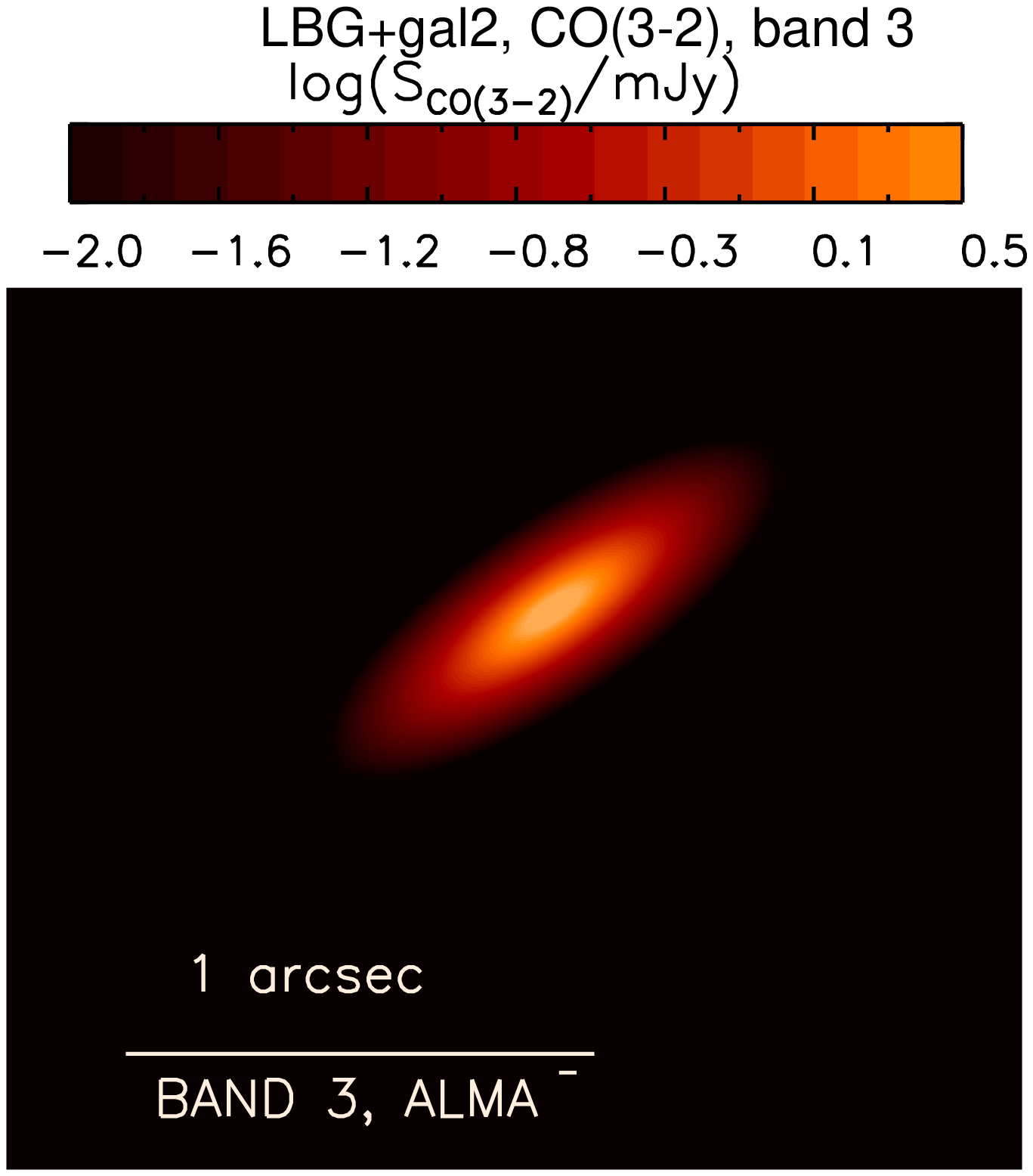}
\includegraphics[width=0.23\textwidth]{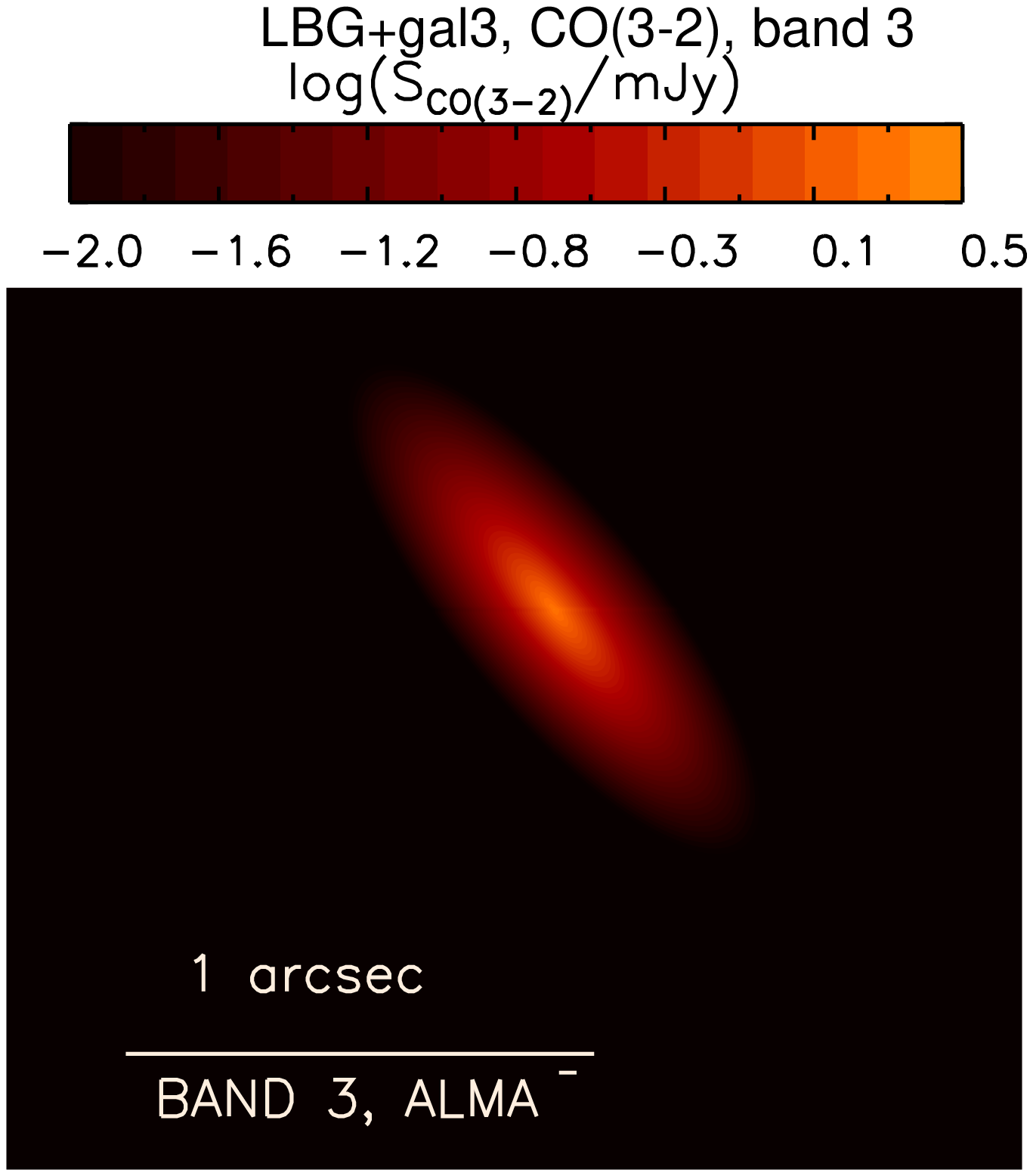}
\vspace{0.2cm}
\includegraphics[width=0.23\textwidth]{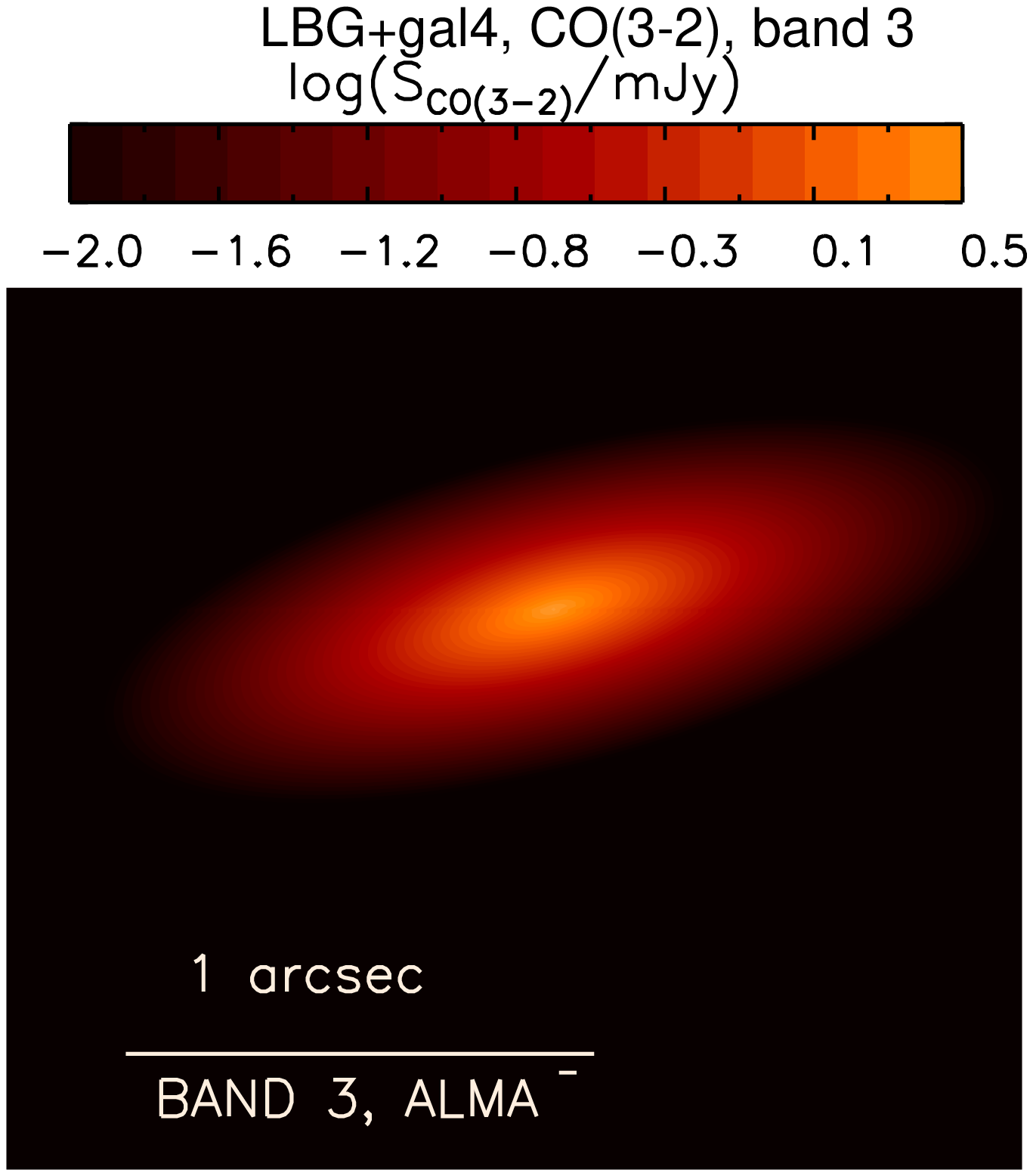}
\includegraphics[width=0.23\textwidth]{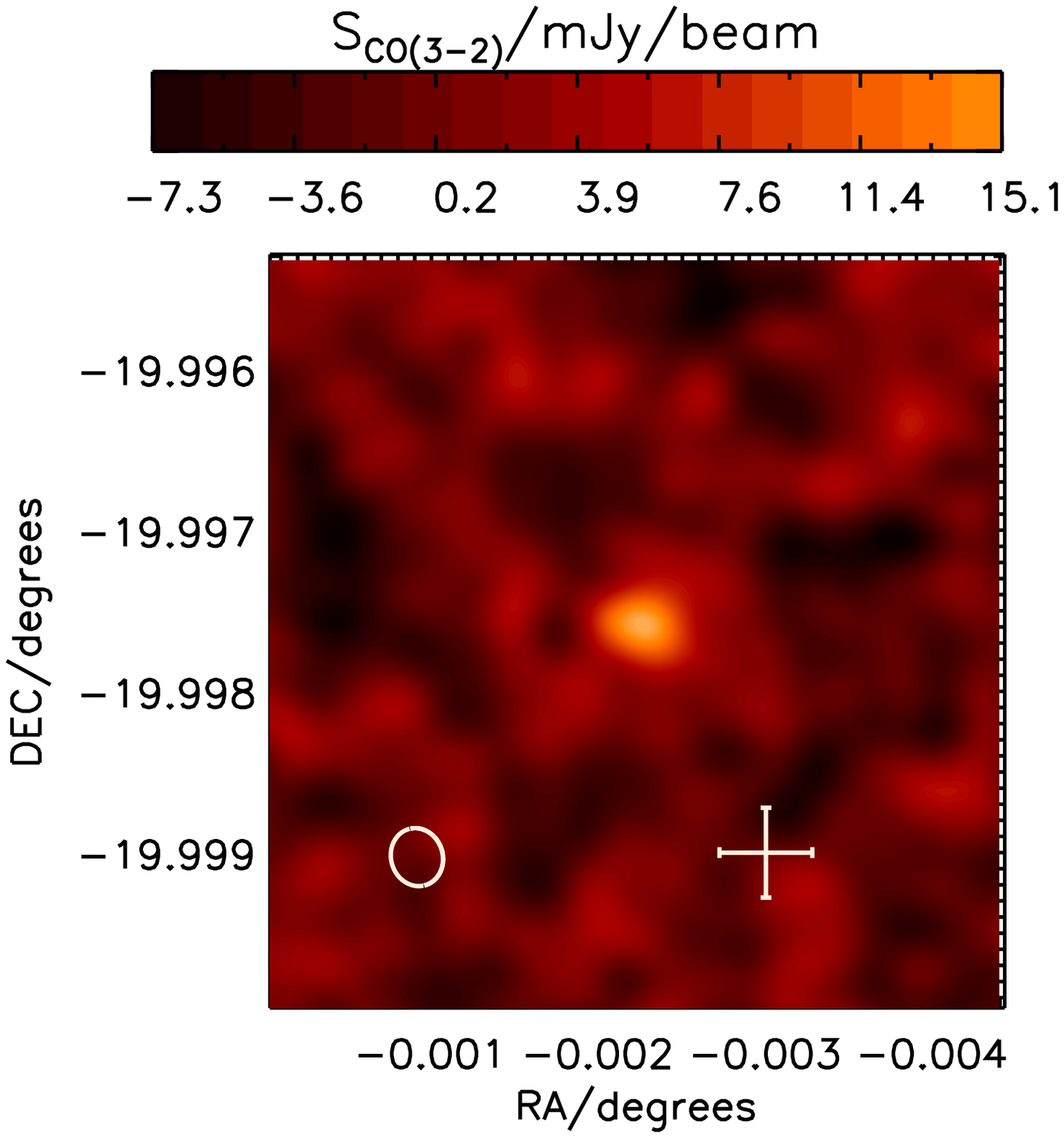}
\includegraphics[width=0.23\textwidth]{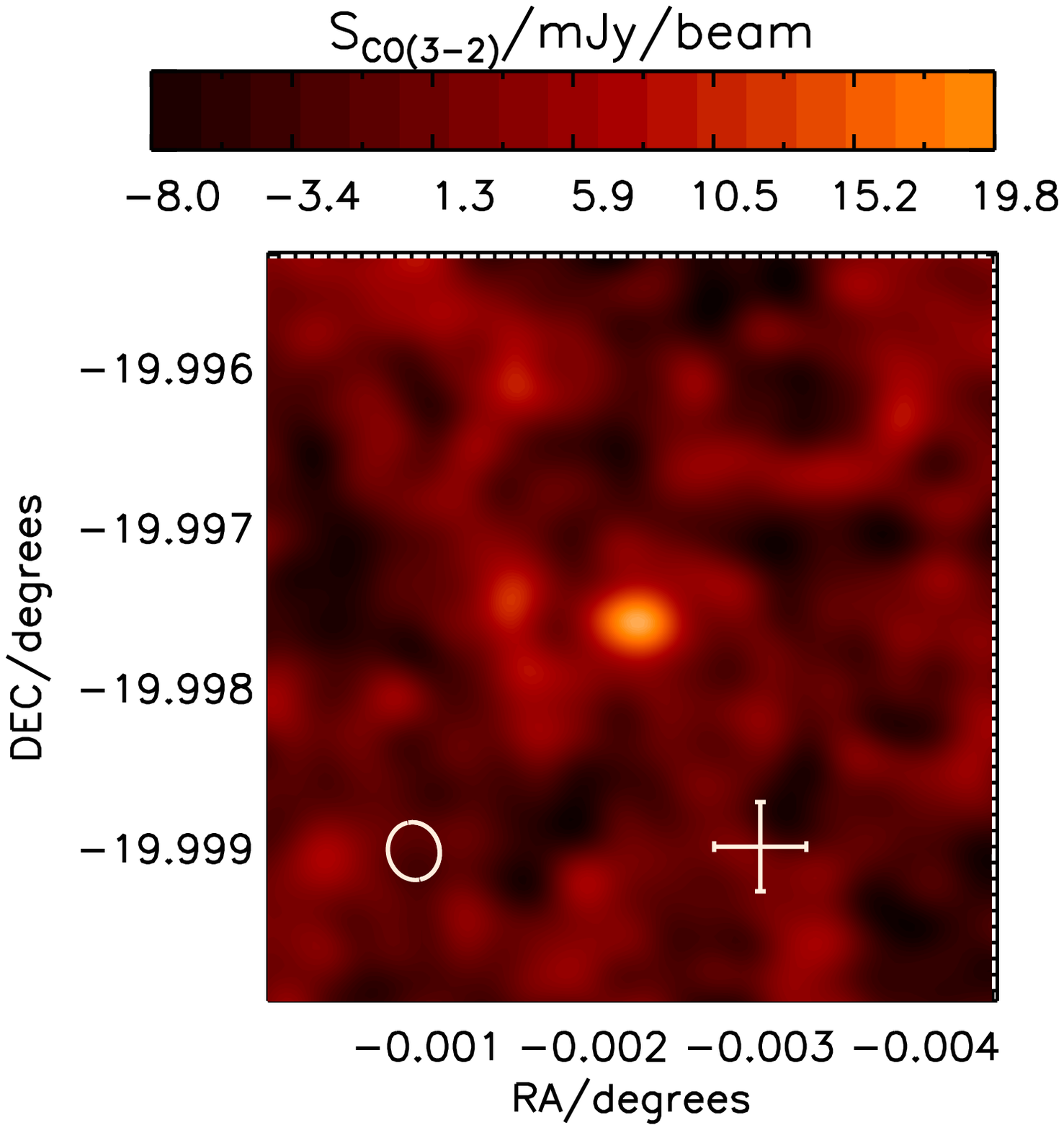}
\includegraphics[width=0.23\textwidth]{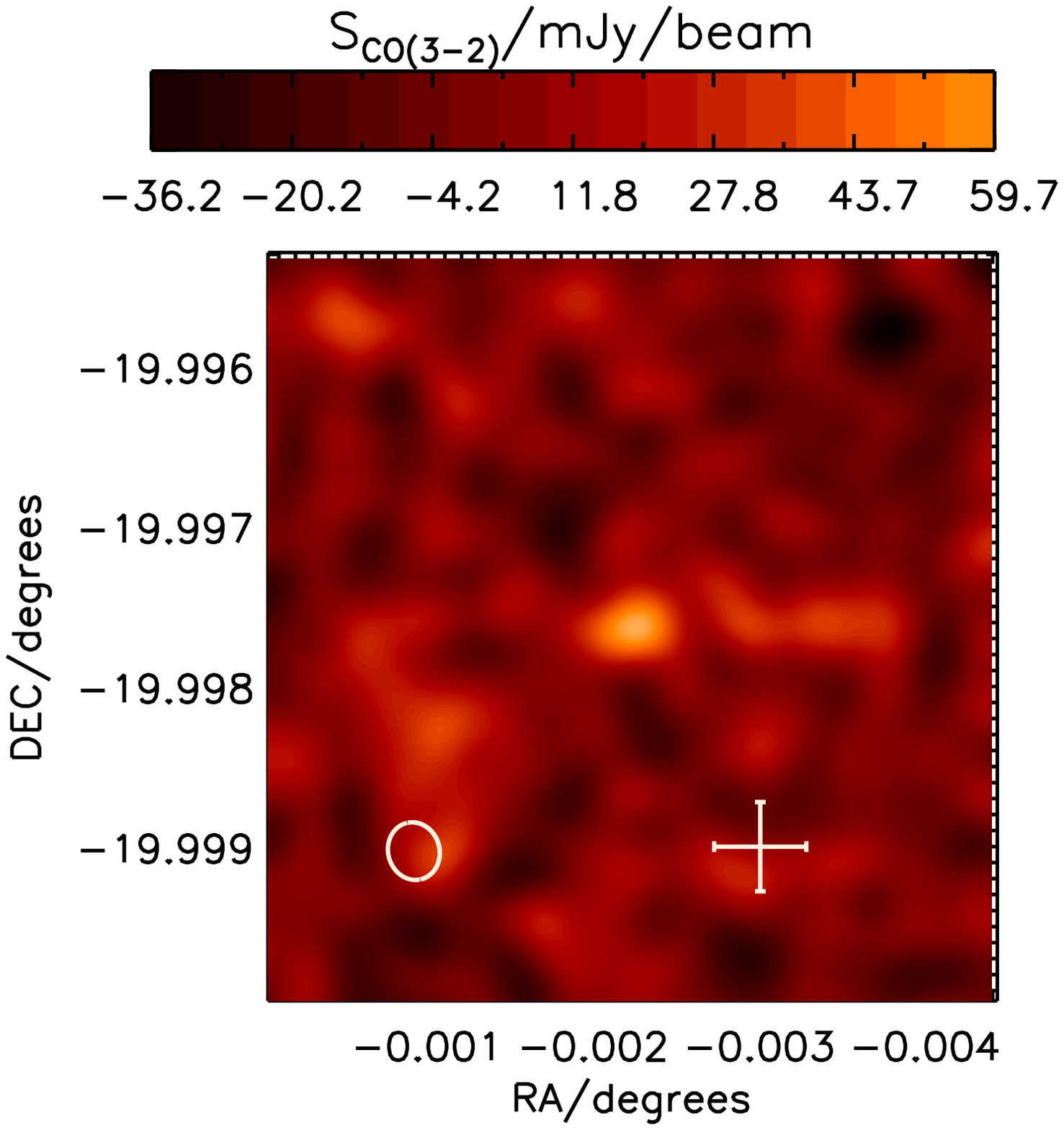}
\includegraphics[width=0.23\textwidth]{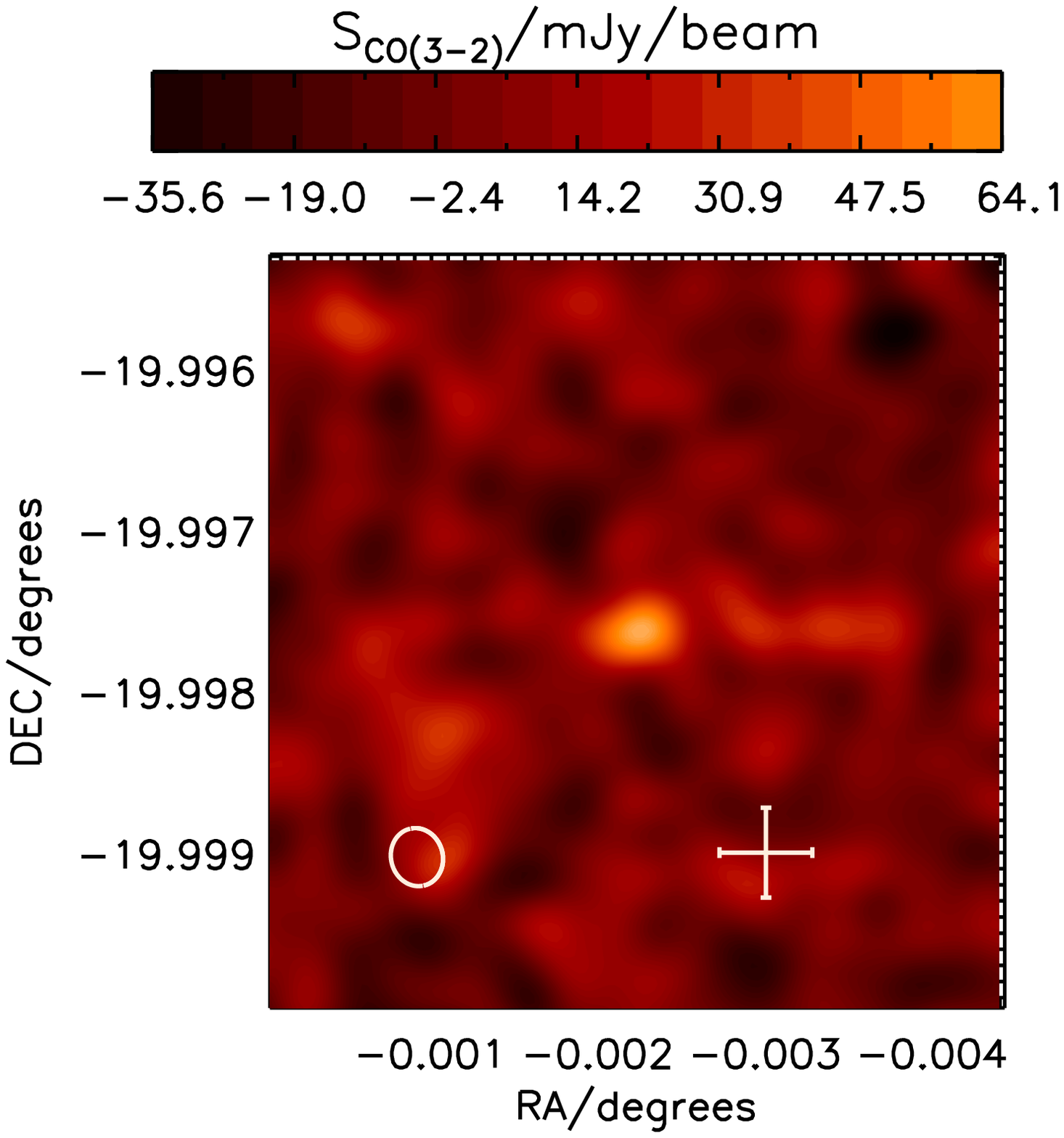}
\caption{Four Lyman-break galaxies at $z=3$. 
{\it Top panels}: the CO$(3-2)$ flux maps in logarithmic units of 
$\rm mJy/pixel$. The horizontal line shows 
1~arcsec. 
{\it Bottom panels:} the observed CO$(3-2)$ flux maps in declination vs. right ascension. 
Flux is in units of $\rm mJy/beam$. 
Maps corresponds to the outputs of the {\tt CASA} software, 
after convolving the original map with the primary beam and 
including an atmospheric model for the background noise. Ellipses in the bottom-left corner 
indicate the beam size and shape and the cross indicates 
a $1\times1$~arcsec$^2$. The flux scale is shown at the top of each figure.
Some relevant properties of these galaxies are listed in Table~\ref{LyBprops}, 
along with the integration time
 used to generate the {\tt CASA} maps of the LBGs.}
\label{Proj31}
\end{center}
\end{figure*}

Fig.~\ref{Proj3b1} shows two of the four BzK galaxies listed in Table~\ref{BzKprops}. 
For each of these galaxies we show mock images with perfect angular resolution 
and no noise (top panels) and the simulated observations (bottom panels) 
for the CO$(3-2)$ and CO$(6-5)$ lines assuming random inclinations and position angles. 
The beam of the instrument\footnote{The beam corresponds to the FWHM of a two-dimensional 
gaussian fitted to the central lobe of the Point Spread Function.} 
is plotted in the panels showing the simulated images.
Note that the angular extents of the output images for the CO$(3-2)$ and CO$(6-5)$ observations 
are different: crosses in the panels show for reference
$1\times1$~arcsec$^2$. This happens because the lower frequencies are observed 
with lower angular resolution than the higher frequencies. 
From the integration times calculated here, it is clear that ALMA can obtain  
$5 \sigma$ detections in relatively short integration times only 
in some of these galaxies.
 Even though sBzKs of Table~\ref{BzKprops} have velocity-integrated CO fluxes that are 
large, these galaxies are sufficiently big so that the peak flux in some cases is faint enough 
 to need large integration times. Note that, if the required $5\sigma$ detection 
is in the integrated flux instead of peak flux, 
the integration times are generally reduced to $\tau_{\rm int}<1$~hour.

This indicates that the sBzK selection could be an effective 
way of constructing a parent catalogue of galaxies 
to follow-up with ALMA\footnote{The ALMA basic specifications are 
described in {\tt https://almascience.nrao.edu/about-alma/full-alma}}. 
However, spatially resolving the ISM of these 
high redshift galaxies will be a very difficult 
task because (1) high order CO lines can be observed at 
better angular resolution, but are at the same time 
fainter and therefore need much longer exposures (see Fig.~\ref{COladderz2}), 
and (2) the decreasing galaxy size with increasing redshift 
predicted by {\tt GALFORM} \citep{Lacey11} and observed by several authors (e.g. 
\citealt{Bouwens04}; \citealt{Oesch10}), imply that 
galaxies at these high redshifts are intrinsically smaller than their local universe
counterparts, and therefore even more difficult to resolve.

In the examples of Fig.~\ref{Proj3b1}, it is possible to observe more than one CO emission line 
in reasonable integration times, which could help to 
constrain the excitation levels of the cold ISM in these galaxies. 
Note that observationally, targeting of sBzK galaxies to study CO has been done 
by \citet{Daddi10} for $5$ very bright galaxies using the PdBI and integration times 
$>10$~hours per source. 
Their 
CO emission is $\approx 2-3$ larger than our brightest example, BzK+gal4, which requires an 
integration time in ALMA of less than $20$~minutes in CO$(3-2)$, indicating again 
that ALMA will be able to detect CO routinely in these galaxies, even with the relatively modest amount 
of emission from the CO$(3-2)$ transition. 

\subsection{Lyman-break galaxies}

Lyman-break galaxies (LBGs) are star-forming galaxies which
are identified through the Lyman-break feature in their spectral energy distributions.
This feature is produced by absorption by neutral hydrogen in the
atmospheres of massive stars, in the ISM of the galaxy and in 
the intergalactic medium (\citealt{Steidel92}; \citealt{Steidel96}). 
Colour selection of these galaxies has been shown to be very efficient and has allowed 
the statistical assessment of their properties (such as the rest-frame UV LF and the size-luminosity relation; 
\citealt{Steidel96}; \citealt{Bouwens04}). 
LBGs are of great interest 
as a tracer of the galaxy 
population at high redshift (see \citealt{Lacey11}). These galaxies 
are at even higher redshifts than BzK galaxies, and are therefore 
key to probing  the evolution in the ISM of galaxies at early epochs.
We show in the following subsection examples of LBGs at $z=3$ and $z=6$.

\subsubsection{Lyman-break galaxies at $z=3$}

Fig.~\ref{Proj31} shows four 
LBGs at $z=3$ from the {\tt GALFORM+UCL$_{-}$PDR} model. In the top panels  
we show mock images of the CO$(3-2)$ emission of the model LBGs, and in the bottom panels we show the 
simulated ALMA observations. 
The intrinsic properties of 
these four galaxies are listed in Table~\ref{LyBprops}, along with other 
relevant information, as discussed below. 
Here, 
the rest-frame UV luminosity includes dust extinction. 
We estimate integration times as in $\S 6.1$ for imaging the CO$(3-2)$, CO$(5-4)$ and CO$(6-5)$ 
transitions, modifying the inputs 
accordingly (e.g. $\nu_{\rm c}$, $\Delta\nu$ , resolution). 
Integration times are also listed in Table~\ref{LyBprops} for the three CO emission lines. 
Note that the CO$(6-5)$ transition in three of the four LBGs shown here needs integration times 
larger than a day to obtain a $5\sigma$ detection. These cases
are not suitable for observation, but it is interesting to 
see how long an integration would be needed to be to 
get a minimum signal for a detection in such a high order CO transition. However, in the four LBGs
 it would be possible to observe more than one CO transition line, which 
would allow the physical conditions in the ISM in these galaxies to be constrained. 

LBGs were randomly chosen from the full sample of LBGs at $z=3$ 
in the {\tt GALFORM} model, in bins of UV rest-frame luminosity.
  The break in the UV LF at $z=3$ is at $M^{*}_{\rm UV}-5\rm log(h)\approx -20.3\,$ \citep{Reddy09}, 
so the LBGs in Fig.~\ref{Proj31} 
 have a UV luminosities covering a large range around $L^{*}_{\rm UV}$. 
In terms of the $M_{\rm stellar}-\rm SFR$ plane (see \citealt{Lagos10}), 
the four LBGs in Table~\ref{LyBprops} lie on the so-called `main' sequence. 
The integration times we calculate for these galaxies indicate 
that imaging $z=3$ LBGs will be an easy task for ALMA, detecting CO$(3-2)$ 
in integrations shorter than few~hours per source. Therefore LBG selection should 
provide a 
promising way of constructing a parent galaxy catalogue to follow up using ALMA. 
Note that imaging of the CO$(3-2)$ line in these LBGs at $z=3$ is easier than 
in the BzKs at $z=2$ shown before. This happens because LBGs are predicted to typically have 
smaller $\sigma_{\rm los}$ than BzKs due to 
their lower baryonic content.

The {\tt GALFORM} model predicts a weak correlation between 
the molecular mass and the UV luminosity, 
$M_{\rm mol}\propto L^{0.5}_{\rm UV}$, while  
the SFR and the gas metallicity have stronger correlations with 
the UV luminosity. Thus, most of the differences in the CO emission between 
LBGs in Fig.~\ref{Proj31} result from the 
different ISM conditions (e.g. gas metallicities, $\Sigma_{\rm SFR}$), 
rather than molecular gas mass. 
In the case of LBG+gal4, the CO$(5-4)$ flux is larger than that of 
CO$(3-2)$. This happens because this LBG 
is undergoing a bright starburst, which leads to a much more excited ISM. 
Its CO SLED peaks at higher J and falls slowly as J increases, compared to the other LBGs shown here.
These differences 
in the excitation levels of CO 
have a big impact on the observability of LBGs in the high-order CO transitions ($J>5$), 
producing large variations 
in the integration times needed  
to get a $\gtrsim 5\sigma$ detection. 
However, it is important to remark that, in this model,  
starburst galaxies constitute only $\approx 10$\% of the galaxies with $M_{\rm UV}\rm -5log(h)< -18$ 
at this redshift, 
even though their number density       
is much higher compared to low redshifts.

\subsubsection{Lyman-break galaxies at $z=6$}
\begin{table*}
\begin{center}
\caption{Properties of the four Lyman-break galaxies studied at $z=6$.% plotted in Figs~\ref{Proj31z6};
Properties are as in Table~\ref{LyBprops}, but at this redshift, the lowest CO 
transitions that fall into ALMA bands are CO$(6-5)$ and CO$(7-6)$, at 
frequencies $\nu^{\rm obs}_{6-5}=98.8\, \rm GHz$ and $\nu^{\rm obs}_{7-6}=115.2\, \rm GHz$, 
respectively.}\label{LyBprops2}
\begin{tabular}{l c c c c }
\\[3pt]
\hline
Properties $z=6$ LBGs                         & LBG+gal5 & LBG+gal6& LBG+gal7& LBG+gal8 \\
\hline 
(1) $M_{\rm rest}(1500$\AA$)-5\rm log(h)$       & -19.45             & -19.6      & -20.3    & -20.9   \\
(2) log($M_{\rm mol}/M_{\odot}$)                & 9.1 & 9.7    & 10  & 9.9    \\
(3) log($M_{\rm stellar}/M_{\odot}$)           & 7.7  & 8.0   & 8.95& 9.2  \\
(4) $\rm SFR/M_{\odot}\, yr^{-1}$               & 6 & 2.9     & 6.3  & 3     \\
(5) $Z_{\rm gas}/Z_{\odot}$                       & 0.03          & 0.05        & 0.1      & 0.3      \\
(6) $r^{\rm mol}_{\rm 50}/\rm kpc$            & 0.27  & 1.1     & 2.1 & 0.5    \\
(7) $\sigma^{\rm los}_{\rm CO}/\rm km\, s^{-1}$      & 104     & 59   & 58 & 270  \\
(8) $S_{\rm CO(2-1),V}/\rm mJy\, km\, s^{-1}$ & 3  & 4.067 & 16.93    & 5.69   \\
(9) $S_{\rm CO(6-5),V}/\rm mJy\, km\, s^{-1}$ & 4.5  & 1.04 & 3.5  & 1.5  \\
(10) $\tau_{\rm int, band 1}$(2-1) ($1\sigma$ noise)  & 3d (0.0057mJy)   & 12.7h(0.014mJy) & 43m (0.057mJy) & 2.9d (0.0058mJy)  \\
(11) $\tau_{\rm int, band 3}$(6-5) ($1\sigma$ noise)  & 179d (0.0009mJy)  & 11d (0.0035mJy) & 1d (0.012mJy) & 101d (0.0011mJy) \\
\hline
band configuration                      & $\nu_{\rm c}$ & $\Delta \nu$ & $R/$arcsec \\
\hline
(13) band~2                             & 32.86~GHz & 0.05~GHz & 4.7 \\
(14) band~3                             & 98.57~GHz & 0.1~GHz & 1.78 \\
\hline
\end{tabular}
\end{center}
\end{table*}

\begin{figure}
\begin{center}
\includegraphics[width=0.23\textwidth]{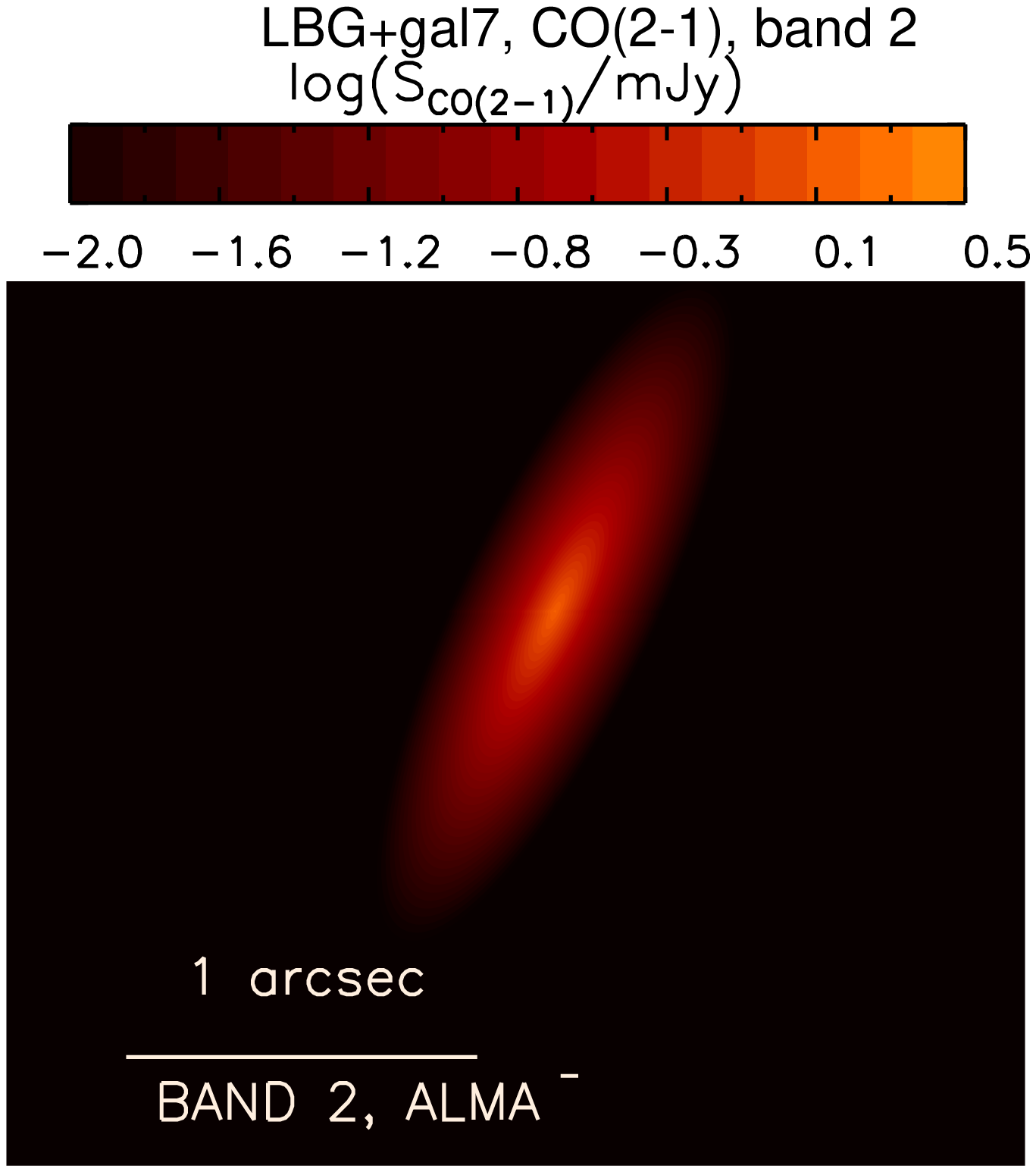}
\vspace{0.2cm}
\includegraphics[width=0.235\textwidth]{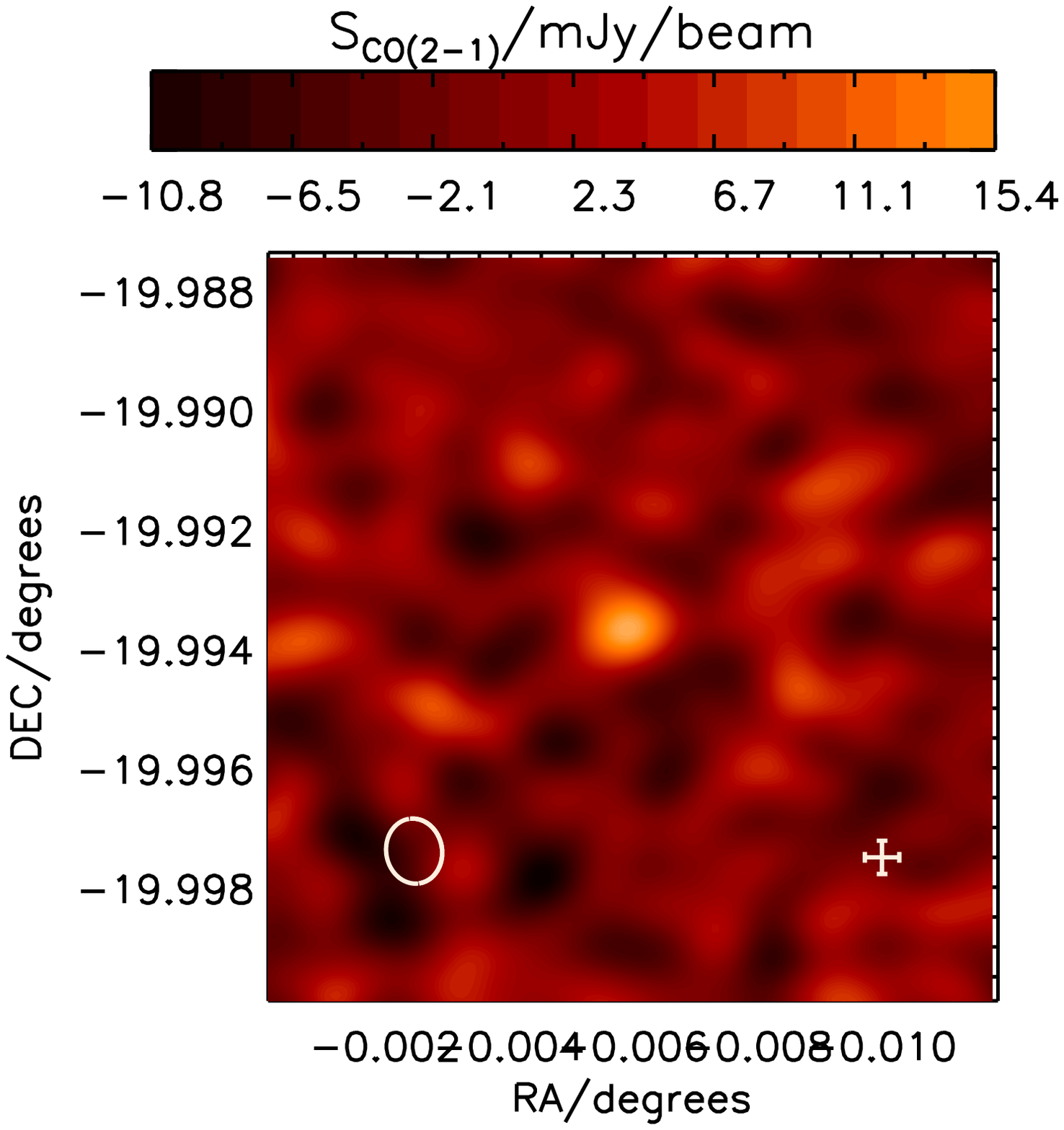}
\caption{As in Fig.~\ref{Proj31} but for a Lyman-break galaxies at $z=6$. At this redshift we focus on the 
CO$(2-1)$ emission line, which falls into the ALMA band 1.
Some relevant properties of this galaxy are listed in Table~\ref{LyBprops2} as LBG+gal7.}
\label{Proj31z6}
\end{center}
\end{figure}

We select four Lyman-break model galaxies at $z=6$ in terms of their UV luminosity.
Given the intrinsically 
faint CO emission of these galaxies, we estimate the 
integration times for these $z=6$ LBGs assuming good weather conditions (i.e. a water 
vapour column density of $0.5$~mm), unlike BzKs and $z=3$ LBGs, for which we assumed 
average weather conditions. Note that only two LBGs out of the four shown in 
Table~\ref{LyBprops2} require an integration time to detect the CO$(2-1)$ emission line 
of $\tau_{\rm int}<1$~day. We show one of these two `observable' LBGs in 
Fig~\ref{Proj31z6}.
At this very high redshift, the detection of 
any CO emission line will be a challenging task even for ALMA. 
Predictions from hydrodynamic simulations of $z=6$ Lyman-$\alpha$ emitters reach 
a similar conclusion. 
 \citet{Vallini12}, using a 
constant $X_{\rm CO(1-0)}$ and assuming local thermodynamic equilibrium to 
estimate the luminosities of higher CO transitions showed that 
only the brightest Lyman-$\alpha$ emitters at $z=6$ would be suitable for observation 
in the full ALMA configuration in 
under $<10$~hours of integration time.
 
A possible solution for the study of these very high redshift galaxies is 
 CO intensity mapping using 
instruments which are designed primarily to detect atomic hydrogen, such as
 the South-African 
SKA pathfinder, MeerKAT\footnote{{\tt http://www.ska.ac.za/meerkat/}}, and in the future, the Square 
Kilometer Array\footnote{{\tt http://www.skatelescope.org/}} (SKA). At 
 such high redshifts these telescopes will also cover the redshifted frequencies of low-J  
CO transitions. Given their larger field-of-view compared to ALMA, 
it is possible to collect the molecular emission of all galaxies in a solid angle  
and to isolate the emission from a narrow redshift range by 
cross-correlating emission maps of different molecules. From this, it is possible to 
construct the emission line power spectrum and its evolution, inferring 
valuable information, such as the total molecular content 
from galaxies that are too faint to be detected individually (Visbal et al. 2010, 2011; see \citealt{Pritchard11}
 for a review). 
 
\section{Discussion and conclusions}

We have presented a new theoretical tool to study the CO emission of galaxies and its
connection to other galaxy properties. { One of the aims of this work is to expand the predictive power
of galaxy formation models. Previously, there was no connection
between the conditions in the ISM in model galaxies
and their CO emission. The CO emission from a galaxy was obtained
from the predicted mass of molecular hydrogen essentially
by adopting an ad-hoc $X_{\rm {CO}}$. In this new hybrid model, 
the value of $X_{\rm {CO}}$ is computed by the PDR model after taking as inputs
selected predicted galaxy properties. A lack of resolution and the
use of simplifying assumptions (which make the calculation
tractable) means that we use proxy properties to describe the conditions
in the ISM. At the end of this exercise, the number of testable predictions
which can be used to reduce the model parameter space has been
considerably increased (e.g. CO luminosity functions, CO-to-IR luminosity ratios and CO SLEDs).}

The hybrid model presented in this work 
combines the galaxy formation model {\tt GALFORM} with the  
Photon Dominated Region code {\tt UCL$_{-}$PDR}, which calculates the chemistry of the cold 
ISM. We use state-of-the-art models: the \citet{Lagos11} galaxy formation model,  
which includes a calculation of the H$_2$ abundance in the ISM of galaxies and self-consistently 
estimates the instantaneous SFR from the H$_2$ surface density, and 
the \citet{Bayet11} PDR model, which models the thermal and chemical states of the ISM 
in galaxies.
The combined code uses the molecular gas mass of galaxies and 
their average ISM properties as predicted by {\tt GALFORM} as inputs to the 
{\tt UCL$_{-}$PDR} model, to estimate the CO emission in several CO transitions 
for each galaxy. 
The average ISM properties required from {\tt GALFORM} 
 are the gas metallicity,  
and the average UV and X-ray radiation fields within galaxies.
The gas metallicity and the X-ray luminosity from AGN are calculated directly in {\tt GALFORM}.
We use a phenomenological approach to estimate the UV 
radiation field 
by assuming a semi-infinite slab 
and a relation between the UV intensity and the SFR surface density in 
galaxies, with and without a correction for the average attenuation of UV photons.
Given that the {\tt GALFORM} model does not produce detailed radial profiles of galaxies, 
the combined {\tt GALFORM+UCL$_{-}$PDR} model focuses on the interpretation of 
global CO luminosities and their relation to other galaxy properties.
 
We show that this hybrid model is able to explain a wide range of the available CO observations of galaxies 
from $z=0$ to $z=6$, including LIRGs, ULIRGs, high redshift normal star-forming
galaxies, SMGs and QSOs. 
Our main conclusion are:

(i) The {\tt GALFORM+UCL$_{-}$PDR} model predicts a $z=0$ CO$(1-0)$ luminosity function 
and CO$(1-0)$-to-IR luminosity relation in good agreement with observations 
(e.g. \citealt{Keres03}; \citealt{Solomon05}). The model favours the inclusion of the 
attenuation of
UV photons due to dust extinction in the estimate of the internal UV radiation field. 

(ii) Starburst galaxies have lower CO$(1-0)$/IR luminosity ratios than
normal star-forming galaxies, which leads to an anti-correlation between 
CO$(1-0)$/IR luminosity ratio and IR luminosity. This is due in part to 
the different SF laws in bursts compared to quiescent SF. 

(iii) The {\tt GALFORM+UCL$_{-}$PDR} model predicts that the 
CO-to-IR luminosity ratio evolves weakly with redshift, regardless of the CO transition, and 
in agreement with 
 local and high redshift observational data. 

(iv) We find 
 that the model is able to explain the 
shape and normalization of the CO SLEDs for local Universe LIRGs and high redshift SMGs. 
The model predicts a peak in their CO SLEDs, on average, at $J=4$ and $J=5$, respectively.
The model predicts that the peak shifts to higher-J values 
with increasing IR luminosity. At a fixed IR luminosity, 
high redshift galaxies are predicted to have ISMs with 
higher gas kinetic temperature than low redshift galaxies, 
a result driven by 
the lower metallicities and higher SFR surface densities in such objects. 
The presence of an AGN affects the emission of 
high CO transitions in galaxies, with galaxies with AGN showing larger CO$(\rm J\rightarrow J-1)$/IR 
luminosity ratios at $J>6$ than 
 galaxies without AGN. The model predicts that observations of these high-J CO transitions 
should provide useful constraints on the heating effects of AGN on the ISM of galaxies.

We have shown that, despite its simplicity, 
this exploratory hybrid model is able to explain the observed CO 
emission of a wide range of galaxy types at low and high redshifts without the need for 
further tuning. This is the first time that a galaxy formation model has been able to 
successfully reproduce
such a wide range of observations of CO along with other galaxy properties. 
This hybrid model can be used to 
predict the observability of galaxies with the current and upcoming generation of millimeter telescopes,
such as LMT, GMT, PdBI and ALMA. This is also applies to radio telescopes 
which can be used to study molecular emission
of high redshift galaxies, such as MeerKAT, and further into the future, the SKA.

As an example of the diagnostic power of the {\tt GALFORM+UCL$_{-}$PDR} model, 
we study the observability of high redshift star-forming galaxies with ALMA, which 
is one of its key science goals. In particular, we focus on colour-selected BzK galaxies 
at $z=2$ and LBGs at $z=3$ and $z=6$. 
We use the ALMA OST software to simulate 
observations of the {\tt GALFORM+UCL$_{-}$PDR} 
model galaxies. For the first time, we present the expected CO fluxes and the 
integration times needed to obtain a 5$\sigma$ detections in the full ALMA configuration.
 We 
find that ALMA should be able to observe star-forming galaxies in low-J CO transitions routinely up to 
$z\approx 3$, with integration times of less than a few hours per source, and in a large 
fraction of the samples, in under $1$~hour. However, 
 for star-forming galaxies at $z=6$, this will be a much more difficult task, given their 
 lower gas masses and metallicities, which lead to lower CO luminosities. 
For these galaxies, future radio telescopes offer a promising alternative 
of intensity mapping of molecular emission lines, from which 
it is possible to learn about the molecular content of faint galaxies. 
Therefore, colour selection of galaxies should be an effective way 
to construct parent samples for follow up with ALMA.

{ Further observational data on the CO SLEDs of galaxies and how these relate to other galaxy
properties will be key to constraining the physical mechanisms included in the model, 
and determine whether our model is sufficient to explain the observations
(particularly of high-redshift galaxies), or whether an improved (more
general) calculation is needed, or indeed if further physical processes
have to be considered.}
The physics included and the simplifications made in this work seem to be good enough to explain  
current observations of CO. The {\tt GALFORM+UCL$_{-}$PDR} 
model will facilitate the interpretation of observations which aim to 
studying the evolution of the mass of molecular gas in galaxies and assist the planning of 
science cases for the new generation of millimeter telescopes,  
and lays the foundation for a new generation of theoretical models.

\nocite{Frayer98}
\nocite{Frayer99}
\nocite{Bell06} 
\nocite{Bell07}
\nocite{Lagos10} 
\nocite{Lagos11}
\nocite{Daddi05}
\nocite{Daddi07}

\nocite{Visbal10}
\nocite{Visbal11}
\nocite{Narayanan09}
\nocite{Narayanan12}

\section*{Acknowledgements}

We thank Serena Viti, Ian Smail, Mark Swinbank, Linda Tacconi, Karin Sandstrom and 
Desika Narayanan 
for useful discussion and comments on this work. 
We thank the anonymous referee for helpful suggestions that improved this work.
We thank the ALMA OT team, and particularly 
 Adam Avinson, Justo Gonz\'alez and Rodrigo Tobar, for helpful remarks on the ALMA OT and the {\tt CASA} software.
CL gratefully acknowledges a STFC Gemini studentship.
EB acknowledges the rolling grants `Astrophysics at Oxford'
PP/EE/E001114/1 and ST/H504862/1 from the UK Research Councils and
the John Fell OUP Research fund, ref 092/267.
JEG is supported by a Banting Fellowship, administered by the Natural Science 
and Engineering Council of Canada. 
The authors benefited from a visit of EB to Durham University supported by
a STFC visitor grant at 
Durham. Calculations for this paper were mainly performed on the ICC 
Cosmology Machine, which is part of the DiRAC Facility jointly funded by STFC, 
the Large Facilities Capital Fund of BIS, and Durham University. 

%----------------------------------------------
\bibliographystyle{mn2e_trunc8}
\bibliography{PdRToGalform}

\begin{thebibliography}{125}
\expandafter\ifx\csname natexlab\endcsname\relax\def\natexlab#1{#1}\fi

\bibitem[{{Aravena} {et~al.}(2012){Aravena}, {Carilli}, {Salvato}, {Tanaka},
  {Lentati}, {Schinnerer}, {Walter}, {Riechers}, {Smolcic}, {Capak}, {Aussel},
  {Bertoldi}, {Chapman}, {Farrah}, {Finoguenov}, {Le Floc'h}, {Lutz}, {Magdis},
  {Oliver}, {Riguccini}, {Berta}, {Magnelli}, \& {Pozzi}}]{Aravena12}
{Aravena} M., {Carilli} C.~L., {Salvato} M., {Tanaka} M., {Lentati} L.,
  {Schinnerer} E., {Walter} F., {Riechers} D. {et~al}, 2012, ApJ in press,
  ArXiv:1207.2795

\bibitem[{{Asplund} {et~al.}(2005){Asplund}, {Grevesse}, \&
  {Sauval}}]{Asplund05}
{Asplund} M., {Grevesse} N., {Sauval} A.~J., 2005, in Astronomical Society of
  the Pacific Conference Series, Vol. 336, Cosmic Abundances as Records of
  Stellar Evolution and Nucleosynthesis, {T.~G.~Barnes III \& F.~N.~Bash}, ed.,
  pp. 25--+

\bibitem[{{Baugh}(2006)}]{Baugh06}
{Baugh} C.~M., 2006, Reports on Progress in Physics, 69, 3101

\bibitem[{{Baugh} {et~al.}(2005){Baugh}, {Lacey}, {Frenk}, {Granato}, {Silva},
  {Bressan}, {Benson}, \& {Cole}}]{Baugh05}
{Baugh} C.~M., {Lacey} C.~G., {Frenk} C.~S., {Granato} G.~L., {Silva} L.,
  {Bressan} A., {Benson} A.~J., {Cole} S., 2005, \mnras, 356, 1191

\bibitem[{{Bayet} {et~al.}(2009{\natexlab{a}}){Bayet}, {Gerin}, {Phillips}, \&
  {Contursi}}]{Bayet09b}
{Bayet} E., {Gerin} M., {Phillips} T.~G., {Contursi} A., 2009{\natexlab{a}},
  \mnras, 399, 264

\bibitem[{{Bayet} {et~al.}(2009{\natexlab{b}}){Bayet}, {Viti}, {Williams},
  {Rawlings}, \& {Bell}}]{Bayet09}
{Bayet} E., {Viti} S., {Williams} D.~A., {Rawlings} J.~M.~C., {Bell} T.,
  2009{\natexlab{b}}, \apj, 696, 1466

\bibitem[{{Bayet} {et~al.}(2011){Bayet}, {Williams}, {Hartquist}, \&
  {Viti}}]{Bayet11}
{Bayet} E., {Williams} D.~A., {Hartquist} T.~W., {Viti} S., 2011, \mnras, 414,
  1583

\bibitem[{{Bell} {et~al.}(2006){Bell}, {Roueff}, {Viti}, \&
  {Williams}}]{Bell06}
{Bell} T.~A., {Roueff} E., {Viti} S., {Williams} D.~A., 2006, \mnras, 371, 1865

\bibitem[{{Bell} {et~al.}(2007){Bell}, {Viti}, \& {Williams}}]{Bell07}
{Bell} T.~A., {Viti} S., {Williams} D.~A., 2007, \mnras, 378, 983

\bibitem[{{Benson}(2010)}]{Benson10b}
{Benson} A.~J., 2010, \physrep, 495, 33

\bibitem[{{Bertram} {et~al.}(2007){Bertram}, {Eckart}, {Fischer}, {Zuther},
  {Straubmeier}, {Wisotzki}, \& {Krips}}]{Bertram07}
{Bertram} T., {Eckart} A., {Fischer} S., {Zuther} J., {Straubmeier} C.,
  {Wisotzki} L., {Krips} M., 2007, \aap, 470, 571

\bibitem[{{Bielby} {et~al.}(2011){Bielby}, {Hudelot}, {McCracken}, {Ilbert},
  {Daddi}, {Le F{\`e}vre}, {Gonzalez-Perez}, {Kneib}, {Marmo}, {Mellier},
  {Salvato}, {Sanders}, \& {Willott}}]{Bielby11}
{Bielby} R., {Hudelot} P., {McCracken} H.~J., {Ilbert} O., {Daddi} E., {Le
  F{\`e}vre} O., {Gonzalez-Perez} V., {Kneib} J.-P. {et~al}, 2011,
  ArXiv:1111.6997

\bibitem[{{Bigiel} {et~al.}(2008){Bigiel}, {Leroy}, {Walter}, {Brinks}, {de
  Blok}, {Madore}, \& {Thornley}}]{Bigiel08}
{Bigiel} F., {Leroy} A., {Walter} F., {Brinks} E., {de Blok} W.~J.~G., {Madore}
  B., {Thornley} M.~D., 2008, \aj, 136, 2846

\bibitem[{{Bigiel} {et~al.}(2011){Bigiel}, {Leroy}, {Walter}, {Brinks}, {de
  Blok}, {Kramer}, {Rix}, {Schruba}, {Schuster}, {Usero}, \&
  {Wiesemeyer}}]{Bigiel11}
{Bigiel} F., {Leroy} A.~K., {Walter} F., {Brinks} E., {de Blok} W.~J.~G.,
  {Kramer} C., {Rix} H.~W., {Schruba} A. {et~al}, 2011, \apjl, 730, L13+

\bibitem[{{Blitz} {et~al.}(2007){Blitz}, {Fukui}, {Kawamura}, {Leroy},
  {Mizuno}, \& {Rosolowsky}}]{Blitz07}
{Blitz} L., {Fukui} Y., {Kawamura} A., {Leroy} A., {Mizuno} N., {Rosolowsky}
  E., 2007, Protostars and Planets V, 81

\bibitem[{{Blitz} \& {Rosolowsky}(2006)}]{Blitz06}
{Blitz} L., {Rosolowsky} E., 2006, \apj, 650, 933

\bibitem[{{Bolatto} {et~al.}(2011){Bolatto}, {Leroy}, {Jameson}, {Ostriker},
  {Gordon}, {Lawton}, {Stanimirovi{\'c}}, {Israel}, {Madden}, {Hony},
  {Sandstrom}, {Bot}, {Rubio}, {Winkler}, {Roman-Duval}, {van Loon},
  {Oliveira}, \& {Indebetouw}}]{Bolatto11}
{Bolatto} A.~D., {Leroy} A.~K., {Jameson} K., {Ostriker} E., {Gordon} K.,
  {Lawton} B., {Stanimirovi{\'c}} S., {Israel} F.~P. {et~al}, 2011, \apj, 741,
  12

\bibitem[{{Bonatto} \& {Bica}(2011)}]{Bonatto11}
{Bonatto} C., {Bica} E., 2011, \mnras, 415, 2827

\bibitem[{{Boselli} {et~al.}(2002){Boselli}, {Lequeux}, \&
  {Gavazzi}}]{Boselli02}
{Boselli} A., {Lequeux} J., {Gavazzi} G., 2002, \aap, 384, 33

\bibitem[{{Bothwell} {et~al.}(2010){Bothwell}, {Chapman}, {Tacconi}, {Smail},
  {Ivison}, {Casey}, {Bertoldi}, {Beswick}, {Biggs}, {Blain}, {Cox}, {Genzel},
  {Greve}, {Kennicutt}, {Muxlow}, {Neri}, \& {Omont}}]{Bothwell10}
{Bothwell} M.~S., {Chapman} S.~C., {Tacconi} L., {Smail} I., {Ivison} R.~J.,
  {Casey} C.~M., {Bertoldi} F., {Beswick} R. {et~al}, 2010, \mnras, 405, 219

\bibitem[{{Bothwell} {et~al.}(2009){Bothwell}, {Kennicutt}, \&
  {Lee}}]{Bothwell09}
{Bothwell} M.~S., {Kennicutt} R.~C., {Lee} J.~C., 2009, \mnras, 400, 154

\bibitem[{{Bothwell} {et~al.}(2012){Bothwell}, {Smail}, {Chapman}, {Genzel},
  {Ivison}, {Tacconi}, {Alaghband-Zadeh}, {Bertoldi}, {Blain}, {Casey}, {Cox},
  {Greve}, {Lutz}, {Neri}, {Omont}, \& {Swinbank}}]{Bothwell12}
{Bothwell} M.~S., {Smail} I., {Chapman} S.~C., {Genzel} R., {Ivison} R.~J.,
  {Tacconi} L.~J., {Alaghband-Zadeh} S., {Bertoldi} F. {et~al}, 2012,
  ArXiv:1205.1511

\bibitem[{{Bournaud} {et~al.}(2010){Bournaud}, {Elmegreen}, {Teyssier},
  {Block}, \& {Puerari}}]{Bournaud10}
{Bournaud} F., {Elmegreen} B.~G., {Teyssier} R., {Block} D.~L., {Puerari} I.,
  2010, \mnras, 409, 1088

\bibitem[{{Bouwens} {et~al.}(2004){Bouwens}, {Illingworth}, {Blakeslee},
  {Broadhurst}, \& {Franx}}]{Bouwens04}
{Bouwens} R.~J., {Illingworth} G.~D., {Blakeslee} J.~P., {Broadhurst} T.~J.,
  {Franx} M., 2004, \apjl, 611, L1

\bibitem[{{Bower} {et~al.}(2006){Bower}, {Benson}, {Malbon}, {Helly}, {Frenk},
  {Baugh}, {Cole}, \& {Lacey}}]{Bower06}
{Bower} R.~G., {Benson} A.~J., {Malbon} R., {Helly} J.~C., {Frenk} C.~S.,
  {Baugh} C.~M., {Cole} S., {Lacey} C.~G., 2006, \mnras, 370, 645

\bibitem[{{Casey} {et~al.}(2009){Casey}, {Chapman}, {Daddi}, {Dannerbauer},
  {Pope}, {Scott}, {Bertoldi}, {Beswick}, {Blain}, {Cox}, {Genzel}, {Greve},
  {Ivison}, {Muxlow}, {Neri}, {Omont}, {Smail}, \& {Tacconi}}]{Casey09}
{Casey} C.~M., {Chapman} S.~C., {Daddi} E., {Dannerbauer} H., {Pope} A.,
  {Scott} D., {Bertoldi} F., {Beswick} R.~J. {et~al}, 2009, \mnras, 400, 670

\bibitem[{{Chang} {et~al.}(2002){Chang}, {Shu}, \& {Hou}}]{Chang02}
{Chang} R.-X., {Shu} C.-G., {Hou} J.-L., 2002, Chin. Journ. Astronomy \&
  Astrophysics, 2, 226

\bibitem[{{Cole} {et~al.}(2000){Cole}, {Lacey}, {Baugh}, \& {Frenk}}]{Cole00}
{Cole} S., {Lacey} C.~G., {Baugh} C.~M., {Frenk} C.~S., 2000, \mnras, 319, 168

\bibitem[{{Combes} {et~al.}(2011){Combes}, {Garc{\'{\i}}a-Burillo}, {Braine},
  {Schinnerer}, {Walter}, \& {Colina}}]{Combes11}
{Combes} F., {Garc{\'{\i}}a-Burillo} S., {Braine} J., {Schinnerer} E., {Walter}
  F., {Colina} L., 2011, \aap, 528, 124

\bibitem[{{Contini} {et~al.}(2002){Contini}, {Treyer}, {Sullivan}, \&
  {Ellis}}]{Contini02}
{Contini} T., {Treyer} M.~A., {Sullivan} M., {Ellis} R.~S., 2002, \mnras, 330,
  75

\bibitem[{{Cook} {et~al.}(2010){Cook}, {Evoli}, {Barausse}, {Granato}, \&
  {Lapi}}]{Cook10}
{Cook} M., {Evoli} C., {Barausse} E., {Granato} G.~L., {Lapi} A., 2010, \mnras,
  402, 941

\bibitem[{{Daddi} {et~al.}(2010){Daddi}, {Bournaud}, {Walter}, {Dannerbauer},
  {Carilli}, {Dickinson}, {Elbaz}, {Morrison}, {Riechers}, {Onodera}, {Salmi},
  {Krips}, \& {Stern}}]{Daddi10}
{Daddi} E., {Bournaud} F., {Walter} F., {Dannerbauer} H., {Carilli} C.~L.,
  {Dickinson} M., {Elbaz} D., {Morrison} G.~E. {et~al}, 2010, \apj, 713, 686

\bibitem[{{Daddi} {et~al.}(2004){Daddi}, {Cimatti}, {Renzini}, {Fontana},
  {Mignoli}, {Pozzetti}, {Tozzi}, \& {Zamorani}}]{Daddi04}
{Daddi} E., {Cimatti} A., {Renzini} A., {Fontana} A., {Mignoli} M., {Pozzetti}
  L., {Tozzi} P., {Zamorani} G., 2004, \apj, 617, 746

\bibitem[{{Daddi} {et~al.}(2007){Daddi}, {Dickinson}, {Morrison}, {Chary},
  {Cimatti}, {Elbaz}, {Frayer}, {Renzini}, {Pope}, {Alexander}, {Bauer},
  {Giavalisco}, {Huynh}, {Kurk}, \& {Mignoli}}]{Daddi07}
{Daddi} E., {Dickinson} M., {Morrison} G., {Chary} R., {Cimatti} A., {Elbaz}
  D., {Frayer} D., {Renzini} A. {et~al}, 2007, \apj, 670, 156

\bibitem[{{Daddi} {et~al.}(2005){Daddi}, {Renzini}, {Pirzkal}, {Cimatti},
  {Malhotra}, {Stiavelli}, {Xu}, {Pasquali}, {Rhoads}, {Brusa}, {di Serego
  Alighieri}, {Ferguson}, {Koekemoer}, {Moustakas}, {Panagia}, \&
  {Windhorst}}]{Daddi05}
{Daddi} E., {Renzini} A., {Pirzkal} N., {Cimatti} A., {Malhotra} S.,
  {Stiavelli} M., {Xu} C., {Pasquali} A. {et~al}, 2005, \apj, 626, 680

\bibitem[{{Danielson} {et~al.}(2010){Danielson}, {Swinbank}, {Smail}, {Cox},
  {Edge}, {Weiss}, {Harris}, {Baker}, {De Breuck}, {Geach}, {Ivison}, {Krips},
  {Lundgren}, {Longmore}, {Neri}, \& {Flaquer}}]{Danielson10}
{Danielson} A.~L.~R., {Swinbank} A.~M., {Smail} I., {Cox} P., {Edge} A.~C.,
  {Weiss} A., {Harris} A.~I., {Baker} A.~J. {et~al}, 2010, \mnras, 1565

\bibitem[{{de Jong}(1996)}]{deJong96}
{de Jong} R.~S., 1996, \aap, 313, 377

\bibitem[{{Dutton} {et~al.}(2010){Dutton}, {van den Bosch}, \&
  {Dekel}}]{Dutton09}
{Dutton} A.~A., {van den Bosch} F.~C., {Dekel} A., 2010, \mnras, 405, 1690

\bibitem[{{Engel} {et~al.}(2010){Engel}, {Tacconi}, {Davies}, {Neri}, {Smail},
  {Chapman}, {Genzel}, {Cox}, {Greve}, {Ivison}, {Blain}, {Bertoldi}, \&
  {Omont}}]{Engel10}
{Engel} H., {Tacconi} L.~J., {Davies} R.~I., {Neri} R., {Smail} I., {Chapman}
  S.~C., {Genzel} R., {Cox} P. {et~al}, 2010, \apj, 724, 233

\bibitem[{{Evans} {et~al.}(2006){Evans}, {Solomon}, {Tacconi}, {Vavilkin}, \&
  {Downes}}]{Evans06}
{Evans} A.~S., {Solomon} P.~M., {Tacconi} L.~J., {Vavilkin} T., {Downes} D.,
  2006, \aj, 132, 2398

\bibitem[{{Fanidakis} {et~al.}(2011){Fanidakis}, {Baugh}, {Benson}, {Bower},
  {Cole}, {Done}, \& {Frenk}}]{Fanidakis10}
{Fanidakis} N., {Baugh} C.~M., {Benson} A.~J., {Bower} R.~G., {Cole} S., {Done}
  C., {Frenk} C.~S., 2011, \mnras, 410, 53

\bibitem[{{Fanidakis} {et~al.}(2012){Fanidakis}, {Baugh}, {Benson}, {Bower},
  {Cole}, {Done}, {Frenk}, {Hickox}, {Lacey}, \& {Del P.~Lagos}}]{Fanidakis10b}
{Fanidakis} N., {Baugh} C.~M., {Benson} A.~J., {Bower} R.~G., {Cole} S., {Done}
  C., {Frenk} C.~S., {Hickox} R.~C. {et~al}, 2012, \mnras, 419, 2797

\bibitem[{{Feldmann} {et~al.}(2012){Feldmann}, {Gnedin}, \&
  {Kravtsov}}]{Feldman12}
{Feldmann} R., {Gnedin} N.~Y., {Kravtsov} A.~V., 2012, \apj, 747, 124

\bibitem[{{Frayer} {et~al.}(1999){Frayer}, {Ivison}, {Scoville}, {Evans},
  {Yun}, {Smail}, {Barger}, {Blain}, \& {Kneib}}]{Frayer99}
{Frayer} D.~T., {Ivison} R.~J., {Scoville} N.~Z., {Evans} A.~S., {Yun} M.~S.,
  {Smail} I., {Barger} A.~J., {Blain} A.~W. {et~al}, 1999, \apjl, 514, L13

\bibitem[{{Frayer} {et~al.}(1998){Frayer}, {Ivison}, {Scoville}, {Yun},
  {Evans}, {Smail}, {Blain}, \& {Kneib}}]{Frayer98}
{Frayer} D.~T., {Ivison} R.~J., {Scoville} N.~Z., {Yun} M., {Evans} A.~S.,
  {Smail} I., {Blain} A.~W., {Kneib} J.-P., 1998, \apjl, 506, L7

\bibitem[{{Fu} {et~al.}(2010){Fu}, {Guo}, {Kauffmann}, \& {Krumholz}}]{Fu10}
{Fu} J., {Guo} Q., {Kauffmann} G., {Krumholz} M.~R., 2010, \mnras, 409, 515

\bibitem[{{Geach} {et~al.}(2011){Geach}, {Smail}, {Moran}, {MacArthur},
  {Lagos}, \& {Edge}}]{Geach11}
{Geach} J.~E., {Smail} I., {Moran} S.~M., {MacArthur} L.~A., {Lagos} C.~d.~P.,
  {Edge} A.~C., 2011, \apjl, 730, L19+

\bibitem[{{Genzel} {et~al.}(2010){Genzel}, {Tacconi}, {Gracia-Carpio},
  {Sternberg}, {Cooper}, {Shapiro}, {Bolatto}, {Bouch{\'e}}, {Bournaud},
  {Burkert}, {Combes}, {Comerford}, {Cox}, {Davis}, {Schreiber},
  {Garcia-Burillo}, {Lutz}, {Naab}, {Neri}, {Omont}, {Shapley}, \&
  {Weiner}}]{Genzel10}
{Genzel} R., {Tacconi} L.~J., {Gracia-Carpio} J., {Sternberg} A., {Cooper}
  M.~C., {Shapiro} K., {Bolatto} A., {Bouch{\'e}} N. {et~al}, 2010, \mnras,
  407, 2091

\bibitem[{{Gnedin} {et~al.}(2009){Gnedin}, {Tassis}, \& {Kravtsov}}]{Gnedin09}
{Gnedin} N.~Y., {Tassis} K., {Kravtsov} A.~V., 2009, \apj, 697, 55

\bibitem[{{Gonz{\'a}lez} {et~al.}(2011){Gonz{\'a}lez}, {Lacey}, {Baugh}, \&
  {Frenk}}]{Gonzalez10}
{Gonz{\'a}lez} J.~E., {Lacey} C.~G., {Baugh} C.~M., {Frenk} C.~S., 2011,
  \mnras, 413, 749

\bibitem[{{Granato} {et~al.}(2000){Granato}, {Lacey}, {Silva}, {Bressan},
  {Baugh}, {Cole}, \& {Frenk}}]{Granato00}
{Granato} G.~L., {Lacey} C.~G., {Silva} L., {Bressan} A., {Baugh} C.~M., {Cole}
  S., {Frenk} C.~S., 2000, \apj, 542, 710

\bibitem[{{Grenier} {et~al.}(2005){Grenier}, {Casandjian}, \&
  {Terrier}}]{Grenier05}
{Grenier} I.~A., {Casandjian} J.-M., {Terrier} R., 2005, Science, 307, 1292

\bibitem[{{Greve} {et~al.}(2005){Greve}, {Bertoldi}, {Smail}, {Neri},
  {Chapman}, {Blain}, {Ivison}, {Genzel}, {Omont}, {Cox}, {Tacconi}, \&
  {Kneib}}]{Greve05}
{Greve} T.~R., {Bertoldi} F., {Smail} I., {Neri} R., {Chapman} S.~C., {Blain}
  A.~W., {Ivison} R.~J., {Genzel} R. {et~al}, 2005, \mnras, 359, 1165

\bibitem[{{Henry} \& {Worthey}(1999)}]{Henry99}
{Henry} R.~B.~C., {Worthey} G., 1999, \pasp, 111, 919

\bibitem[{{Hitschfeld} {et~al.}(2008){Hitschfeld}, {Aravena}, {Kramer},
  {Bertoldi}, {Stutzki}, {Bensch}, {Bronfman}, {Cubick}, {Fujishita}, {Fukui},
  {Graf}, {Honingh}, {Ito}, {Jakob}, {Jacobs}, {Klein}, {Koo}, {May}, {Miller},
  {Miyamoto}, {Mizuno}, {Onishi}, {Park}, {Pineda}, {Rabanus}, {R{\"o}llig},
  {Sasago}, {Schieder}, {Simon}, {Sun}, {Volgenau}, {Yamamoto}, \&
  {Yonekura}}]{Hitschfeld08}
{Hitschfeld} M., {Aravena} M., {Kramer} C., {Bertoldi} F., {Stutzki} J.,
  {Bensch} F., {Bronfman} L., {Cubick} M. {et~al}, 2008, \aap, 479, 75

\bibitem[{{Ivison} {et~al.}(2011){Ivison}, {Papadopoulos}, {Smail}, {Greve},
  {Thomson}, {Xilouris}, \& {Chapman}}]{Ivison11}
{Ivison} R.~J., {Papadopoulos} P.~P., {Smail} I., {Greve} T.~R., {Thomson}
  A.~P., {Xilouris} E.~M., {Chapman} S.~C., 2011, \mnras, 46

\bibitem[{{Kauffmann} {et~al.}(2012){Kauffmann}, {Li}, {Fu}, {Saintonge},
  {Catinella}, {Tacconi}, {Kramer}, {Genzel}, {Moran}, \&
  {Schiminovich}}]{Kauffmann12}
{Kauffmann} G., {Li} C., {Fu} J., {Saintonge} A., {Catinella} B., {Tacconi}
  L.~J., {Kramer} C., {Genzel} R. {et~al}, 2012, \mnras, 2601

\bibitem[{{Kennicutt}(1998)}]{Kennicutt98}
{Kennicutt} Jr. R.~C., 1998, \apj, 498, 541

\bibitem[{{Kennicutt} {et~al.}(2003){Kennicutt}, {Bresolin}, \&
  {Garnett}}]{Kennicutt03}
{Kennicutt} Jr. R.~C., {Bresolin} F., {Garnett} D.~R., 2003, \apj, 591, 801

\bibitem[{{Keres} {et~al.}(2003){Keres}, {Yun}, \& {Young}}]{Keres03}
{Keres} D., {Yun} M.~S., {Young} J.~S., 2003, \apj, 582, 659

\bibitem[{{Krumholz} {et~al.}(2009{\natexlab{a}}){Krumholz}, {McKee}, \&
  {Tumlinson}}]{Krumholz09b}
{Krumholz} M.~R., {McKee} C.~F., {Tumlinson} J., 2009{\natexlab{a}}, \apj, 693,
  216

\bibitem[{{Krumholz} {et~al.}(2009{\natexlab{b}}){Krumholz}, {McKee}, \&
  {Tumlinson}}]{Krumholz09}
---, 2009{\natexlab{b}}, \apj, 699, 850

\bibitem[{{Lacey} {et~al.}(2012){Lacey}, {Baugh}, \& {Frenk}}]{Lacey11b}
{Lacey} C.~G., {Baugh} C.~M., {Frenk} C.~S., 2012. In preparation

\bibitem[{{Lacey} {et~al.}(2011){Lacey}, {Baugh}, {Frenk}, \&
  {Benson}}]{Lacey11}
{Lacey} C.~G., {Baugh} C.~M., {Frenk} C.~S., {Benson} A.~J., 2011, \mnras, 45

\bibitem[{{Lacey} {et~al.}(2008){Lacey}, {Baugh}, {Frenk}, {Silva}, {Granato},
  \& {Bressan}}]{Lacey08}
{Lacey} C.~G., {Baugh} C.~M., {Frenk} C.~S., {Silva} L., {Granato} G.~L.,
  {Bressan} A., 2008, \mnras, 385, 1155

\bibitem[{{Lagos} {et~al.}(2011{\natexlab{a}}){Lagos}, {Baugh}, {Lacey},
  {Benson}, {Kim}, \& {Power}}]{Lagos11}
{Lagos} C.~D.~P., {Baugh} C.~M., {Lacey} C.~G., {Benson} A.~J., {Kim} H.-S.,
  {Power} C., 2011{\natexlab{a}}, \mnras, 418, 1649

\bibitem[{{Lagos} {et~al.}(2011{\natexlab{b}}){Lagos}, {Lacey}, {Baugh},
  {Bower}, \& {Benson}}]{Lagos10}
{Lagos} C.~D.~P., {Lacey} C.~G., {Baugh} C.~M., {Bower} R.~G., {Benson} A.~J.,
  2011{\natexlab{b}}, \mnras, 416, 1566

\bibitem[{{Lamareille} {et~al.}(2004){Lamareille}, {Mouhcine}, {Contini},
  {Lewis}, \& {Maddox}}]{Lamareille04}
{Lamareille} F., {Mouhcine} M., {Contini} T., {Lewis} I., {Maddox} S., 2004,
  \mnras, 350, 396

\bibitem[{{Lara-L{\'o}pez} {et~al.}(2010){Lara-L{\'o}pez}, {Cepa},
  {Bongiovanni}, {P{\'e}rez Garc{\'{\i}}a}, {Ederoclite}, {Casta{\~n}eda},
  {Fern{\'a}ndez Lorenzo}, {Povi{\'c}}, \& {S{\'a}nchez-Portal}}]{Lara-Lopez10}
{Lara-L{\'o}pez} M.~A., {Cepa} J., {Bongiovanni} A., {P{\'e}rez Garc{\'{\i}}a}
  A.~M., {Ederoclite} A., {Casta{\~n}eda} H., {Fern{\'a}ndez Lorenzo} M.,
  {Povi{\'c}} M. {et~al}, 2010, \aap, 521, L53

\bibitem[{{Leroy} {et~al.}(2007){Leroy}, {Bolatto}, {Stanimirovic}, {Mizuno},
  {Israel}, \& {Bot}}]{Leroy07}
{Leroy} A., {Bolatto} A., {Stanimirovic} S., {Mizuno} N., {Israel} F., {Bot}
  C., 2007, \apj, 658, 1027

\bibitem[{{Leroy} {et~al.}(2011){Leroy}, {Bolatto}, {Gordon}, {Sandstrom},
  {Gratier}, {Rosolowsky}, {Engelbracht}, {Mizuno}, {Corbelli}, {Fukui}, \&
  {Kawamura}}]{Leroy11}
{Leroy} A.~K., {Bolatto} A., {Gordon} K., {Sandstrom} K., {Gratier} P.,
  {Rosolowsky} E., {Engelbracht} C.~W., {Mizuno} N. {et~al}, 2011, \apj, 737,
  12

\bibitem[{{Leroy} {et~al.}(2008){Leroy}, {Walter}, {Brinks}, {Bigiel}, {de
  Blok}, {Madore}, \& {Thornley}}]{Leroy08}
{Leroy} A.~K., {Walter} F., {Brinks} E., {Bigiel} F., {de Blok} W.~J.~G.,
  {Madore} B., {Thornley} M.~D., 2008, \aj, 136, 2782

\bibitem[{{Lin} {et~al.}(2011){Lin}, {Dickinson}, {Jian}, {Merson}, {Baugh},
  {Scott}, {Foucaud}, {Wang}, {Yan}, {Yan}, {Cheng}, {Guo}, {Helly}, {Kirsten},
  {Koo}, {Lagos}, {Meger}, {Pope}, {Simard}, {Grogin}, {Messias}, \&
  {Wang}}]{Lin11}
{Lin} L., {Dickinson} M., {Jian} H.-Y., {Merson} A.~I., {Baugh} C.~M., {Scott}
  D., {Foucaud} S., {Wang} W.-H. {et~al}, 2011, ArXiv:1111.2135

\bibitem[{{Lisenfeld} {et~al.}(2011){Lisenfeld}, {Espada}, {Verdes-Montenegro},
  {Kuno}, {Leon}, {Sabater}, {Sato}, {Sulentic}, {Verley}, \&
  {Yun}}]{Lisenfeld11}
{Lisenfeld} U., {Espada} D., {Verdes-Montenegro} L., {Kuno} N., {Leon} S.,
  {Sabater} J., {Sato} N., {Sulentic} J. {et~al}, 2011, \aap, 534, A102

\bibitem[{{Mac Low} \& {Glover}(2012)}]{MacLow12}
{Mac Low} M.-M., {Glover} S.~C.~O., 2012, \apj, 746, 135

\bibitem[{{Magdis} {et~al.}(2011){Magdis}, {Daddi}, {Elbaz}, {Sargent},
  {Dickinson}, {Dannerbauer}, {Aussel}, {Walter}, {Hwang}, {Charmandaris},
  {Hodge}, {Riechers}, {Rigopoulou}, {Carilli}, {Pannella}, {Mullaney},
  {Leiton}, \& {Scott}}]{Magdis11}
{Magdis} G.~E., {Daddi} E., {Elbaz} D., {Sargent} M., {Dickinson} M.,
  {Dannerbauer} H., {Aussel} H., {Walter} F. {et~al}, 2011, \apjl, 740, L15

\bibitem[{{Mannucci} {et~al.}(2010){Mannucci}, {Cresci}, {Maiolino}, {Marconi},
  \& {Gnerucci}}]{Mannucci10}
{Mannucci} F., {Cresci} G., {Maiolino} R., {Marconi} A., {Gnerucci} A., 2010,
  \mnras, 408, 2115

\bibitem[{{Marconi} {et~al.}(2004){Marconi}, {Risaliti}, {Gilli}, {Hunt},
  {Maiolino}, \& {Salvati}}]{Marconi04}
{Marconi} A., {Risaliti} G., {Gilli} R., {Hunt} L.~K., {Maiolino} R., {Salvati}
  M., 2004, \mnras, 351, 169

\bibitem[{{Meijerink} {et~al.}(2007){Meijerink}, {Spaans}, \&
  {Israel}}]{Meijerink07}
{Meijerink} R., {Spaans} M., {Israel} F.~P., 2007, \aap, 461, 793

\bibitem[{{Meijerink} {et~al.}(2011){Meijerink}, {Spaans}, {Loenen}, \& {van
  der Werf}}]{Meijerink11}
{Meijerink} R., {Spaans} M., {Loenen} A.~F., {van der Werf} P.~P., 2011, \aap,
  525, A119

\bibitem[{{Melbourne} \& {Salzer}(2002)}]{Melbourne02}
{Melbourne} J., {Salzer} J.~J., 2002, \aj, 123, 2302

\bibitem[{{Merson} {et~al.}(2012){Merson}, {Baugh}, {Helly}, {Gonzalez-Perez},
  {Cole}, {Bielby}, {Norberg}, {Frenk}, {Benson}, {Bower}, {Lacey}, \&
  {Lagos}}]{Merson12}
{Merson} A.~I., {Baugh} C.~M., {Helly} J.~C., {Gonzalez-Perez} V., {Cole} S.,
  {Bielby} R., {Norberg} P., {Frenk} C.~S. {et~al}, 2012, ArXiv:1206.4049

\bibitem[{{Nagy} {et~al.}(2012){Nagy}, {van der Tak}, {Fuller}, {Spaans}, \&
  {Plume}}]{Nagy12}
{Nagy} Z., {van der Tak} F.~F.~S., {Fuller} G.~A., {Spaans} M., {Plume} R.,
  2012, \aap, 542, A6

\bibitem[{{Narayanan} {et~al.}(2009){Narayanan}, {Cox}, {Hayward}, {Younger},
  \& {Hernquist}}]{Narayanan09}
{Narayanan} D., {Cox} T.~J., {Hayward} C.~C., {Younger} J.~D., {Hernquist} L.,
  2009, \mnras, 400, 1919

\bibitem[{{Narayanan} {et~al.}(2005){Narayanan}, {Groppi}, {Kulesa}, \&
  {Walker}}]{Narayanan05}
{Narayanan} D., {Groppi} C.~E., {Kulesa} C.~A., {Walker} C.~K., 2005, \apj,
  630, 269

\bibitem[{{Narayanan} {et~al.}(2012){Narayanan}, {Krumholz}, {Ostriker}, \&
  {Hernquist}}]{Narayanan12}
{Narayanan} D., {Krumholz} M.~R., {Ostriker} E.~C., {Hernquist} L., 2012,
  \mnras, 421, 3127

\bibitem[{{Neri} {et~al.}(2003){Neri}, {Genzel}, {Ivison}, {Bertoldi}, {Blain},
  {Chapman}, {Cox}, {Greve}, {Omont}, \& {Frayer}}]{Neri03}
{Neri} R., {Genzel} R., {Ivison} R.~J., {Bertoldi} F., {Blain} A.~W., {Chapman}
  S.~C., {Cox} P., {Greve} T.~R. {et~al}, 2003, \apjl, 597, L113

\bibitem[{{Obreschkow} {et~al.}(2009){Obreschkow}, {Heywood}, {Kl{\"o}ckner},
  \& {Rawlings}}]{Obreschkow09d}
{Obreschkow} D., {Heywood} I., {Kl{\"o}ckner} H.-R., {Rawlings} S., 2009, \apj,
  702, 1321

\bibitem[{{Oesch} {et~al.}(2010){Oesch}, {Bouwens}, {Carollo}, {Illingworth},
  {Trenti}, {Stiavelli}, {Magee}, {Labb{\'e}}, \& {Franx}}]{Oesch10}
{Oesch} P.~A., {Bouwens} R.~J., {Carollo} C.~M., {Illingworth} G.~D., {Trenti}
  M., {Stiavelli} M., {Magee} D., {Labb{\'e}} I. {et~al}, 2010, \apjl, 709, L21

\bibitem[{{Papadopoulos}(2010)}]{Papadopoulos10}
{Papadopoulos} P.~P., 2010, \apj, 720, 226

\bibitem[{{Papadopoulos} {et~al.}(2011){Papadopoulos}, {van der Werf},
  {Xilouris}, {Isaak}, {Gao}, \& {Muehle}}]{Papadopoulos11}
{Papadopoulos} P.~P., {van der Werf} P., {Xilouris} E.~M., {Isaak} K.~G., {Gao}
  Y., {Muehle} S., 2011, MNRAS in press, ArXiv:1109.4176

\bibitem[{{Peletier} {et~al.}(1990){Peletier}, {Davies}, {Illingworth},
  {Davis}, \& {Cawson}}]{Peletier90}
{Peletier} R.~F., {Davies} R.~L., {Illingworth} G.~D., {Davis} L.~E., {Cawson}
  M., 1990, \aj, 100, 1091

\bibitem[{{Pelupessy} \& {Papadopoulos}(2009)}]{Pelupessy09}
{Pelupessy} F.~I., {Papadopoulos} P.~P., 2009, \apj, 707, 954

\bibitem[{{Pelupessy} {et~al.}(2006){Pelupessy}, {Papadopoulos}, \& {van der
  Werf}}]{Pelupessy06}
{Pelupessy} F.~I., {Papadopoulos} P.~P., {van der Werf} P., 2006, \apj, 645,
  1024

\bibitem[{{Pritchard} \& {Loeb}(2011)}]{Pritchard11}
{Pritchard} J.~R., {Loeb} A., 2011, ArXiv:1109.6012

\bibitem[{{Rahman} {et~al.}(2012){Rahman}, {Bolatto}, {Xue}, {Wong}, {Leroy},
  {Walter}, {Bigiel}, {Rosolowsky}, {Fisher}, {Vogel}, {Blitz}, {West}, \&
  {Ott}}]{Rahman11}
{Rahman} N., {Bolatto} A.~D., {Xue} R., {Wong} T., {Leroy} A.~K., {Walter} F.,
  {Bigiel} F., {Rosolowsky} E. {et~al}, 2012, \apj, 745, 183

\bibitem[{{Reach} {et~al.}(1994){Reach}, {Koo}, \& {Heiles}}]{Reach94}
{Reach} W.~T., {Koo} B.-C., {Heiles} C., 1994, \apj, 429, 672

\bibitem[{{Reddy} \& {Steidel}(2009)}]{Reddy09}
{Reddy} N.~A., {Steidel} C.~C., 2009, \apj, 692, 778

\bibitem[{{Riechers}(2011)}]{Riechers10}
{Riechers} D.~A., 2011, \apj, 730, 108

\bibitem[{{R{\"o}llig} {et~al.}(2007){R{\"o}llig}, {Abel}, {Bell}, {Bensch},
  {Black}, {Ferland}, {Jonkheid}, {Kamp}, {Kaufman}, {Le Bourlot}, {Le Petit},
  {Meijerink}, {Morata}, {Ossenkopf}, {Roueff}, {Shaw}, {Spaans}, {Sternberg},
  {Stutzki}, {Thi}, {van Dishoeck}, {van Hoof}, {Viti}, \&
  {Wolfire}}]{Rollig07}
{R{\"o}llig} M., {Abel} N.~P., {Bell} T., {Bensch} F., {Black} J., {Ferland}
  G.~J., {Jonkheid} B., {Kamp} I. {et~al}, 2007, \aap, 467, 187

\bibitem[{{Saintonge} {et~al.}(2011){Saintonge}, {Kauffmann}, {Kramer},
  {Tacconi}, {Buchbender}, {Catinella}, {Fabello}, {Graci{\'a}-Carpio}, {Wang},
  {Cortese}, {Fu}, {Genzel}, {Giovanelli}, {Guo}, {Haynes}, {Heckman},
  {Krumholz}, {Lemonias}, {Li}, {Moran}, {Rodriguez-Fernandez}, {Schiminovich},
  {Schuster}, \& {Sievers}}]{Saintonge11}
{Saintonge} A., {Kauffmann} G., {Kramer} C., {Tacconi} L.~J., {Buchbender} C.,
  {Catinella} B., {Fabello} S., {Graci{\'a}-Carpio} J. {et~al}, 2011, \mnras,
  415, 32

\bibitem[{{Schruba} {et~al.}(2011){Schruba}, {Leroy}, {Walter}, {Bigiel},
  {Brinks}, {de Blok}, {Dumas}, {Kramer}, {Rosolowsky}, {Sandstrom},
  {Schuster}, {Usero}, {Weiss}, \& {Wiesemeyer}}]{Schruba11}
{Schruba} A., {Leroy} A.~K., {Walter} F., {Bigiel} F., {Brinks} E., {de Blok}
  W.~J.~G., {Dumas} G., {Kramer} C. {et~al}, 2011, \aj, 142, 37

\bibitem[{{Scoville} {et~al.}(2003){Scoville}, {Frayer}, {Schinnerer}, \&
  {Christopher}}]{Scoville03}
{Scoville} N.~Z., {Frayer} D.~T., {Schinnerer} E., {Christopher} M., 2003,
  \apjl, 585, L105

\bibitem[{{Sheth} {et~al.}(2004){Sheth}, {Blain}, {Kneib}, {Frayer}, {van der
  Werf}, \& {Knudsen}}]{Sheth04}
{Sheth} K., {Blain} A.~W., {Kneib} J.-P., {Frayer} D.~T., {van der Werf} P.~P.,
  {Knudsen} K.~K., 2004, \apjl, 614, L5

\bibitem[{{Silva} {et~al.}(1998){Silva}, {Granato}, {Bressan}, \&
  {Danese}}]{Silvia98}
{Silva} L., {Granato} G.~L., {Bressan} A., {Danese} L., 1998, \apj, 509, 103

\bibitem[{{Solomon} {et~al.}(1997){Solomon}, {Downes}, {Radford}, \&
  {Barrett}}]{Solomon97}
{Solomon} P.~M., {Downes} D., {Radford} S.~J.~E., {Barrett} J.~W., 1997, \apj,
  478, 144

\bibitem[{{Solomon} \& {Vanden Bout}(2005)}]{Solomon05}
{Solomon} P.~M., {Vanden Bout} P.~A., 2005, \araa, 43, 677

\bibitem[{{Springel} {et~al.}(2005){Springel}, {White}, {Jenkins}, {Frenk},
  {Yoshida}, {Gao}, {Navarro}, {Thacker}, {Croton}, {Helly}, {Peacock}, {Cole},
  {Thomas}, {Couchman}, {Evrard}, {Colberg}, \& {Pearce}}]{Springel05}
{Springel} V., {White} S.~D.~M., {Jenkins} A., {Frenk} C.~S., {Yoshida} N.,
  {Gao} L., {Navarro} J., {Thacker} R. {et~al}, 2005, \nat, 435, 629

\bibitem[{{Steidel} {et~al.}(1996){Steidel}, {Giavalisco}, {Pettini},
  {Dickinson}, \& {Adelberger}}]{Steidel96}
{Steidel} C.~C., {Giavalisco} M., {Pettini} M., {Dickinson} M., {Adelberger}
  K.~L., 1996, \apjl, 462, L17

\bibitem[{{Steidel} \& {Hamilton}(1992)}]{Steidel92}
{Steidel} C.~C., {Hamilton} D., 1992, \aj, 104, 941

\bibitem[{{Tacconi} {et~al.}(2010){Tacconi}, {Genzel}, {Neri}, {Cox}, {Cooper},
  {Shapiro}, {Bolatto}, {Bouch{\'e}}, {Bournaud}, {Burkert}, {Combes},
  {Comerford}, {Davis}, {Schreiber}, {Garcia-Burillo}, {Gracia-Carpio}, {Lutz},
  {Naab}, {Omont}, {Shapley}, {Sternberg}, \& {Weiner}}]{Tacconi10}
{Tacconi} L.~J., {Genzel} R., {Neri} R., {Cox} P., {Cooper} M.~C., {Shapiro}
  K., {Bolatto} A., {Bouch{\'e}} N. {et~al}, 2010, \nat, 463, 781

\bibitem[{{Tacconi} {et~al.}(2006){Tacconi}, {Neri}, {Chapman}, {Genzel},
  {Smail}, {Ivison}, {Bertoldi}, {Blain}, {Cox}, {Greve}, \&
  {Omont}}]{Tacconi06}
{Tacconi} L.~J., {Neri} R., {Chapman} S.~C., {Genzel} R., {Smail} I., {Ivison}
  R.~J., {Bertoldi} F., {Blain} A. {et~al}, 2006, \apj, 640, 228

\bibitem[{{Tremonti} {et~al.}(2004){Tremonti}, {Heckman}, {Kauffmann},
  {Brinchmann}, {Charlot}, {White}, {Seibert}, {Peng}, {Schlegel}, {Uomoto},
  {Fukugita}, \& {Brinkmann}}]{Tremonti04}
{Tremonti} C.~A., {Heckman} T.~M., {Kauffmann} G., {Brinchmann} J., {Charlot}
  S., {White} S.~D.~M., {Seibert} M., {Peng} E.~W. {et~al}, 2004, \apj, 613,
  898

\bibitem[{{Vallini} {et~al.}(2012){Vallini}, {Dayal}, \& {Ferrara}}]{Vallini12}
{Vallini} L., {Dayal} P., {Ferrara} A., 2012, \mnras, 2445

\bibitem[{{van der Werf} {et~al.}(2011){van der Werf}, {Berciano Alba},
  {Spaans}, {Loenen}, {Meijerink}, {Riechers}, {Cox}, {Weiss}, \&
  {Walter}}]{VanderWerf11}
{van der Werf} P.~P., {Berciano Alba} A., {Spaans} M., {Loenen} E., {Meijerink}
  R., {Riechers} D., {Cox} P., {Weiss} A. {et~al}, 2011, ArXiv:1106.4825

\bibitem[{{van der Werf} {et~al.}(2010){van der Werf}, {Isaak}, {Meijerink},
  {Spaans}, {Rykala}, {Fulton}, {Loenen}, {Walter}, {Wei{\ss}}, {Armus},
  {Fischer}, {Israel}, {Harris}, {Veilleux}, {Henkel}, {Savini}, {Lord},
  {Smith}, {Gonz{\'a}lez-Alfonso}, {Naylor}, {Aalto}, {Charmandaris}, {Dasyra},
  {Evans}, {Gao}, {Greve}, {G{\"u}sten}, {Kramer}, {Mart{\'{\i}}n-Pintado},
  {Mazzarella}, {Papadopoulos}, {Sanders}, {Spinoglio}, {Stacey}, {Vlahakis},
  {Wiedner}, \& {Xilouris}}]{VanderWerf10}
{van der Werf} P.~P., {Isaak} K.~G., {Meijerink} R., {Spaans} M., {Rykala} A.,
  {Fulton} T., {Loenen} A.~F., {Walter} F. {et~al}, 2010, \aap, 518, L42+

\bibitem[{{Visbal} \& {Loeb}(2010)}]{Visbal10}
{Visbal} E., {Loeb} A., 2010, J. Cosmol. Astropart. Phys., 11, 16

\bibitem[{{Visbal} {et~al.}(2011){Visbal}, {Trac}, \& {Loeb}}]{Visbal11}
{Visbal} E., {Trac} H., {Loeb} A., 2011, J. Cosmol. Astropart. Phys., 8, 10

\bibitem[{{Wei{\ss}} {et~al.}(2007){Wei{\ss}}, {Downes}, {Neri}, {Walter},
  {Henkel}, {Wilner}, {Wagg}, \& {Wiklind}}]{Weiss07}
{Wei{\ss}} A., {Downes} D., {Neri} R., {Walter} F., {Henkel} C., {Wilner}
  D.~J., {Wagg} J., {Wiklind} T., 2007, \aap, 467, 955

\bibitem[{{Wei{\ss}} {et~al.}(2005){Wei{\ss}}, {Downes}, {Walter}, \&
  {Henkel}}]{Weiss05}
{Wei{\ss}} A., {Downes} D., {Walter} F., {Henkel} C., 2005, \aap, 440, L45

\bibitem[{{Wolfire} {et~al.}(2010){Wolfire}, {Hollenbach}, \&
  {McKee}}]{Wolfire10}
{Wolfire} M.~G., {Hollenbach} D., {McKee} C.~F., 2010, \apj, 716, 1191

\bibitem[{{Wolfire} {et~al.}(2003){Wolfire}, {McKee}, {Hollenbach}, \&
  {Tielens}}]{Wolfire03}
{Wolfire} M.~G., {McKee} C.~F., {Hollenbach} D., {Tielens} A.~G.~G.~M., 2003,
  \apj, 587, 278

\bibitem[{{Wong} \& {Blitz}(2002)}]{Wong02}
{Wong} T., {Blitz} L., 2002, \apj, 569, 157

\bibitem[{{Yao} {et~al.}(2003){Yao}, {Seaquist}, {Kuno}, \& {Dunne}}]{Yao03}
{Yao} L., {Seaquist} E.~R., {Kuno} N., {Dunne} L., 2003, \apj, 588, 771

\bibitem[{{Young} \& {Scoville}(1991)}]{Young91}
{Young} J.~S., {Scoville} N.~Z., 1991, \araa, 29, 581

\end{thebibliography}
%---------------------------------------------------------------------
\label{lastpage}

\appendix
\section[]{The CO line and IR luminosity}\label{App:COIR}

In this appendix we explain in more detail how we calculate CO luminosities 
in the different units used in the paper and also the total IR luminosity. 

We express the CO luminosity in three different ways: 
(i) as a line luminosity, $L_{\rm CO}$,
typically 
expressed in solar luminosities, (ii) as a velocity-integrated CO luminosity, $L_{\rm CO,V}$, which is typically expressed
in units of $\rm Jy\, km\, s^{-1}\, Mpc^{-2}$, and (iii) 
as a brightness temperature luminosity, $L^{\prime}_{\rm CO}$, in units of
$\rm K\, km\, s^{-1}\, pc^{-2}$. We estimate these luminosities from the
molecular hydrogen mass and the value of $X_{\rm CO}$ for each galaxy.

The spectral energy distribution of a source is characterised by the monochromatic 
luminosity, $l_{\nu}(\nu_{\rm rest})$, where $\nu_{\rm rest}$ is the rest-frame frequency. 
The total luminosity of the emission line we are interested in is simply the integral 
of $l_{\nu}(\nu_{\rm rest})$ 
over the frequency width of the line, 

\begin{equation}
L_{\rm CO}=\int l_{\nu}(\nu_{\rm rest})\, {\rm d} \nu_{\rm rest}.
\label{Lcodef}
\end{equation}

\noindent The units of $L_{\rm CO}$ are proportional to $\rm erg\, s^{-1}$.
Observationally, what is measured is the 
monochromatic flux, $s_{\nu}(\nu_{\rm obs})$, where $\nu_{\rm obs}=\nu_{\rm rest}/(1+z)$ is the 
observed frequency.
The flux density of this emitter is simply 
the frequency-integrated flux, $S=\int s_{\nu}(\nu_{\rm obs}){\rm d}\nu_{\rm obs}$. 
The frequency-integrated 
flux can be calculated from the intrinsic luminosity, which
is what we predict in the {\tt GALFORM+UCL$_{-}$PDR} model, and the 
luminosity distance, $D_{\rm L}$, 

\begin{equation}
S_{\rm CO}=\frac{L_{\rm CO}}{4\pi D^2_{\rm L}}.
\label{Scodef}
\end{equation}

A widely used luminosity in radio observations 
is the velocity-integrated CO luminosity, $L_{\rm CO,V}$. This  
 corresponds to the integral of the monochromatic 
luminosity over velocity  

\begin{equation}
L_{\rm CO,V}=\int l_{\nu}(\nu_{\rm rest})\, {\rm d} V=\frac{c}{\nu_{\rm rest}}\, L_{\rm CO}.
\label{Lcodef2}
\end{equation}

\noindent Here $c$ is the speed of light 
and ${\rm d} V$ is the 
differential velocity, which is related to $\nu_{\rm rest}$ and $\nu_{\rm obs}$ as 
${\rm d} V=c\, ({\rm d} \nu_{\rm rest}/\nu_{\rm rest})=c\, ({\rm d} \nu_{\rm 
obs}/\nu_{\rm obs})$. 
Observationally, the 
velocity-integrated luminosity is calculated from the 
velocity-integrated flux, $S_{\rm CO,V}=\int s_{\nu}(\nu_{\rm obs}) {\rm d} V$.
 We can estimate the observable quantity, $S_{\rm CO,V}$, from our predicted 
$L_{\rm CO,V}$ as, 

\begin{eqnarray}
S_{\rm CO,V} &=& (1+z)\, \frac{L_{\rm CO,V}}{4\pi\,D^2_{\rm L}}.
\label{EqLCOV}
\end{eqnarray} 

%\noindent Note that $L_{\rm CO,V}=(c/\nu_{\rm rest})\, L_{\rm CO}$. 
The third widely used luminosity in radio observations is the 
brightness temperature luminosity. 
The definition of the rest-frame brightness temperature of an emitting source is  

\begin{equation}
T_{\rm B}(\nu_{\rm obs})=\frac{c^2}{2\, k_{\rm B}}\, \frac{s_{\nu}(\nu_{\rm obs})\, (1+z)}{\nu^{2}_{\rm obs}\, \Omega}.
\label{EqTb}
\end{equation}

\noindent Here $k_{\rm B}$ is Boltzmann's constant and
$\Omega$ is the
solid angle subtended by the source. The brightness temperature in Eq.~\ref{EqTb}
 is an intrinsic quantity, given that the factor $(1+z)$ converts the brightness temperature 
from the observer-frame  to the rest-frame.
In the regime of low frequencies (the Rayleigh-Jeans regime), 
such for as the rotational transitions of CO, in an optically thick medium and with 
thermalised CO transitions, 
the brightness temperature corresponds to the true temperature of the gas. 
The integrated CO line intensity is defined as the velocity-integrated 
brightness temperature,  $I_{\rm CO}=\int T_{\rm B}(\nu_{\rm obs})\, {\rm d} V$. 
The brightness temperature luminosity is then defined as 

\begin{equation}
L^{\prime}_{\rm CO}=I_{\rm CO}\, \Omega\, D^2_{\rm A},
\label{Lprime}
\end{equation}

\noindent where $D_{\rm A}=D_{\rm L}(1+z)^2$ is the angular diameter distance and therefore 
$\Omega\, D^2_{\rm A}$ is the area of the source. From Eqs.~\ref{Scodef}, \ref{EqTb} and \ref{Lprime} 
it is possible to relate $L^{\prime}_{\rm CO}$ to $L_{\rm CO}$, 

\begin{equation}
L^{\prime}_{\rm CO}=\frac{c^3}{8\pi k_{\rm B}\, \nu^3_{\rm rest}}\, L_{\rm CO}.
\label{LprimeLco}
\end{equation}
 
By definition, the relation between $L^{\prime}_{\rm CO}$ and the molecular hydrogen mass 
is parametrised by the factor $\alpha_{\rm CO}$, 

\begin{equation}
L^{\prime}_{\rm CO}=\frac{M_{\rm H_2}}{\alpha_{\rm CO}}.
\label{EqLprime1}
\end{equation}

\noindent Note that here we define $\alpha_{\rm CO}$ in terms of molecular hydrogen mass, 
as done for example by \citet{Tacconi10} and \citet{Genzel10}. However, 
  other authors define $\alpha_{\rm CO}$ in terms of the total molecular gas mass (e.g. \citealt{Solomon05}). 
These two definitions differ by a factor $X_{\rm H}$, the hydrogen mass fraction. 

In Eq.~$6$ in $\S 2.2$ we introduced the relation between 
$I_{\rm CO}$ and the molecular hydrogen column density
$N_{\rm H_2}$, $X_{\rm CO}=N_{\rm H_2}/I_{\rm CO}$. 
Given that $M_{\rm H_2}=m_{\rm H_2}\, N_{\rm H_2}\, \Omega\, D^2_{\rm A}$, 
where $m_{\rm H_2}$ is the mass of a hydrogen molecule, the relation between 
$\alpha_{\rm CO}$ and $X_{\rm CO}$ is simply

\begin{equation}
\alpha_{\rm CO}=m_{\rm H_2}\, X_{\rm CO}.
\end{equation}

\noindent We can therefore estimate the brightness temperature CO luminosity introduced above 
from the molecular hydrogen mass, calculated in 
{\tt GALFORM}, and the $X_{\rm CO}$ conversion factor calculated in the $\tt UCL_{-}PDR$ model 
as,

\begin{eqnarray}
L^{\prime}_{\rm CO}&=&\frac{M_{\rm H_2}}{m_{\rm H_2}\, X_{\rm CO}}.
\label{TbL}
\end{eqnarray}

\noindent $L_{\rm CO}$ and $L_{\rm CO,V}$ are also estimated from $M_{\rm H_2}$ and $X_{\rm CO}$ using 
Eqs~\ref{Lcodef}, \ref{Lcodef2}, \ref{LprimeLco} and \ref{TbL}. 
For a more extended review of all the conversions between units and from CO luminosity 
to molecular mass, see Appendices A~and~B in \citet{Obreschkow09d}. 

To facilitate the comparison with observations, we use $L_{\rm CO,V}$ to
construct the CO luminosity function and $L_{\rm CO}$ to compare against IR luminosity. 
To construct CO flux density maps in $\S 5$, we use the above relations to 
determine the velocity-integrated line flux, $S_{\rm V}$, from $M_{\rm H_2}$ and $X_{\rm CO}$.

Throughout this paper we make extensive comparisons between the CO and IR 
luminosities. In {\tt GALFORM}, we define the total IR luminosity 
to be an integral over the rest-frame wavelength range 8$-$1000~$\mu$m, 
which approximates the total luminosity emitted by interstellar
dust, free from contamination by starlight. To estimate the IR luminosity, we use the method described in \citet{Lacey11}
and \citet{Gonzalez10} (see also \citealt{Lacey11b}, in prep.), which uses a
physical model for the dust extinction at each wavelength to calculate
the total amount of stellar radiation absorbed by dust in each galaxy,
which is then equal to its total IR luminosity. The dust model assumes
a two-phase interstellar medium, with star-forming clouds embedded in
a diffuse medium. The total mass of dust is predicted by  
{\texttt{GALFORM}} self-consistently from the cold gas mass and  
metallicity, assuming a dust-to-gas ratio which is proportional to the
gas metallicity, while the radius of the diffuse dust component is
assumed to be equal to that of the star-forming component, which corresponds to the disk or the bulge half-mass radius depending on whether 
the galaxy is a quiescent disk or a starburst, respectively. 
This dust model successfully explains the Lyman-break galaxy LF up to
$z\sim 10$ (see \citealt{Lacey11}).

\end{document}